\documentclass[
 reprint,
 superscriptaddress,
 amsmath,amssymb,
 aps,
 longbibliography,
 prb
]{revtex4-2}

\usepackage[caption=false]{subfig}
\usepackage{graphicx}
\usepackage{dcolumn}
\usepackage{bm}
\usepackage[dvipsnames]{xcolor}
\usepackage[normalem]{ulem}
\usepackage{comment}
\includecomment{Luca}
\usepackage{comment}
\usepackage {hyperref}
\hypersetup{
    colorlinks = true,
    linkcolor = blue,
    anchorcolor = blue,
    citecolor = blue,
    filecolor = blue,
    urlcolor = blue
    }
\DeclareMathAlphabet{\mathpzc}{OT1}{pzc}{m}{it}

\newcommand{\suppname}{Supplemental Material}

\begin{document}

\preprint{DOI: 10.1103/wpsj-rqsl}

\title{Structural and physical properties of gyromorphs and disordered stealthy hyperuniform media}


\author{Murray Skolnick}
\affiliation{Princeton Materials Institute, Princeton University, Princeton, New Jersey 08544, USA}%

\author{Riccardo Franchi}
\affiliation{Department of Electrical \& Computer Engineering, Boston University, Boston, Massachusetts 02215, USA}

\author{Luca Dal Negro}
\email{dalnegro@bu.edu}
\affiliation{Department of Electrical \& Computer Engineering, Boston University, Boston, Massachusetts 02215, USA}
\affiliation{Department of Physics, Boston University, Boston, Massachusetts 02215, USA}
\affiliation{Division of Materials Science \& Engineering, Boston University, Brookline, Massachusetts 02446, USA}

\author{Paul J. Steinhardt}
\affiliation{Department of Physics, Princeton University, Princeton, New Jersey 08544, USA}

\author{Salvatore Torquato}
\email{torquato@electron.princeton.edu}
\affiliation{Department of Chemistry, Princeton University, Princeton, New Jersey 08544, USA}
\affiliation{Department of Physics, Princeton University, Princeton, New Jersey 08544, USA}
\affiliation{Princeton Materials Institute, Princeton University, Princeton, New Jersey 08544, USA}
\affiliation{Program in Applied and Computational Mathematics, Princeton University, Princeton, New Jersey 08544, USA}

\date{Received 25 June 2026; revised 27 July 2026; accepted 28 July 2026}

\begin{abstract}

Disordered stealthy hyperuniform materials exhibit the statistical isotropy of a liquid combined with some of the structural properties of a crystal---including homogeneity, suppressed density fluctuations at large length scales, bounded holes, and a structure factor $S(k)$ that vanishes for a finite range of wave vectors, $0<k\le K$.
Numerous studies have shown that this combination leads to highly unusual physical properties for a disordered medium, including optical transparency, effective delocalization, ultrafast spreadability, optimal diffusion coefficients and conductivities, and complete isotropic photonic band gaps.
Gyromorphs, point patterns whose structure factor includes rings of intense Bragg-like peaks arranged with discrete $G$-fold rotational symmetry, were recently introduced as counterexamples: disordered media that can somehow achieve the same physical properties, in some cases with higher performance, without stealthiness or hyperuniformity.
In this paper, we resolve the puzzle of how gyromorphs fit consistently with the stealthy hyperuniform studies.
We first show that gyromorphs are actually hyperuniform and, in the large-$G$ limit where they become nearly isotropic, they belong to the weakest form of hyperuniformity, known as class III.
Thus, gyromorphs should have comparatively degraded physical properties compared to stealthy hyperuniform media, which belong to the strongest form of hyperuniformity, known as class I.
We verify this expectation using the rigorous spectral Green's-matrix method for the calculation of the density of states and Purcell factors in large arrays of electric dipoles.
We find that gyromorphs display size-dependent pseudogaps richly populated by localized states rather than smooth band gaps like those found for highly stealthy hyperuniform (class I) materials or in deterministic structures such as Vogel spirals and triangular lattices.
Furthermore, we predict similar disorder-induced degradation relative to stealthy hyperuniformity with regard to transparency, spreadability, and diffusion properties.

\end{abstract}

\maketitle


\section{\label{sec:intro}Introduction}

Disordered hyperuniform materials that combine the statistical isotropy of a liquid
with selected structural and physical properties of a crystal have
emerged as a promising frontier for discovering novel media with unique functional capabilities. Such materials
lack the long-range translational order and discrete rotational
symmetry of crystals and quasicrystals, yet can exhibit suppressed
long-wavelength density fluctuations, bounded holes, and other structural
features that distinguish them sharply from a Poisson distribution
of scatterers \cite{To03a,To18a}. The resulting physical properties---including optical
transparency \cite{Le16,Fr17,To21a,Ki23,Ki24,Ri25,Kl26}, effective delocalization of waves \cite{Fr17,Va26}, ultrafast
spreadability \cite{To21d}, optimal effective diffusion coefficients \cite{Zh16b} and conductivities \cite{To18c}, and complete isotropic
photonic band gaps \cite{Fl09b,Kl22}---have motivated a sustained effort to classify
disordered media according to the strength and character of their
structural correlations, and to understand how those correlations
translate into the fidelity of the physical properties they
enable~\cite{To18a,Fl09b}.

A central organizing framework for this classification is provided by
\textcolor{black}{the quantification of large-scale density fluctuations of disordered hyperuniform or nonhyperuniform systems.}
Point patterns are hyperuniform if their structure factor $S(\bf{k})$ vanishes as  $|\mathbf{k}|\equiv k$ tends to zero. If the structure factor approaches zero as a power law, this corresponds to $S(k) \sim k^{\alpha}$ as $k \to
0$, where $\alpha>0$~\cite{To03a,To18a}. \textcolor{black}{(The case $\alpha=0$ corresponds to typical disordered nonhyperuniform systems \cite{To18a,To21c}.)} Hyperuniform
systems are conventionally sorted into three classes of increasing
disorder: class~I ($\alpha > 1$), the strongest form, with the slowest
possible growth of number variance and the most crystal-like
suppression of long-wavelength fluctuations; class~II ($\alpha = 1$),
with logarithmic corrections; and class~III ($0 < \alpha < 1$), the
weakest form, in which long-wavelength fluctuations are suppressed
relative to a Poisson process but considerably less so than in
class~I. Various physical
properties of disordered media depend sensitively on $\alpha$ and  improve monotonically as $\alpha$ increases 
\cite{To21a,To21d,Ki23}.

Stealthy hyperuniformity is a special case within class~I in
which $S(k)$ is constrained to vanish exactly over a continuous range
of small wave vectors $0 < k \le K$; the parameter $\chi$ quantifies
the fraction of wave-vector modes constrained to have $S(k) = 0$ relative to the total number of degrees of freedom, and
hence the strength of the long-wavelength
constraint~\cite{Uc04b,Ba08,To15,To18a}. 
Disordered stealthy ground states can be obtained as  highly degenerate ground states
of a certain long-ranged pair potential, provided
that $\chi <1/2$ in two and three dimensions \cite{To15}. 
\textcolor{black}{When disordered stealthy hyperuniform systems are realized as heterogeneous media, their resulting physical properties have been found to be far superior to those of other correlated disordered media, including nonstealthy hyperuniform media~\cite{Ki24b}. In $d$ dimensions, their physical properties—including spreadability, diffusion, conduction, mechanical response, and transparency~\cite{Ki23,To21d,Zh16b,To18c,Sk23,Sk25a,Va25}—improve monotonically as $\chi$ approaches its upper limit within the disordered regime. Furthermore, these properties are generally superior to those of nonstealthy hyperuniform systems characterized by a positive and bounded scaling exponent $\alpha$.}

The connection between this structural classification and photonic
band-gap formation in two-dimensional \textcolor{black}{(2D)} disordered network solids was
investigated systematically by Klatt, Steinhardt, and
Torquato~\cite{Kl22}, who compared a broad
range of disordered networks with varying degrees of local and global
order---including random sequential addition, equiluminous, stealthy
nonhyperuniform, antihyperuniform, and perfect glass structures---
and established a framework for ranking them according to the width
and functional form of their photonic band tails. They found that the
gap in the density of states exhibits exponential tails and the
apparent photonic band gaps close rapidly as the system size increases
for nearly all disordered networks considered. The only exceptions
were sufficiently stealthy hyperuniform cases, for which the band gaps
remained open and the band tails exhibited a power-law scaling
reminiscent of the photonic band-gap behavior of photonic crystals in
the thermodynamic limit. The picture that emerges is one in which
disordered media form a graded hierarchy: stealthy hyperuniform
systems sit at the top, with true band gaps and crystal-like band-tail
behavior, while weaker forms of structural order yield progressively
shallower pseudogaps that vanish in the large-system limit.

Against this backdrop, Casiulis, Shih, and
Martiniani~\cite{Ca25} recently introduced
gyromorphs: disordered point patterns whose structure factor includes
rings of intense, Bragg-like peaks arranged with discrete $G$-fold
rotational symmetry, becoming effectively isotropic in the large-$G$
limit. They reported that, in two dimensions and at low refractive
index contrast, gyromorphs outperform stealthy hyperuniform media,
quasicrystals, and Vogel spirals in the formation of isotropic
photonic band gaps. Taken at face value, this claim is striking: 
gyromorphs are not stealthy, and their structure-factor constraints
are imposed at specific Bragg-like wave vectors rather than over a
continuous range at small $k$. How, then, do they fit consistently
within the structural hierarchy established by the hyperuniformity
classification and by the band-tail analysis of
Ref.~\cite{Kl22}?

In this paper, we address that question directly. \textcolor{black}{We first characterize 2D gyromorphs structurally. Importantly, in contrast to disordered stealthy hyperuniform systems, gyromorphs
are not generated under periodic boundary conditions; rather
these nonperiodic point configurations are created within a 
square under free boundary conditions. We begin by examining the large-$R$ asymptotic scaling of their local number variance $\sigma^2_N(R)$.
We find that gyromorphs are in fact class III hyperuniform (not noticed in Ref.~\cite{Ca25}), the weakest form of hyperuniformity,
which enables us to extract the hyperuniformity scaling exponent $\alpha$ that lies between zero and unity.} Furthermore, in the large-$G$ limit in which they become effectively isotropic---the regime relevant to the comparisons of Ref.~\cite{Ca25}---we find \textcolor{black}{statistically that their ensemble-averaged values of $\alpha$ generally decrease as $G$ increases.} 
\textcolor{black}{We also provide numerical evidence that such high-$G$ gyromorphs do not possess the bounded hole property, which is central to opening robust photonic band gaps \cite{To18a,Kl22}}. 
This operating regime is precisely that of the effectively isotropic gyromorph examples emphasized in Ref.~\cite{Ca25}. 
Within the hyperuniformity framework, this classification implies that gyromorphs should exhibit physical properties degraded relative to class~I stealthy hyperuniform media. 
For example, \textcolor{black}{it follows} that gyromorphs cannot exhibit the ultrafast \textcolor{black}{spreadability}, transparency, optimal diffusion constants, and conductivity of stealthy hyperuniform systems, as detailed in Sec. \ref{sec:other-properties}. 

\textcolor{black}{The photonic consequences of this structural classification of gyromorphs motivate us
to study their band-gap formation using two complementary optical
settings and methods. First, we employ a well-established full-wave density-of-states methodology via the MIT Photonic Bands code (mpb)~\cite{MPB}
to analyze two different 2D
structures under periodic boundary conditions:
a high-$\chi$ stealthy hyperuniform system and a class III random organization system \cite{Chaik08, He15, Ma19} whose hyperuniformity exponent is comparable to that of the effectively isotropic $G=60$ gyromorphs. 
In each case, we consider high-dielectric disks in air
with disk centers taken from the aforementioned systems. 
We find that the stealthy hyperuniform system supports a \textcolor{black}{robust} transverse magnetic (TM) band gap, whereas the class III random organization system exhibits only a shallow pseudogap.}

\textcolor{black}{Second, because plane-wave methods like mpb cannot be used to investigate
nonperiodic media, we analyze gyromorphs directly via powerful spectral Green's-matrix methods correct to all orders of multiple scattering for finite open arrays of resonant point dipoles
as well as Purcell-factor calculations. 
This methodology is perfectly suitable to study the nonperiodic gyromorph samples because it does not impose periodic boundary conditions and naturally probes the collective scattering resonances of finite open arrays. 
The features identified in Ref.~\cite{Ca25} as isotropic band gaps in gyromorphs are found here to be fully resolved pseudogaps filled with localized states rather than true band gaps. 
Thus, both optical settings and methods support the same qualitative conclusion: Class III hyperuniform structures, whether random organization models or gyromorphs, offer shallow pseudogaps at best, in vivid contrast to high-$\chi$ stealthy hyperuniform structures with robust gap formation.}


For completeness, we also consider in the appendixes triangular lattices and Vogel spirals in our comparison, since they were benchmarks in Ref.~\cite{Ca25}, even though they lie outside the framework of statistically homogeneous disordered media: a Vogel spiral has a distinguished central point and a radially varying local environment, and so is not statistically homogeneous in the strict sense. We deliberately do not include quasicrystals in our comparisons, as they possess long-range quasiperiodic translational order and exact discrete rotational symmetry, placing them in a structurally distinct category from the disordered isotropic and effectively isotropic media considered here.  

The remainder of the paper is organized as follows: In Sec.~\ref{sec:background} we summarize the structural descriptors used in this work. The model systems and optical methods are described in Sec.~\ref{sec:background_model_systems_and_optical_methods}. In Sec.~\ref{sec:structural_results}, \textcolor{black}{the large-scale density fluctuations and hole-size statistics of gyromorphs across $G$ values are compared to those of stealthy hyperuniform systems using the local number variance $\sigma_N^2(R)$ and void nearest-neighbor probability density $H_V(r)$, respectively.} 
Section~\ref{sec:optical_results} presents the optical calculations in two separate settings: first, full-wave mpb density-of-states calculations \textcolor{black}{for finite-size dielectric disk arrays derived from random organization and stealthy hyperuniform systems, and second, Green's-matrix calculations for finite open gyromorphic and stealthy point-dipole arrays}. 
In Sec. \ref{sec:other-properties}, we compare and contrast other physical properties of gyromorphs and disordered stealthy 
hyperuniform media. 
We summarize and discuss implications \textcolor{black}{of our work} in Sec.~\ref{sec:discussion}. \textcolor{black}{In the appendixes, we further compare
the optical properties of gyromorphs and stealthy systems
to those of Vogel spirals and the triangular lattice.}


\section{\label{sec:background} Structural Descriptors: Definitions and Preliminaries}

Here, we provide basic definitions of the structural 
descriptors that we employ in this work: local number variance
and the void nearest-neighbor functions. We compute
these descriptors to vividly distinguish the 
differences in the structural characteristics of the models considered in this work.
\smallskip

\subsection{\label{sec:hyperuniformity}Hyperuniformity}

\textcolor{black}{For a large class of many-particle disordered and ordered systems, hyperuniformity can be defined by the condition that the structure factor $S(k)$ vanishes as the wave number $k$ tends to zero.
More generally, a hyperuniform system
is one in which the local number variance $\sigma_N^2(R) \equiv \langle N(R)^2\rangle-\langle N(R)\rangle^2 $
associated with a $d$-dimensional spherical window of radius $R$
grows more slowly than $R^d$ \cite{To03a,To18a},
where $N(R)$ is the number of points within the window
and angular brackets denote an ensemble average.
For statistically homogeneous point patterns in $d$ dimensions, garden-variety disordered nonhyperuniform systems have $\sigma_N^2(R)\sim R^d$.}

\textcolor{black}{
Consider systems in which the structure factor has a power-law form in the vicinity of the origin, i.e., $S(k)\sim k^\alpha$.
For hyperuniform systems, the exponent $\alpha$ is positive ($\alpha >0$) and its value determines three different large-$R$ scaling behaviors of the number variance  \cite{To03a,To18a}:
\begin{equation}
\sigma_N^2(R)\sim
\begin{cases}
R^{d-1}, & \alpha>1 \quad \text{(class I)},\\
R^{d-1}\ln R, & \alpha=1 \quad \text{(class II)},\\
R^{d-\alpha}, & 0<\alpha<1 \quad \text{(class III)}.
\end{cases}
\end{equation}\label{eqn:asymptotics}
Classes I and III describe the strongest
and weakest forms of hyperuniformity, respectively.
We note that all stealthy hyperuniform systems belong to class I \cite{To15}. It is seen that while the value of $\alpha$ can be extracted from the large-$R$ scalings of $\sigma_N^2(R)$ for classes
II and III, this is not the case for class I. By contrast, 
the cases $\alpha=0$ and $-d < \alpha <0$ respectively correspond to
typical disordered nonhyperuniform systems (e.g., ordinary liquids
and glasses) and antihyperuniform systems (e.g., thermal critical points) \cite{To18a,To21c}.}





\subsection{\label{sec:hole_statistics}Void nearest-neighbor statistics}

\textcolor{black}{To characterize the distribution of spherical 
``holes'' (spherical regions empty of points) in the point configurations considered here, we compute the void nearest-neighbor probability density function $H_V(r)$ \cite{To02a,To10d}. 
For a statistically homogeneous point pattern in $d$-dimensional Euclidean space $\mathbb{R}^d$, $H_V(r)dr$ is the probability that, from an arbitrarily placed test point, the nearest particle center lies at a distance between $r$ and $r+dr$ \cite{To02a}. 
Because $H_V(r)$ has dimensions of inverse length, we report the dimensionless quantity $\rho^{-1/2}H_V(r)$, where $\rho$ is the
number density.}

\textcolor{black}{Systems in which $H_V(r)$ has infinite support in the thermodynamic limit must possess arbitrarily large holes \cite{To02a,Ki19b}. On the other hand, $H_V(r)$ for infinite systems 
with bounded holes have compact support; i.e., 
there exists a finite covering radius $R_c$ such that every point in space lies within distance $R_c$ of some particle center and hence $H_V(r)=0$ for all $r>R_c$ \cite{To10d}.  
All crystals have bounded holes as do  disordered stealthy hyperuniform systems  \cite{Zh17a,Gh18}.
By contrast, nonstealthy hyperuniformity does not guarantee the bounded-hole property \cite{Zh17a,To18a}. }


\section{\label{sec:background_model_systems_and_optical_methods}{Model Systems 
and Optical Methods}}


\textcolor{black}{In this section, we describe how to generate 2D stealthy hyperuniform and gyromorph point patterns. Subsequently, we summarize the two complementary methods employed to compute their optical properties: full-wave density-of-states calculations for periodic media and Green's-matrix calculations for finite open point-dipole arrays.}
\smallskip

\subsection{\label{sec:models}Models}

\textcolor{black}{
All 2D model structures are rescaled to the same number density $\rho$, and lengths are reported in units of the characteristic interparticle spacing $\rho^{-1/2}$.} 
\smallskip

\subsubsection{Disordered stealthy hyperuniform systems} 

We generate disordered stealthy hyperuniform point patterns via the collective coordinates optimization technique using Poisson distribution initial conditions \cite{Uc04b,Ba08,To15}. 
In this scheme, we employ a limited-memory Broyden-Fletcher-Goldfarb-Shanno (L-BFGS) optimization algorithm. 
The fidelity of our stealthy hyperuniform samples is ensured by checking that the collective coordinate optimization is run until the energy above the ground state is less than $10^{-19}$ times what it was for the initial condition \cite{Uc04b,To15,Zh15a}. 
To remain consistent with the case considered in Ref. \cite{Ca25}, we consider disordered stealthy hyperuniform point patterns with stealthiness parameter $\chi\approx0.49$, where in two dimensions $\chi\equiv K^2/(16\pi\rho)$ in the thermodynamic limit \cite{To15}. 
For the full-wave density-of-states calculations with mpb, we use ensembles of 100 configurations with $N=4000$ particles. 
An image of a circular cutout of such a stealthy hyperuniform point pattern is shown in Fig. \ref{fig:shu_cutout}. 

\subsubsection{Gyromorphs}

\begin{figure}
    \centering
    \subfloat[\label{fig:shu_cutout}]{\includegraphics[width=0.5\linewidth]{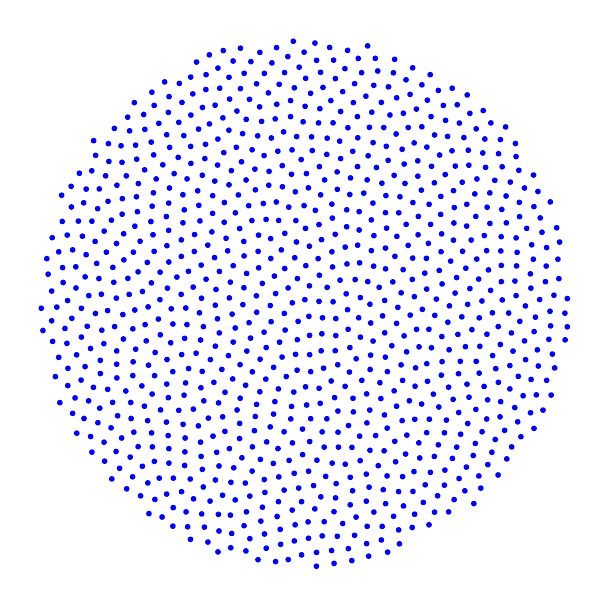}}%
    \subfloat[\label{fig:gyr_cutout}]{\includegraphics[width=0.5\linewidth]{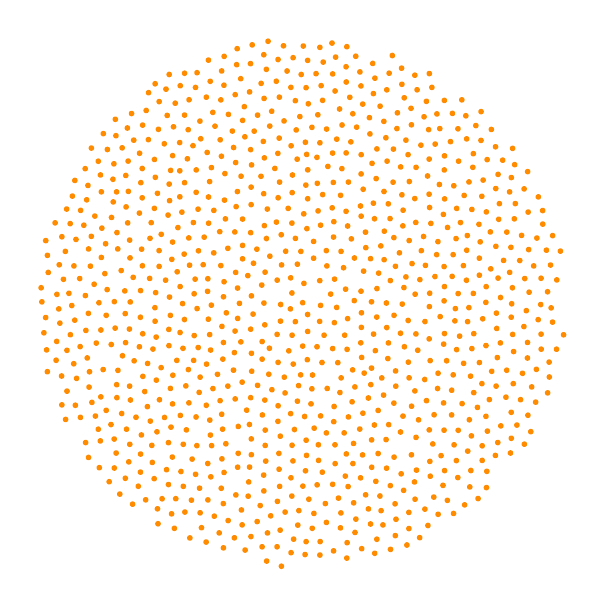}}%
    \caption{Representative images of (a) a stealthy hyperuniform point pattern with $\chi\approx0.49$ and (b) a $G=60$ gyromorph with Fourier ring radius $KL/(2\pi)=100$. The structures in (a) and (b) each contain $N=1000$ points and were prepared by taking circular cutouts of larger systems with $N\approx10^4$. As discussed in the main text, the sample sizes used to ascertain their optical properties are an order of magnitude larger.}
    \label{fig:structures}
\end{figure}

We generate \textcolor{black}{all} gyromorph point patterns using the FReSCo algorithm \cite{Sh24} and following the \textcolor{black}{same optimization procedure} described in Ref.~\cite{Ca25}. 
Gyromorphs are defined by imposing intense Bragg-like peaks on a ring of radius $K$ in reciprocal space, with the peaks arranged according to a prescribed rotational order $G$. 
For the optical calculations, we use the same structural parameters as those employed in the coupled-dipole simulations of Ref.~\cite{Ca25}: rotational order $G=60$, Fourier ring radius $KL/(2\pi)=100$, where $L$ is the system side length, and systems containing $N\approx10^4$ points before taking the finite arrays used in the Green's-matrix calculations. 
For the structural analyses, we consider ensembles (50 individual configurations) of larger gyromorphs with $N\approx10^5$ points and rotational orders $G=8$, $10$, $14$, $18$, $24$, $60$, $100$, $120$, and $200$ at fixed ring radius $KL/(2\pi)=300$. 
Detailed validation of our generated gyromorph structures, including comparison of their structure factors $S(k)$, pair-correlation functions $g_2(r)$, ``gyromorphic'' correlation functions $g_G(r)$, and coupled-dipole calculations with those presented in Ref.~\cite{Ca25}, is provided in the \suppname{}~\cite{note:SI}. 
An image of a circular cutout of a $G=60$ and $KL/(2\pi)=100$ gyromorph is shown in Fig. \ref{fig:gyr_cutout}. 
We show in Sec. \ref{sec:large-scale} that gyromorphs
 are generally hyperuniform of class III.



\subsection{\label{sec:mpb_dos}Full-wave density of states for periodic dielectric arrays of disks in air}

We compute the full-wave photonic density of states for periodic 
two-phase dielectric media (dielectric circular disks in air) by solving Maxwell's equations using mpb \cite{MPB}.
Following Ref.~\cite{Kl22}, the area-normalized density of states is defined as
\begin{equation}
\mathcal{D}(\omega)=\frac{1}{A}\sum_i \delta(\omega-\omega_i),
\end{equation}
where $A$ is the supercell area and $\omega_i$ are the eigenfrequencies computed at the $\Gamma$ point of the periodic supercell.
For finite periodic samples, $\mathcal{D}(\omega)$ is approximated by a histogram and then averaged over an ensemble of statistically independent configurations.

Because gyromorph samples do not have natural periodic boundary conditions, we cannot apply mpb directly to them.
Instead, we use mpb to compare two systems generated under periodic boundary conditions: a high-$\chi$ stealthy hyperuniform point pattern and a class III random organization disk pattern with a hyperuniformity exponent comparable to that of the $G=60$ gyromorphs.
To construct the corresponding photonic structures, we place identical dielectric circular disks (phase 2) at the points of each configuration in an air background (phase 1), which is equivalent
to transversely isotropic high-dielectric aligned
cylinders in air.  The dielectric contrast is $\epsilon_2/\epsilon_1=9$ and the 
disk (cylinder) packing fraction is $\phi_2=0.05$, both values of which match the dilute-scatterer parameters used in Ref.~\cite{Ca25}; all other numerical simulation parameters follow Ref.~\cite{Kl22}.
\color{black}

\subsection{\label{sec:greens_dos}Green's-matrix \textcolor{black}{analysis} for finite open point-dipole arrays}

\textcolor{black}{In order to analyze the optical properties of finite nonperiodic point patterns, we utilize powerful Green's-matrix methods. Specifically, we} 
characterize band-gap formation by computing the density of states (DOS) of large 2D arrays of electric dipoles using the Green's-matrix spectral method for both out-of-plane [transverse magnetic (TM)] and 
in-plane [transverse electric (TE)] dipole resonance \cite{skipetrov2016finite,skipetrov2020finite,monsarrat2022pseudogap}. This approach enables the understanding of the multiple scattering
behavior of general (periodic and nonperiodic) arrays of electric dipoles through the spectral properties of the corresponding Green's matrix \cite{dal2016structural,Sg20,Sg22,Sg21}. In particular, the Green's matrix of open systems is non-Hermitian and the real and imaginary parts of its complex eigenvalues $\Lambda_{m}$ ($m\in 1,2,\ldots,N$) correspond to the frequency detuning $(\omega_{m}-\omega_{0})$ and decay rate $\Gamma_{m}$ (both normalized to the resonant width $\Gamma_{0}$ of an isolated dipole) of the scattering resonances of the investigated system. This approach rigorously accounts for all the multiple scattering orders and the unambiguous determination of the DOS for both TM and TE scenarios. In addition, the optical response of finite-size arrays was also investigated by their local density of states (LDOS) obtained by solving the multiple scattering Foldy-Lax equations within the Green's-matrix method under dipole excitation \cite{Sg19,Sg22,Da21,caze2013strong}. 
In this approximation, each scattering element is treated as a subwavelength resonant electric dipole, valid in the Rayleigh regime $k_0 a_{\rm sc}\ll 1$, where $a_{\rm sc}$ is the physical scatterer radius and $k_0$ is the free-space wave number \cite{Sg19,Sg22,Da21,Ca25,caze2013strong}. 
The material and size dependence of each scatterer is absorbed into its resonant polarizability, while the collective optical response is determined by the multiple electromagnetic coupling among the dipoles. 
To avoid artifacts induced by sharp corners in the spectrum, we consider circular cutouts of these structures (e.g., see those in Fig. \ref{fig:structures}). 

This framework directly probes the scattering resonances of finite open arrays without imposing periodic boundary conditions, making it well suited to gyromorphs, Vogel spirals, and other aperiodic structures \cite{Sg19,Sg22,Ca25}. 
For Vogel spiral arrays, Green's-matrix calculations have been shown to recover the spatial structure of localized band-edge modes previously obtained using full-wave finite-element LDOS simulations \cite{Li11,Sg19,Tr12}, supporting the reliability of the Green's-matrix approach for diagnosing spectral depletion and localized resonances in finite aperiodic arrays. 
For TM polarization, we use the scalar Green's matrix, while for TE polarization we use the corresponding dyadic vector Green's matrix \cite{monsarrat2022pseudogap,Sg19,Sg22}.
Additional technical details concerning the Green's-matrix method are provided in Appendix~\ref{app:DOS_gmatrix}.
The complex eigenvalues of these non-Hermitian matrices encode the normalized detuned frequencies and decay rates of the scattering resonances, from which we compute the optical density of states $\mathcal{N}(\omega)$ \cite{skipetrov2020finite,Sg19,Sg22,monsarrat2022pseudogap}.

\textcolor{black}{Because of the technological relevance of controlling vacuum fluctuations in complex optical nanostructures, we complement the spectral analysis of the DOS with the calculation of the LDOS for both polarizations \cite{Li11,monsarrat2022pseudogap,Kl22}.} 
Specifically, we compute the Purcell factor $\rho({\bf r}_s,\omega)/\rho_0$, i.e., the LDOS at a source position ${\bf r}_s$ normalized by the corresponding homogeneous-medium value, which is enhanced by long-lived localized resonances inside depleted spectral regions \cite{Li11,Da21,monsarrat2022pseudogap} (see Appendix~\ref{app:LDOS} for additional details). 
This combined global and local spectral analysis distinguishes robust gap formation from pseudogap behavior caused by finite residual DOS and localized in-gap resonances. 
\color{black}

\section{Structural Analysis\label{sec:structural_results}}

\textcolor{black}{We now examine the structural properties of gyromorphs relative to disordered $\chi=0.49$ stealthy hyperuniform systems. 
We first use local number statistics to determine the hyperuniformity class of gyromorphs, then analyze their hole-size statistics through the void nearest-neighbor function $H_V(r)$, and finally discuss the implications for spreadability and related diffusion properties. 
This structural characterization sets the stage for the optical comparisons in Sec.~\ref{sec:optical_results}.} 


\subsection{\label{sec:large-scale} Large-scale density fluctuations}

\textcolor{black}{Because gyromorphs are generated as finite nonperiodic samples,
the positions of the observation-window centers must be restricted so that each sampling window remains entirely inside the gyromorph boundary. The number variance provides a boundary-unbiased estimate of the large-scale scaling exponent.
By contrast, the structure-factor plots reported in Ref.~\cite{Ca25} were computed from circular cutouts with a Hamming window; this procedure is useful for visualizing the imposed Bragg-like rings, but it produces an effectively inhomogeneous weighted density field and therefore cannot be used to diagnose hyperuniformity or extract $\alpha$ (see the \suppname~\cite{note:SI}).}

\begin{figure}
    \centering
    \includegraphics[width=1.0\linewidth]{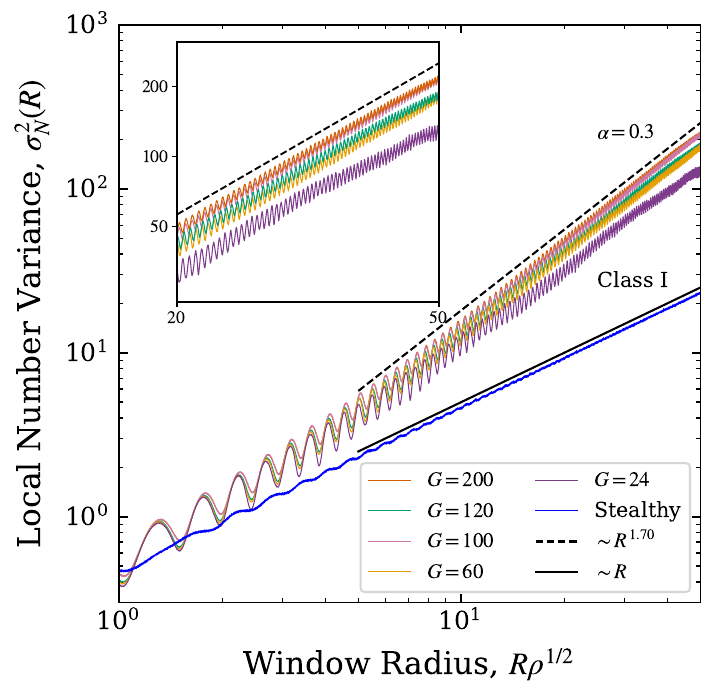}
    \caption{\textcolor{black}{Log-log plots of the local number variance $\sigma_N^2(R)$ for $\chi=0.49$ stealthy hyperuniform and $24\leq G \leq200$ gyromorph systems with Fourier ring radius $KL/(2\pi)=300$. \textcolor{black}{The inset highlights the large-$R$ class III hyperuniform scaling of the gyromorphs observed for $5\leq R\rho^{1/2} \leq50$.} Note that only the local variance $\sigma_N^2(R)$ curves for higher-$G$ gyromorphs, which are effectively isotropic, are plotted here in order to keep them on equal footing with the truly isotropic stealthy hyperuniform systems.}}
    \label{fig:variance}
\end{figure}

The ensemble-averaged local number variance $\sigma_N^2(R)$ of $\chi=0.49$ stealthy hyperuniform and $24\leq G\leq200$ gyromorph systems with $N\approx10^5$ particles are plotted in Fig.~\ref{fig:variance}. 
\textcolor{black}{Over the accessible large-$R$ fitting range, $5\leq R\rho^{1/2}\leq50$}, the stealthy hyperuniform system exhibits class I window perimeter scaling, i.e., $\sigma_N^2(R)\sim R$, whereas the ensemble-averaged gyromorph number variances exhibit weaker class III hyperuniform scaling with $\sigma_N^2(R)\sim R^{2-\alpha}$, where $0<\alpha<1$. 
\textcolor{black}{We separately fit the number variance of each realization to obtain a finite-size estimate of $\alpha$. Table~\ref{tab:hu_exponents} reports the mean and standard deviation of these estimates across the ensemble.} 
While we also attempted reciprocal-space estimates of $\alpha$ from gyromorph configurations without a Hamming window or the circular cutout construction, the small-$k$ behavior was not stable enough to yield an unambiguous exponent, which we attribute to the absence of periodic boundary conditions in gyromorphs. 
The lower-rotational-order gyromorphs, such as $G=8$ and $G=10$, have the largest \textcolor{black}{ensemble-averaged} exponents, $\alpha=0.89\pm0.11$ and $\alpha=0.88\pm0.11$, respectively. 
By contrast, the higher-rotational-order gyromorphs most relevant to the effectively isotropic optical comparisons have smaller exponents \textcolor{black}{on average}, with $0.32\lesssim\alpha\lesssim0.45$ for $G=60$, $100$, $120$, and $200$. 
Overall, increasing the rotational order $G$ improves the effective angular isotropy of the gyromorphs, but is \textcolor{black}{generally} accompanied by a substantially weaker suppression of large-scale density fluctuations. 
By contrast, the stealthy hyperuniform system combines strong class I suppression of large-scale fluctuations with statistical isotropy controlled independently by the stealthiness parameter $\chi$.


\textcolor{black}{The realization-to-realization fluctuations in gyromorph $\alpha$ exponents are appreciable. Moreover, while all fitted values of $\alpha$ through $G=24$ are positive, a minority of those in the $G\geq60$ ensembles are negative. Such configurations exhibit effectively antihyperuniform scaling over the accessible fitting range, but a negative finite-size fit does not by itself establish antihyperuniformity in the infinite-system limit.} 
We attribute \textcolor{black}{such ensemble fluctuations in large-scale gyromorph structure} to the fact that the small-$k$ region of gyromorphs, and hence their large-scale density fluctuations, are not explicitly constrained by the imposed ring of Bragg-like peaks during generation. 
While $\alpha$ cannot be extracted from the variance for class I systems, we verified the class I hyperuniformity of the stealthy systems by fits to the power-law $R^{\beta}$ which yielded $\beta=1.00\pm0.00$. 
The small ensemble standard deviation of $\beta$ for the stealthy systems is attributable to the fact that the small-$k$ structure of such systems is explicitly constrained when generating them. Overall, these findings highlight the importance of considering ensemble statistics when diagnosing hyperuniformity---especially when considering weakly hyperuniform class III systems. 
\textcolor{black}{As we will show in Appendix \ref{app:optics}, these realization-to-realization fluctuations in the large-scale structure of gyromorphs translate directly into appreciable fluctuations in their TM optical spectra, whereas only small fluctuations are observed \textcolor{black}{in the TM spectra} for stealthy systems.}
While the extracted exponents should be interpreted as finite-size estimates, the large system sizes considered here, $N\approx10^5$, and the ensemble averaging provide strong evidence for class III scaling \textcolor{black}{in the present systems and likely even larger realizations of gyromorphs.} 


\begin{table}
\caption{\textcolor{black}{Mean values and standard deviations calculated from the set of hyperuniformity exponents $\alpha$ obtained by separately fitting the large-$R$ scaling of the number variance $\sigma_N^2(R)$ for each realization over the fitting interval $5\leq R\rho^{1/2} \leq50$. For each gyromorph realization, $\alpha$ is obtained by fitting the power law $CR^{\beta}$ and then taking $\alpha=2-\beta$. \textcolor{black}{For all systems, we found that changing the fitting interval within reasonable bounds changes the ensemble-averaged exponent by less than $5\%$.}}}
\label{tab:hu_exponents}
\centering
\begin{tabular*}{\columnwidth}{@{\extracolsep{\fill}}lc}
\hline
System & $\alpha$ \\
\hline
Stealthy, $\chi=0.49$ & $-$ \\
Gyromorph, $G=8$ & $0.89\pm0.11$\\
Gyromorph, $G=10$ & $0.88\pm0.11$\\
Gyromorph, $G=14$ & $0.56\pm0.21$\\
Gyromorph, $G=18$ & $0.63\pm0.16$\\
Gyromorph, $G=24$ & $0.49\pm0.19$\\
Gyromorph, $G=60$ & $0.42\pm0.31$\\
Gyromorph, $G=100$ & $0.34\pm0.25$\\
Gyromorph, $G=120$ & $0.43\pm0.27$\\
Gyromorph, $G=200$ & $0.38\pm0.30$\\
\hline
\end{tabular*}
\end{table}

\subsection{Hole-size statistics\label{sec:holes_results}}

\begin{figure}
    \centering
    \includegraphics[width=1.0\linewidth]{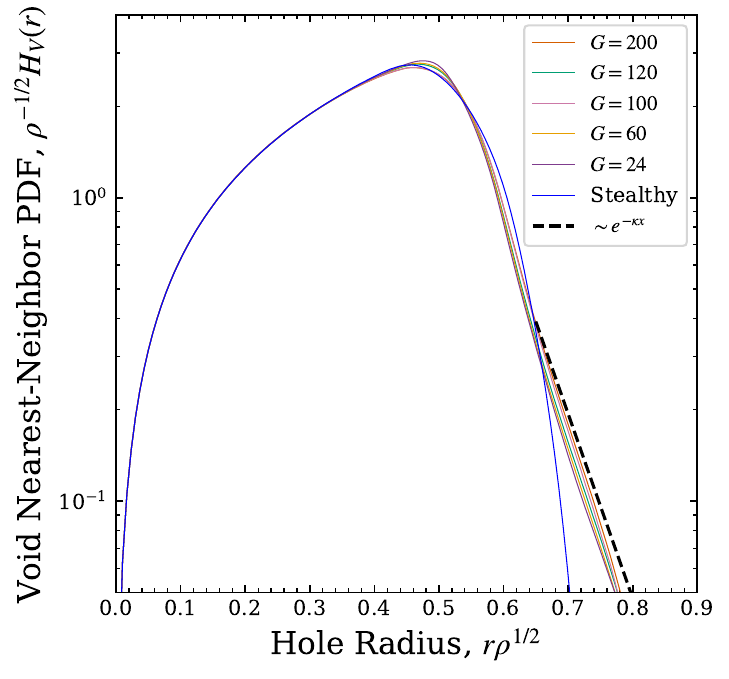}
    \caption{\textcolor{black}{Semilogarithmic plots of the void nearest-neighbor probability density $H_V(r)$ for stealthy hyperuniform systems with $\chi=0.49$, and $24\leq G\leq 200$ gyromorphs with Fourier ring radius $KL/(2\pi)=300$. The plotted quantity is $\rho^{-1/2}H_V(r)$. The dashed black lines highlight the exponential-like large-$r$ tails of the $G=24,60,120$ gyromorphs. Note that only the void nearest-neighbor probability density function $H_V(r)$ curves for higher-$G$ gyromorphs, which are effectively isotropic, are plotted here in order to keep them on equal footing with the truly isotropic stealthy hyperuniform systems.}}
    \label{fig:Hv}
\end{figure}

The ensemble-averaged void nearest-neighbor probability density functions $H_V(r)$ for the stealthy hyperuniform system and representative $24\leq G\leq 200$ gyromorphs are plotted in Fig.~\ref{fig:Hv}.
While $H_V(r)$ for the stealthy systems rapidly terminates at a finite radius, consistent with the bounded-hole property of such systems \cite{Gh18}, $H_V(r)$ for the gyromorphs exhibits an exponential-like large-$r$ tail for $r\rho^{1/2}\gtrsim0.7$ before abruptly dropping to zero due to finite-size effects. 
Thus, the exponential-like large-$r$ tails in $H_V(r)$ for the large gyromorph samples ($N\approx10^5$) provide strong numerical evidence against compact support over the length scales probed, as is common in nonstealthy hyperuniform structures \cite{Zh17a,To18a,Ki19b,note:poisson_Hv}.
The observed weak hyperuniformity and apparent lack of bounded holes in gyromorphs motivate further examination of their optical and other physical properties.

\color{black}


\section{\label{sec:optical_results} Optical Properties}

\subsection{\label{sec:band_tails} Density of states for periodic dielectric arrays of disks in air}

We use the mpb code described in Sec.~\ref{sec:mpb_dos} to compute the density of states for stealthy hyperuniform disk patterns with $\chi \approx 0.49$ and class III random organization disk patterns at {\it criticality}, whose hyperuniformity scaling exponent $\alpha \approx 0.42$. 
The packing fraction $\phi_2$ in all cases is taken to be 0.05
to match the gyromorph value. 
We emphasize that both models  are prepared under periodic boundary conditions. 
The critical random organization model provides a useful class III benchmark because its hyperuniformity exponent, $\alpha \approx 0.42$, is the same as the ensemble-averaged $\alpha$ found for the $G=60$ gyromorphs. 
In the random-organization model, overlapping disks at a given time step are identified as active and displaced randomly, while nonoverlapping disks remain fixed. For the monodisperse disk random organization model considered here, the critical packing fraction is $\phi_{2c} \approx 0.4065$ \cite{He15,Ma19}. 
The diameters of these disks are shrunk so that
 $\phi_2=0.05.$

\begin{figure}
    \centering
    \includegraphics[width=1.0\linewidth]{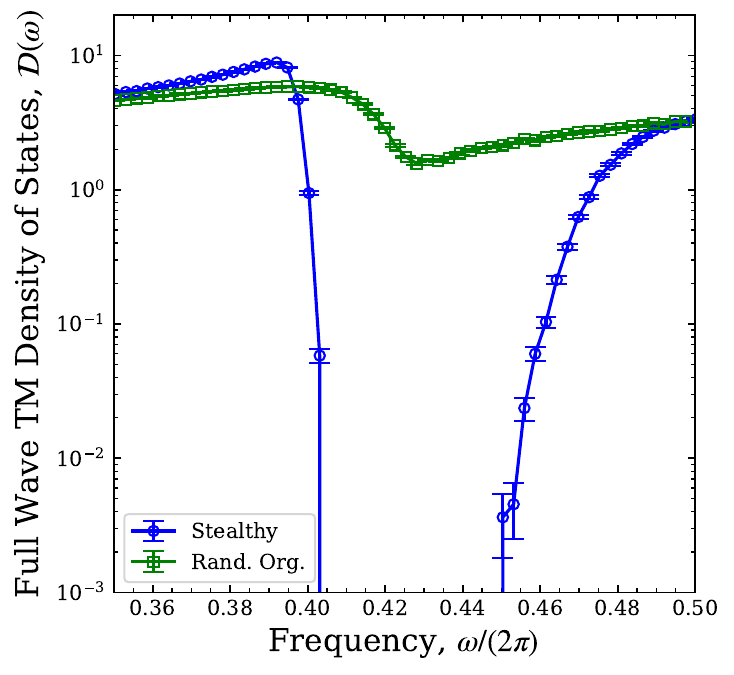}
    \caption{\textcolor{black}{Area-normalized full-wave TM density of states, $\mathcal{D}(\omega)$, at the $\Gamma$ point for periodic arrays of identical high-dielectric disks in air generated from stealthy hyperuniform systems with $\chi\simeq0.49$ and class III random-organization disk patterns with scaling exponent $\alpha=0.42$. The dielectric contrast is $\epsilon_2/\epsilon_1=9$ and the disk packing fraction is $\phi_2=0.05$. The stealthy hyperuniform system exhibits a TM band gap, with $\mathcal{D}(\omega)=0$ over the indicated frequency interval at the resolution considered, whereas the random organization system exhibits only a shallow pseudogap with finite $\mathcal{D}(\omega)$.}}\label{fig:mpb_dos}
\end{figure}

The results in Fig.~\ref{fig:mpb_dos} show 
that the periodic disordered stealthy hyperuniform disk arrays exhibit a clear TM band gap, with a finite frequency interval over which no eigenstates are found and hence $\mathcal{D}(\omega)=0$ at the system size and resolution considered. 
By contrast, the periodic class III random organization disk arrays exhibit only a shallow pseudogap: $\mathcal{D}(\omega)$ is depressed for nearly the same frequency range, but it remains finite. 
This comparison establishes that while weak class III hyperuniform media can produce shallow spectral depletion, 
they are far inferior to the robust TM gap achieved by  high-$\chi$ stealthy hyperuniform systems.

\color{black}

\subsection{\label{sec:greens_matrix} Finite open point-dipole arrays}

\begin{figure*}[t]
    \centering
    \includegraphics[width=\linewidth]{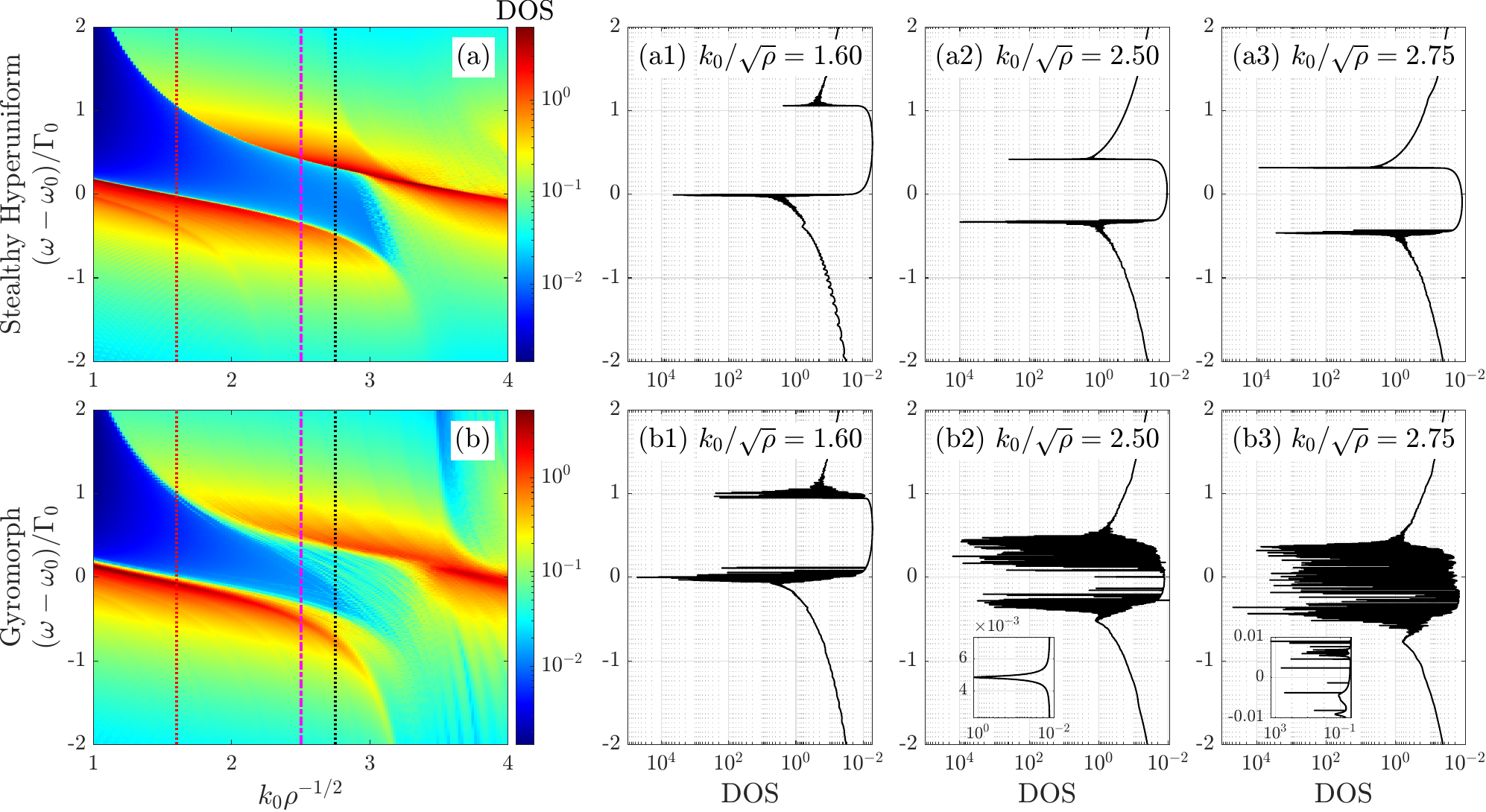}
    \caption{TM DOS maps for (a) stealthy hyperuniform and (b) gyromorph. The maps are plotted as a function of the normalized wave number $k_{0}\rho^{-1/2}$ and the normalized angular frequency $(\omega - \omega_{0})/\Gamma_{0}$. Corresponding line cuts are extracted from the DOS maps at [(a1), (b1)] $k_0\rho^{-1/2} = 1.60$ (dotted red line), [(a2), (b2)] $2.50$ (dash-dotted magenta line), and [(a3), (b3)] $2.75$ (dotted black line).
    Calculations are performed for arrays of $N=6500$ dipoles, with a normalized angular frequency resolution of $8\times 10^{-3}$ in the DOS maps and $10^{-7}$ in the line cuts.
    }
    \label{fig:TM_DOS_map}
\end{figure*}

We now turn to the photonic setting directly relevant to gyromorphs: finite open arrays of resonant point dipoles. 
Here, we use the accurate Green's-matrix method to compare gyromorphs directly with finite circular cutouts of stealthy hyperuniform point patterns under the same point-dipole framework. 
In order to systematically investigate the band-gap properties of the structures, we computed the phase diagrams of the DOS as a function of frequency and normalized wave numbers $k_{0}\rho^{-1/2}$ over a broad range of values. Here $k_{0}$ is the free-space wave number \cite{monsarrat2022pseudogap}. 
We remark that the normalized wave number is inversely related to the optical density $OD=\rho\lambda^{2}$ of the arrays according to: $OD=4\pi^{2}/\gamma^{2}$ and $\gamma=k_{0}\rho^{-1/2}$. 

\begin{figure}
    \centering
    \includegraphics[width=\linewidth]{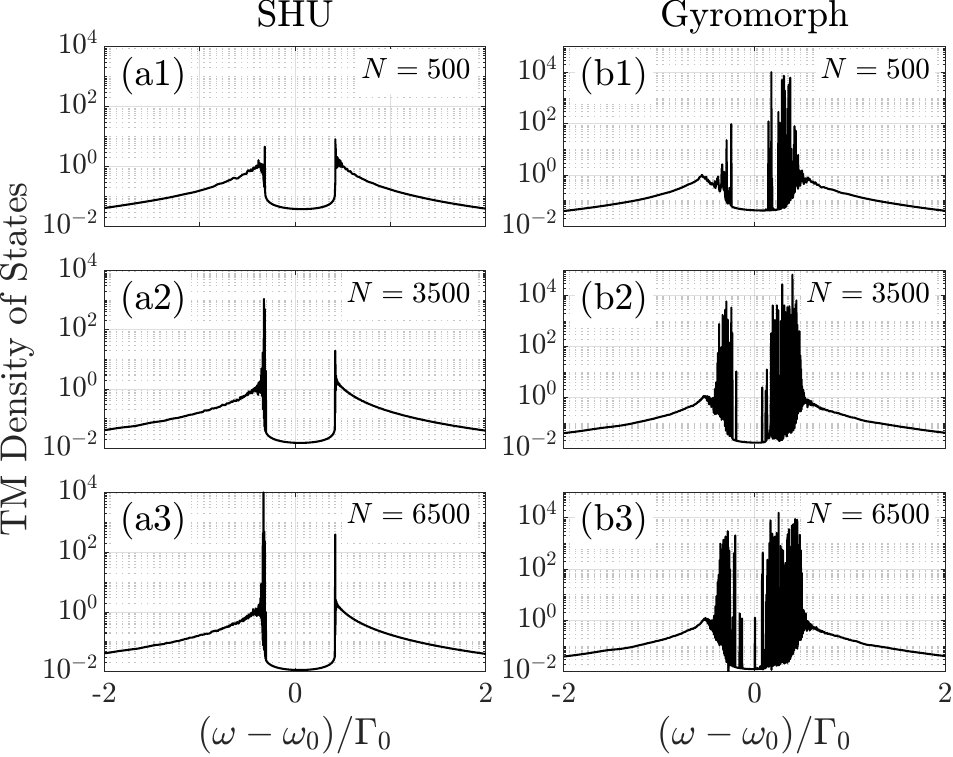}
    \caption{TM DOS as a function of the normalized angular frequency $(\omega-\omega_0)/\Gamma_0$ for different numbers of dipoles, $N$: [(a1)--(a3)] stealthy hyperuniform (SHU) and [(b1)--(b3)] gyromorph structures, respectively, evaluated at $N = 500$, $3500$, and $6500$. The resolution in $(\omega-\omega_0)/\Gamma_0$ is $10^{-7}$. The normalized wave number is set to $k_0\rho^{-1/2} = 2.5$.}
    \label{fig:TEandTM_DOSvsN_SHU_G}
\end{figure}

The TM optical densities of states for the stealthy hyperuniform and gyromorph point-dipole arrays are shown in Figs.~\ref{fig:TM_DOS_map}(a) and \ref{fig:TM_DOS_map}(b), respectively.
The stealthy hyperuniform array exhibits a smooth and wide TM spectral gap (free of any localized states), whereas the gyromorph array exhibits a much narrower and highly fragmented TM pseudogap. 
The high-resolution line cuts across detuned frequency at selected optical densities in Figs.~\ref{fig:TM_DOS_map}(b1--b3) show that the gyromorph pseudogap is populated by a large density of states near its fragmented band edges and deep inside the depleted spectral region, in stark contrast to the smooth TM spectral gap of the stealthy hyperuniform array [see Figs.~\ref{fig:TM_DOS_map}(a1--a3)]. 
As highlighted in the inset panels of Figs.~\ref{fig:TM_DOS_map}(b2) and \ref{fig:TM_DOS_map}(b3), the dense black regions consist of many individually resolved Lorentzian lineshapes, confirming that the gyromorph pseudogap is densely populated by narrow resonant modes. 
The finite-size scaling of the TM DOS for $k_0\rho^{-1/2}=2.5$, shown in Fig.~\ref{fig:TEandTM_DOSvsN_SHU_G}, further confirms the distinct optical properties of stealthy and gyromorphic media: The stealthy TM gap remains robust as the number of dipoles increases, whereas the gyromorph TM pseudogap remains filled with residual states. 

\textcolor{black}{Additionally, a comprehensive analysis of deterministic structures, Vogel spirals, and the triangular lattice, is provided in Appendix~\ref{app:optics}.
We show both TM and TE spectra for Vogel spirals and compare them to stealthy hyperuniform and gyromorphic systems. 
We study Vogel spirals because it was claimed in Ref.~\cite{Ca25} that the quality of their photonic band gaps was inferior to that of gyromorphs. 
On the contrary, we found that the Vogel spirals exhibit cleaner and broader TM gaps than those of gyromorphs. The TE DOS of gyromorphs is likewise filled with localized resonances rather than a clean gap, in contrast to what was observed in Ref.~\cite{Ca25}. 
We also compare these optical spectra to those of the triangular lattice, which provides an optimal benchmark for photonics applications.}


\color{black}

\section{\label{sec:other-properties}Comparison of Other Physical Properties of Gyromorphs and Stealthy Hyperuniform Media}

It has been known for some time now that high-$\chi$ disordered stealthy hyperuniform systems are endowed with physical properties
that are superior to nonhyperuniform 
and even nonstealthy hyperuniform systems, including optical transparency \cite{Le16,Fr17,To21a,Ki23,Ki24,Ki24b,Ri25,Kl26}, ultrafast spreadability \cite{To21d}, effective delocalization of waves \cite{Fr17,Va26}, optimal effective diffusion coefficients of two-phase media \cite{Zh16b}, and effective conductivities of networks \cite{To18c}. 
These qualities are due to the fact that while
high-$\chi$ disordered stealthy hyperuniform structures
are statistically isotropic with liquidlike short-range
order, they are simultaneously characterized
by certain crystal structural properties, such as no single scattering from intermediate to
infinite wavelengths and bounded holes \cite{To18a},
that often endow materials with enhanced physical properties.
 
For nonstealthy hyperuniform media in which the structure obeys a power-law form $S(k) \sim k^\alpha$, a larger scaling exponent $\alpha$ generally leads to improved physical
properties. For example, using an accurate strong-contrast
formula for the effective dynamic dielectric
constant of $d$-dimensional 
two-phase media \cite{To21a} , it was shown that its imaginary part scales as $k^{d+\alpha}$ as $k$ tends to zero, implying a greater degree of transparency at long wavelengths as $\alpha$ increases.
In this regard, it is clear that any class III hyperuniform system, including gyromorphs, cannot perform as well
as any class I system in which $\alpha$ remains bounded.
Moreover, unlike gyromorphs, disordered stealthy hyperuniform
systems are transparent for a continuous  band of 
frequencies around the origin for $d \ge 2$ \cite{Le16,To21a}
and effectively transparent for large one-dimensional (1D) samples \cite{Ki23,Ki24,Kl26}.

Another example that vividly illustrates
the differences in the materials performance
of stealthy hyperuniform systems and gyromorphs is offered by the time-dependent diffusion spreadability ${\cal S}(t)$, which gives the fraction
of solute that diffuses in a two-phase system that is originally
all in one of the two phases at time $t$ equal to zero.
The excess spreadability ${\cal S}(\infty) - {\cal S}(t) $
at long times is directly related to the exponent $\alpha$;
specifically,  ${\cal S}(\infty) - {\cal S}(t) \sim  t^{-(d+\alpha)/2}$ for large $t$. We see that typical nonhyperuniform media \textcolor{black}{($\alpha=0$)} have slower spreadability than any hyperuniform one. Thus, the excess spreadability
for a 2D gyromorph with $\alpha =0.42$ scales, for large $t$, like
${\cal S}(\infty) - {\cal S}(t) \sim t^{-1.21}$. This decay rate is to be contrasted with the {\it ultrafast} decay for stealthy hyperuniform media or, more precisely, exponentially fast decay \cite{To21d}.
The rate of energy dissipation in such structures follows a similar ranking: Class III gyromorph-derived materials dissipate by a slower power law, rather than by the exponentially fast decay associated with stealthy hyperuniform materials \cite{Sk23}.

\section{\label{sec:discussion} Conclusions and Discussion}
\color{black}
In conclusion, we have systematically investigated the structural and physical properties of gyromorphs and compared them with those of disordered stealthy hyperuniform media.
We considered gyromorphs over a broad range of rotational orders $G$ and showed that the effectively isotropic large-$G$ gyromorphs are weakly class III hyperuniform rather than nonhyperuniform.
In contrast to stealthy hyperuniform media, which 
have bounded holes and robust photonic band gaps, we have provided strong evidence that the large holes present in gyromorphs result in large realization-to-realization fluctuations in their optical spectra. 
Class III hyperuniformity can produce shallow pseudogaps, but is insufficient to produce the robust gap formation associated with high-$\chi$ stealthy hyperuniformity. 
This conclusion about class III hyperuniformity applies in both optical platforms considered here: dielectric disk arrays derived from random-organization models exhibit only a shallow TM pseudogap and gyromorphic point-dipole arrays exhibit fragmented, size-dependent pseudogaps populated by localized states \textcolor{black}{similar to \textcolor{black}{nonstealthy} disordered systems}. 

We also demonstrated that this structural hierarchy is reflected in 
other physical properties of gyromorphs and disordered
stealthy hyperuniform media.  
Specifically, we showed that, unlike gyromorphs or any class III hyperuniform medium, disordered stealthy hyperuniform
systems are transparent or effectively transparent
for a continuous  band of  frequencies around the origin.
Due to their weak class III hyperuniformity, 
we also specifically revealed that gyromorph-derived heterogeneous materials must exhibit long-time power-law decay
of the spreadability and energy dissipation rather than the ultrafast exponential rates of these characteristics for stealthy hyperuniform media. 

In summary, high-$\chi$ disordered stealthy hyperuniform media remain the gold standard among amorphous  structures in that they
rival the materials performance of crystal structures 
with the advantages of directionally independent physical properties. They are a remarkable class of materials whose physical properties frequently outperform those of both nonhyperuniform and conventional hyperuniform systems. Their distinctive advantage arises from a unique structural duality: They retain the statistical isotropy and liquidlike short-range order characteristic of ordinary disordered materials while simultaneously exhibiting crystal-like features. Notably, these systems eliminate single scattering from intermediate to infinite wavelengths and possess bounded holes, attributes typically associated with crystalline order. This unprecedented combination of disorder and order is responsible for many of the exceptional physical properties observed in stealthy hyperuniform materials.

In the present work, we considered only a single stealthiness value, $\chi\approx0.49$, and intentionally used a dilute disk packing fraction $\phi_2$ in the mpb calculations to match the structural parameters used in Ref.~\cite{Ca25}.
\textcolor{black}{For $0<\chi<0.49$, the stealthy exclusion region will still produce ultrafast spreadability in the long-time asymptotic limit \cite{To21d}, although the associated decay rate decreases with $\chi$ \cite{To21d,Va25}. Previous continuum-dielectric studies show that stealthy photonic band gaps also degrade as $\chi$ decreases and may give way to pseudogaps below a threshold $\chi$ value \cite{Fl09b,Fr17,Kl22}. We therefore expect the TM gap in stealthy point-dipole arrays to narrow as $\chi$ decreases. Determining the extent of this degradation across $\chi$ values remains an interesting problem for future work.}

Previous work also indicates that the low volume fractions considered here are far from optimal \cite{Fl09b,Kl22}.
It would therefore be desirable to optimize the TM band-gap properties via mpb over the wide range of packing fractions achievable using the recently introduced ``ultradense'' stealthy hyperuniform particle packings and across $\chi$ values in the disordered regime.
Such packings can attain substantially larger maximum packing fractions than conventional stealthy configurations for $0<\chi<0.5$, while also exhibiting distinct short-range order induced by their soft-core constraint \cite{To25,Ki25}.
A corresponding Green's-matrix study could determine how $\chi$ and ultradense local structure affect spectral gaps, localized resonances, and finite-size scaling throughout the disordered stealthy regime.

A natural extension of the present work is the comparison of gyromorphs and stealthy hyperuniform systems in three dimensions, where TM and TE polarizations are coupled. 
Prior Green's-matrix studies of three-dimensional (3D) stealthy hyperuniform systems showed that increasing stealthiness can suppress proximity resonances, open spectral gaps, and drive a transition toward weak localization \cite{Sg22}. 
We would expect that the 3D gyromorphs are also class III hyperuniform. 
If this turns out to be the case, a corresponding comprehensive study of the optical density of states of 3D gyromorphs and stealthy structures with the accurate Green's-matrix method might again reveal that the former yield shallow pseudogaps without the robust localization behavior associated with high-$\chi$ stealthy hyperuniformity.

\color{black}

\begin{acknowledgments}
\textcolor{black}{M.S., P.J.S., and S.T. were supported by the U.S. Army Research Office under Cooperative Agreement No. W911NF-22-2-0103, and acknowledge the Princeton Institute for Computational Science and Engineering for the computational resources used in this work.
R.F. and L.D.N. are pleased to acknowledge that the computational work on the Green's-matrix analysis reported in this paper was performed on the Shared Computing Cluster, which is administered by Boston University's Research Computing Services.} 

\end{acknowledgments}

\section*{Data Availability}
There are no publicly available research data or software supporting this article.
Requests for further information or data should be sent to the authors.

\appendix

\section{Calculation of the Two-Dimensional Density of States via the Green's-Matrix Method\label{app:DOS_gmatrix}}

\subsection{Theoretical framework}
To compute the DOS for open systems of $N$ point scatterers, we employ an effective non-Hermitian Hamiltonian approach based on the Green's-matrix formalism. This method allows us to extract the collective scattering modes (quasimodes) and their respective lifetimes by solving an eigenvalue problem. The system is physically characterized by the coordinates of the scatterers $\mathbf{r}_j$, an effective area $A$, and a point density $\rho = N/A$. The characteristic wave vector in vacuum is denoted as $k_0$.

\subsection{Transverse magnetic polarization}
For TM-polarized light, where the electric field is strictly out of plane, the wave propagation is treated as a 2D scalar problem. The system is thus fully described by an $N \times N$ non-Hermitian matrix $\hat{G}^{\text{TM}}$. The off-diagonal elements govern the propagation of scalar waves between scatterer $i$ and scatterer $j$ in free space, defined as \cite{caze2013strong, monsarrat2022pseudogap}:
\begin{equation} \label{eq:G_TM_offdiag}
    G_{ij}^{\text{TM}} = i H_0^{(1)}(k_0 r_{ij}) \quad \text{for } i \neq j,
\end{equation}
where $r_{ij} = |\mathbf{r}_i - \mathbf{r}_j|$ is the Euclidean distance between the scatterers, and $H_0^{(1)}$ is the zero-order Hankel function of the first kind.

To account for the natural radiative decay and the unshifted resonance frequency of an isolated scatterer, the diagonal elements are enforced to be purely imaginary:
\begin{equation} \label{eq:G_TM_diag}
    G_{jj}^{\text{TM}} = i.
\end{equation}
It is important to note a fundamental normalization difference relative to the physical 2D free-space Green's function, which typically carries a prefactor of $i/4$ \cite{caze2013strong, monsarrat2022pseudogap}. To map the physical propagator to the dimensionless matrix formalism established by Skipetrov \cite{skipetrov2020finite}---where the self-energy imaginary part must strictly evaluate to $1$ ($\text{Im}(G_{jj}) = 1$) to yield the correct single-dipole decay rate $\Gamma_0$---we must scale the entire physical Green's function by a factor of $4$. This mathematical scaling ensures that the eigenvalues directly provide the dimensionless physical quantities without requiring additional adjustments.

\subsection{Transverse electric polarization}
For TE-polarized light, the electric field lies in the plane of the scatterers, necessitating a fully vectorial treatment. The system is therefore described by a larger $2N \times 2N$ dyadic Green's matrix $\hat{G}^{\text{TE}}$, which accounts for the coupled in-plane field components (e.g., $x$ and $y$).

The interaction between scatterer $i$ and scatterer $j$ is captured by a $2 \times 2$ off-diagonal block defined as \cite{monsarrat2022pseudogap}:
\begin{equation} \label{eq:G_TE_offdiag}
    \mathbf{G}_{ij}^{\text{TE}} = 2i\left[F_1(k_0 r_{ij}) \mathbf{I}_{2\times 2} - F_2(k_0 r_{ij}) (\mathbf{\hat{r}}_{ij} \otimes \mathbf{\hat{r}}_{ij})\right],
\end{equation}
where $\mathbf{I}_{2\times 2}$ is the $2 \times 2$ identity matrix, $\mathbf{\hat{r}}_{ij} = (\mathbf{r}_i - \mathbf{r}_j)/r_{ij}$ is the in-plane unit vector connecting the two scatterers, and $\otimes$ denotes the outer product. The scalar functions $F_1(\xi)$ and $F_2(\xi)$ dictate the vector wave scattering dynamics and are explicitly given by:
\begin{equation} \label{eq:F1_func}
    F_1(\xi) = H_0^{(1)}(\xi) - \frac{H_1^{(1)}(\xi)}{\xi},
\end{equation}
\begin{equation} \label{eq:F2_func}
    F_2(\xi) = H_0^{(1)}(\xi) - 2\frac{H_1^{(1)}(\xi)}{\xi},
\end{equation}
with $H_1^{(1)}$ being the first-order Hankel function of the first kind.

Consistent with the scalar TM case, the self-interaction terms are normalized to preserve the fundamental resonance features of the isolated scatterer. Thus, the diagonal $2 \times 2$ blocks are defined as:
\begin{equation} \label{eq:G_TE_diag}
    \mathbf{G}_{jj}^{\text{TE}} = i \mathbf{I}_{2\times 2}.
\end{equation}
This rigorous vectorial approach accurately models the complex dipole-dipole interactions for TE waves, and has been recently utilized to compute the inverse participation ratio (IPR) and map the TE DOS for stealthy hyperuniform and triangular structural arrays \cite{monsarrat2022pseudogap}.

\subsection{Spectral analysis and density-of-states normalization}
The complex eigenvalues $\Lambda_m$ are obtained by diagonalizing the corresponding Green's matrix:
\begin{equation} \label{eq:eigenproblem}
    \hat{G} \mathbf{\psi}_m = \Lambda_m \mathbf{\psi}_m,
\end{equation}
where the total number of modes is $M = N$ for TM polarization and $M = 2N$ for TE polarization. These eigenvalues encode both the resonance frequencies $\omega_m$ and the decay rates $\Gamma_m$ of the system's quasimodes. Following the standard dimensionless notation \cite{skipetrov2020finite}, we define the normalized frequency shift $\nu_m$ and the normalized decay rate $\gamma_m$ as:
\begin{equation} \label{eq:nu_m}
    \nu_m = \frac{\omega_m - \omega_0}{\Gamma_0} = -0.5 \, \text{Re}(\Lambda_m),
\end{equation}
\begin{equation} \label{eq:gamma_m}
    \gamma_m = \frac{\Gamma_m}{\Gamma_0} = \text{Im}(\Lambda_m).
\end{equation}
Finally, the total normalized density of states, $\mathcal{N}(\nu)$, is computed by superimposing the Lorentzian line shapes corresponding to all independent quasimodes \cite{skipetrov2020finite}:
\begin{equation} \label{eq:dos_nu}
    \mathcal{N}(\nu) = \frac{1}{M\pi} \sum_{m=1}^{M} \frac{(\gamma_m / 2)}{(\nu - \nu_m)^2 + (\gamma_m / 2)^2}.
\end{equation}
Because the area under each Lorentzian curve with respect to the normalized frequency $\nu$ is exactly $\pi$, the prefactor $1/(M\pi)$ strictly normalizes the distribution such that the continuous integral over all normalized frequencies intrinsically equals unity:
\begin{equation} \label{eq:integral}
    \int_{-\infty}^{\infty} \mathcal{N}(\nu) \, d\nu = 1.
\end{equation}
Consequently, further numerical normalization is unnecessary \cite{skipetrov2020finite}.

\section{Calculation of the Two-Dimensional Local Density of States\label{app:LDOS}}

\subsection{Theoretical framework and scatterer properties}
To calculate the LDOS at a specific source position $\mathbf{r}_S$ within a two-dimensional open array of $N$ point scatterers, we evaluate the interaction between a source dipole $\mathbf{p}$ located at $\mathbf{r}_S$ and the surrounding scattering environment. Within the Foldy-Lax multiple scattering framework, each discrete point scatterer is mathematically characterized by its electric polarizability $\alpha(\omega)$ \cite{caze2013strong}:
\begin{equation} \label{eq:polarizability}
    \alpha(\omega) = \frac{1}{k^2} \frac{2\Gamma_0}{\omega_0 - \omega - i\Gamma_0/2},
\end{equation}
where $k = \omega/c$ is the free-space wave number, $\omega_0$ is the resonance angular frequency, and $\Gamma_0$ is the natural radiative linewidth. This formulation describes nonabsorbing scatterers and strictly satisfies the optical theorem and energy conservation \cite{caze2013strong}. To ensure that the scattering cross section of the individual particles remains essentially constant across the broad spectral range investigated in our numerical simulations \cite{caze2013strong}, we impose a strongly overdamped condition, $\Gamma_0 = 100\,\omega_0 \gg \omega_0$.

\subsection{Transverse magnetic polarization}
For TM-polarized light, the electric field is treated as a scalar quantity oscillating perpendicular to the 2D plane. The propagation of this scalar field in free space is governed by the physical 2D vacuum Green's function \cite{caze2013strong,monsarrat2022pseudogap}:
\begin{equation} \label{eq:G0_TM}
    G_0^{\text{TM}}(\mathbf{r}_i, \mathbf{r}_j, \omega) = \frac{i}{4} H_0^{(1)}(k_0|\mathbf{r}_i - \mathbf{r}_j|).
\end{equation}
Note that this physical propagator differs from the dimensionless matrix elements defined in Eq.~\eqref{eq:G_TM_offdiag} by a factor of $1/4$, which is necessary here to properly recover the physical scattered fields.

To evaluate the LDOS, the local exciting field $E_i$ on each scatterer $i$ is obtained by rigorously solving the system of $N$ self-consistent Foldy-Lax multiple scattering equations \cite{caze2013strong}:
\begin{equation} \label{eq:Ei_TM}
    E_i = \mu_0 \omega^2 G_0^{\text{TM}}(\mathbf{r}_i, \mathbf{r}_S, \omega) p + k^2 \alpha(\omega) \sum_{j \neq i} G_0^{\text{TM}}(\mathbf{r}_i, \mathbf{r}_j, \omega) E_j.
\end{equation}
After solving this linear system for the local fields on all individual scatterers, the total scattered field $E_{sca}$ returning to the origin $\mathbf{r}_S$ is computed as:
\begin{equation} \label{eq:Esca_TM}
    E_{sca} = k^2 \alpha(\omega) \sum_{i} G_0^{\text{TM}}(\mathbf{r}_S, \mathbf{r}_i, \omega) E_i.
\end{equation}
The normalized TM LDOS $\rho^{\text{TM}}(\mathbf{r}_S, \omega)/\rho^{\text{TM}}_0$, which corresponds to the Purcell factor, is then extracted directly from the imaginary part of the scattered field evaluated at the source position:
\begin{equation} \label{eq:rho_TM}
    \frac{\rho^{\text{TM}}(\mathbf{r}_S, \omega)}{\rho^{\text{TM}}_0} = \frac{\frac{1}{4}+\Im\left[\frac{E_{sca}}{\mu_0 \omega^2 p}\right]}{\frac{1}{4}},
\end{equation}
where $\rho^{\text{TM}}_0$ represents the 2D vacuum TM LDOS.

\subsection{Transverse electric polarization}
In the TE polarization case, the electric \textcolor{black}{dipoles} oscillate within the scattering plane, requiring a fully vectorial treatment. The scalar Green's function is systematically replaced by the $2 \times 2$ dyadic vacuum Green's tensor $\mathbf{G}_0^{\text{TE}}(\mathbf{r}_i, \mathbf{r}_j, \omega)$. The spatial coupling between a scatterer at $\mathbf{r}_i$ and one at $\mathbf{r}_j$ is given by:
\begin{equation} \label{eq:G0_TE}
    \mathbf{G}_0^{\text{TE}}(\mathbf{r}_i, \mathbf{r}_j, \omega) = \frac{i}{4} \left[ F_1(k_0 r_{ij}) \mathbf{I}_{2\times 2} - F_2(k_0 r_{ij}) (\mathbf{\hat{r}}_{ij} \otimes \mathbf{\hat{r}}_{ij}) \right].
\end{equation}
The scalar functions $F_1(\xi)$ and $F_2(\xi)$, which map the vectorial wave propagation dynamics, are explicitly defined in Eqs.~\eqref{eq:F1_func} and \eqref{eq:F2_func}. Note that this physical propagator differs from the dimensionless matrix elements defined in Eq.~\eqref{eq:G_TE_offdiag} by a factor of $1/8$, which is necessary here to properly recover the physical scattered fields.

The self-consistent multiple scattering equations are correspondingly generalized to account for these vectorial interactions:
\begin{equation} \label{eq:Ei_TE}
    \mathbf{E}_i = \mu_0 \omega^2 \mathbf{G}_0^{\text{TE}}(\mathbf{r}_i, \mathbf{r}_S, \omega) \mathbf{p} + k^2 \alpha(\omega) \sum_{j \neq i} \mathbf{G}_0^{\text{TE}}(\mathbf{r}_i, \mathbf{r}_j, \omega) \mathbf{E}_j.
\end{equation}
After solving this $2N \times 2N$ linear system for the coupled in-plane field components, the scattered field returning to $\mathbf{r}_S$ is evaluated:
\begin{equation} \label{eq:Esca_TE}
    \mathbf{E}_{sca} = k^2 \alpha(\omega) \sum_{i} \mathbf{G}_0^{\text{TE}}(\mathbf{r}_S, \mathbf{r}_i, \omega) \mathbf{E}_i.
\end{equation}
Finally, the normalized TE LDOS $\rho^{\text{TE}}(\mathbf{r}_S, \omega)/\rho^{\text{TE}}_0$ is extracted from the projection of the scattered field onto the source dipole orientation:
\begin{equation} \label{eq:rho_TE}
    \frac{\rho^{\text{TE}}(\mathbf{r}_S, \omega)}{\rho^{\text{TE}}_0} = \frac{\frac{1}{8}+\Im\left[\frac{\mathbf{E}_{sca}\cdot\mathbf{p}}{\mu_0 \omega^2 |p|^2}\right]}{\frac{1}{8}},
\end{equation}
where $\rho^{\text{TE}}_0$ is the 2D vacuum TE LDOS.
\color{black}

\section{Generation of Vogel Spirals\label{app:vogel_gen}}

Vogel spirals are generated in polar coordinates according to \cite{Ad11}
\begin{equation}
r_n=a_0\sqrt{n},\qquad \theta_n=n\theta_0,
\end{equation}
where $\theta_0=2\pi(1-\tau^{-1})\approx137.508^\circ$ is the golden angle and $\tau=(1+\sqrt{5})/2$ is the golden ratio \cite{Ad11}. 
A Vogel spiral differs from the statistically homogeneous point patterns considered in the main text because it is generated about a preferred origin, so its local statistics depend systematically on radial position. 
For this reason, we use Vogel spirals as optical benchmarks but do not treat them as statistically homogeneous systems for the purpose of estimating a hyperuniformity exponent. However, Vogel spirals have been extensively investigated in photonics as they feature broad photonic band gaps that coexist with distinctively localized modes with multifractal properties that are of interest in a number of engineering applications \cite{pollard2009low,Tr12,Li11b,trevino2011circularly,dal2012analytical}.

\section{Additional Optical Analysis\label{app:optics}}

\begin{figure*}[t]
    \centering
    \includegraphics[width=\linewidth]{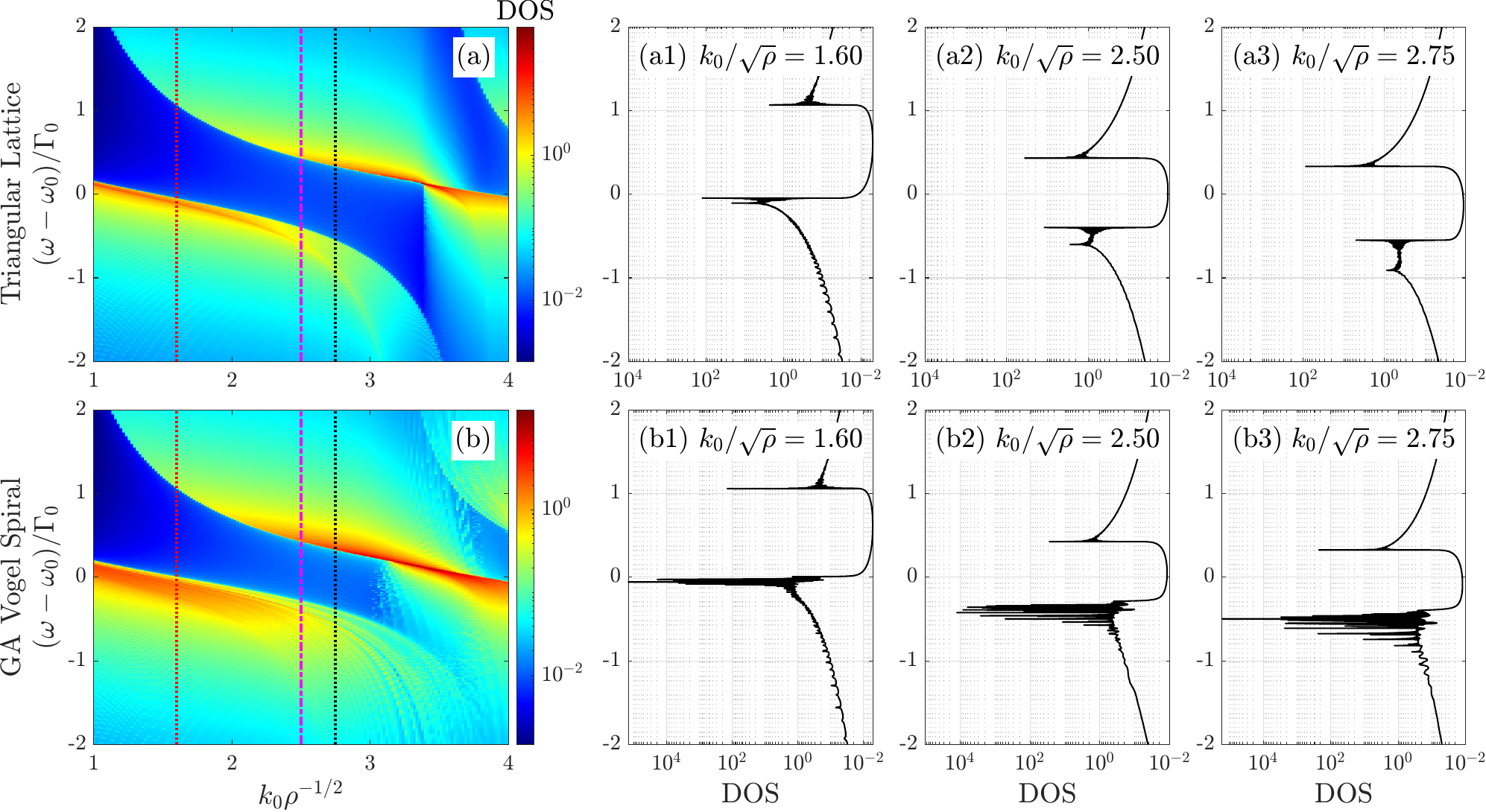}
    \caption{TM DOS maps for (a) triangular lattice and (b) golden-angle Vogel spiral. The maps are plotted as a function of the normalized wave number $k_{0}\rho^{-1/2}$ and the normalized angular frequency $(\omega - \omega_{0})/\Gamma_{0}$. Corresponding line cuts are extracted from the DOS maps at [(a1), (b1)] $k_0\rho^{-1/2} = 1.60$ (dotted red line), [(a2), (b2)] $2.50$ (dash-dotted magenta line), and [(a3), (b3)] $2.75$ (dotted black line).
    Calculations are performed for arrays of $N=6500$ dipoles, with a normalized angular frequency resolution of $8\times 10^{-3}$ in the DOS maps and $10^{-7}$ in the line cuts.
    }
    \label{fig:TM_DOS_map_Tri_Vogel}
\end{figure*}

\begin{figure}
    \centering
    \includegraphics[width=\linewidth]{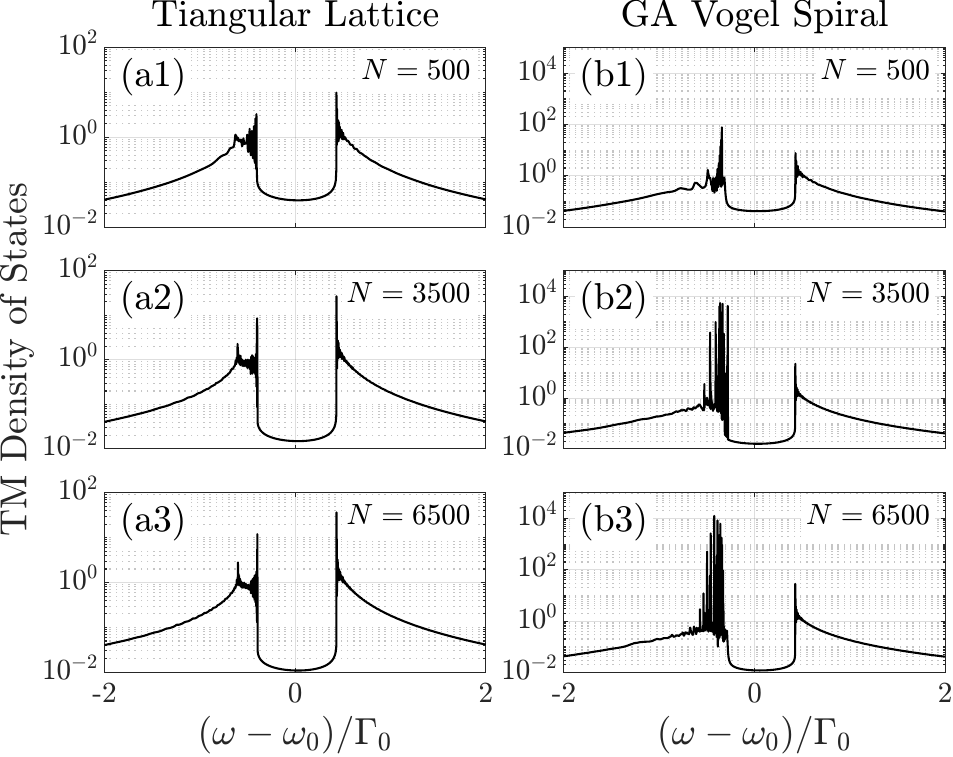}
    \caption{TM DOS as a function of the normalized angular frequency $(\omega-\omega_0)/\Gamma_0$ for different numbers of dipoles $N$: [(a1)--(a3)] triangular lattice and [(b1)--(b3)] golden-angle Vogel spiral structures, evaluated at $N = 500$, $3500$, and $6500$. The resolution in $(\omega-\omega_0)/\Gamma_0$ is $10^{-7}$. The normalized wave number is set to $k_0\rho^{-1/2} = 2.5$.}
    \label{fig:TM_DOSvsN_Tri_Vogel}
\end{figure}

\textcolor{black}{The TM DOS maps and line cuts for the triangular lattice and golden-angle Vogel spiral are presented in Fig.~\ref{fig:TM_DOS_map_Tri_Vogel}, with their corresponding finite-size scaling shown in Fig.~\ref{fig:TM_DOSvsN_Tri_Vogel}.
Notably, the smooth band edges of the stealthy array (Fig.~\ref{fig:TM_DOS_map}) closely resemble those of the triangular lattice and the golden-angle Vogel spiral.
Furthermore, comparing the spectral evolution across all four geometries reveals a distinct hierarchy in the onset of the band gap as a function of the normalized wave number $k_0\rho^{-1/2}$.
The triangular lattice opens a gap first, at the highest $k_0\rho^{-1/2}$ values, followed almost simultaneously by the SHU system and the golden-angle Vogel spiral.
By contrast, the gyromorph array is the last to exhibit spectral depletion, forming its fragmented pseudogap at comparatively lower values of $k_0\rho^{-1/2}$.}

\begin{figure*}[t]
    \centering
    \includegraphics[width=\linewidth]{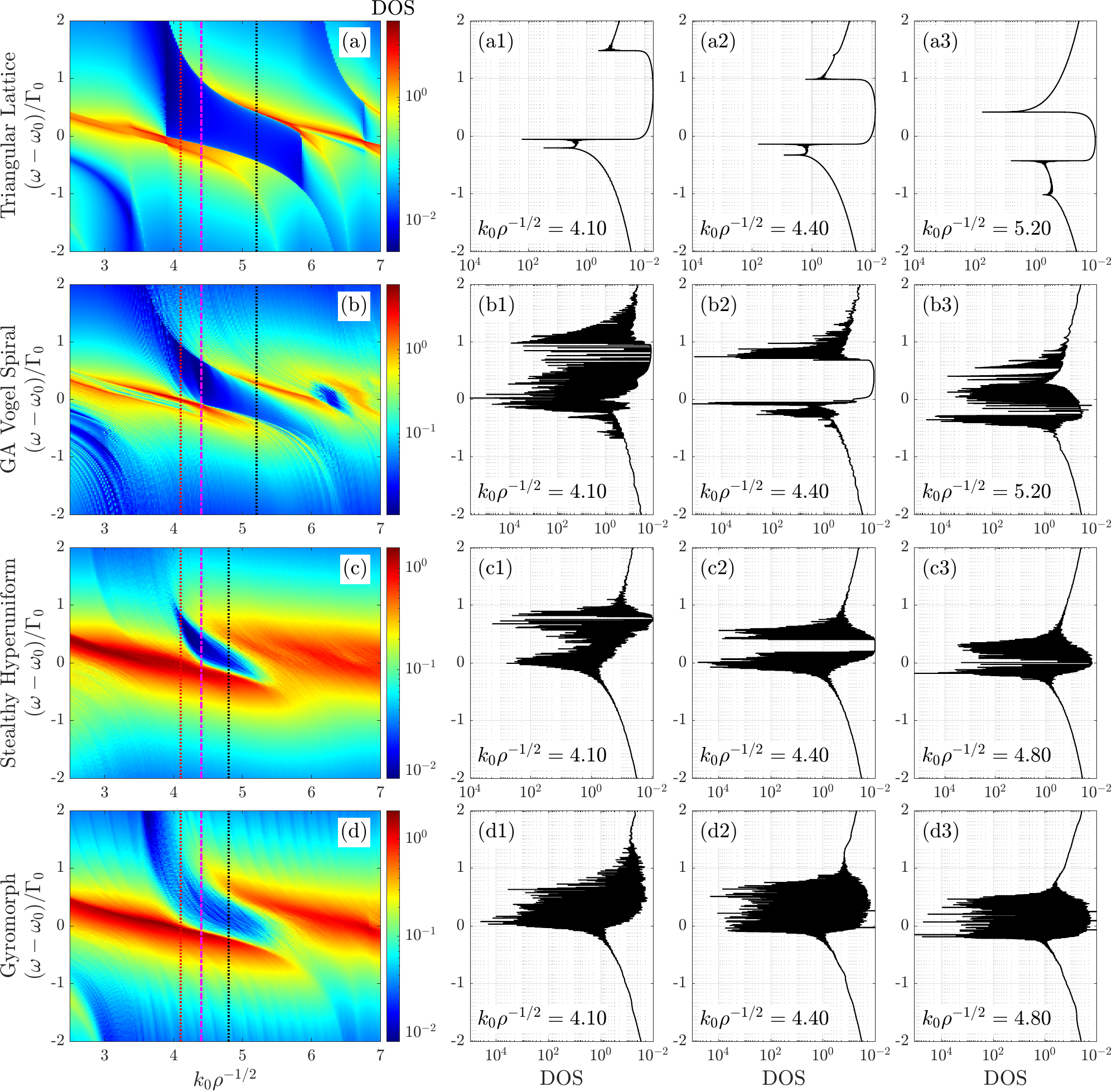}
    \caption{TE DOS maps for (a) triangular lattice, (b) golden-angle Vogel spiral, (c) SHU, and (d) gyromorph. The maps are plotted as a function of the normalized wave number $k_{0}\rho^{-1/2}$ and the normalized angular frequency $(\omega - \omega_{0})/\Gamma_{0}$. Corresponding line cuts are extracted from the DOS maps at [(a1), (b1), (c1), (d1)] $k_0\rho^{-1/2} = 4.1$ (dotted red line), [(a2), (b2), (c2), (d2)] $4.4$ (dash-dotted magenta line), and [(a3), (b3), (c3), (d3)] $k_0\rho^{-1/2} = 5.2, 5.2, 4.8, 4.8$ (dotted black line). Calculations are performed for arrays of $N=6500$ dipoles, with a normalized angular frequency resolution of $8\times 10^{-3}$ in the DOS maps and $10^{-7}$ in the line cuts.
    }
    \label{fig:TE_DOS_map}
\end{figure*}

\begin{figure*}
    \centering
    \includegraphics[width=\linewidth]{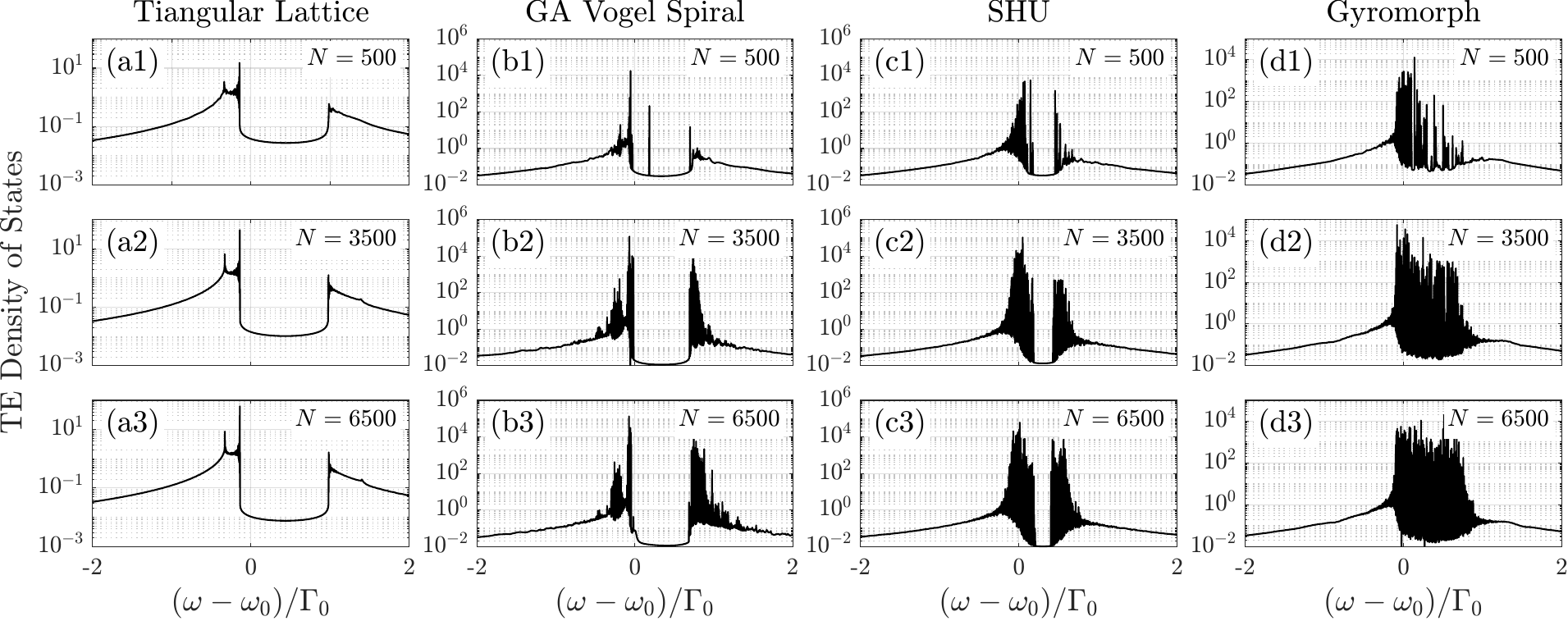}
    \caption{TE DOS as a function of the normalized angular frequency $(\omega-\omega_0)/\Gamma_0$ for different numbers of dipoles, $N$, for the [(a1)--(a3)] triangular lattice, [(b1)--(b3)] golden-angle Vogel spiral, [(c1)--(c3)] SHU, and [(d1)--(d3)] gyromorph structures, respectively, evaluated at $N = 500$, $3500$, and $6500$. The resolution in $(\omega-\omega_0)/\Gamma_0$ is $10^{-7}$. The normalized wave number is set to $k_0\rho^{-1/2} = 4.4$.}
    \label{fig:TE_DOSvsN_Tri_Vogel_SHU_G}
\end{figure*}

\begin{figure*}
    \centering
    \includegraphics[width=\linewidth]{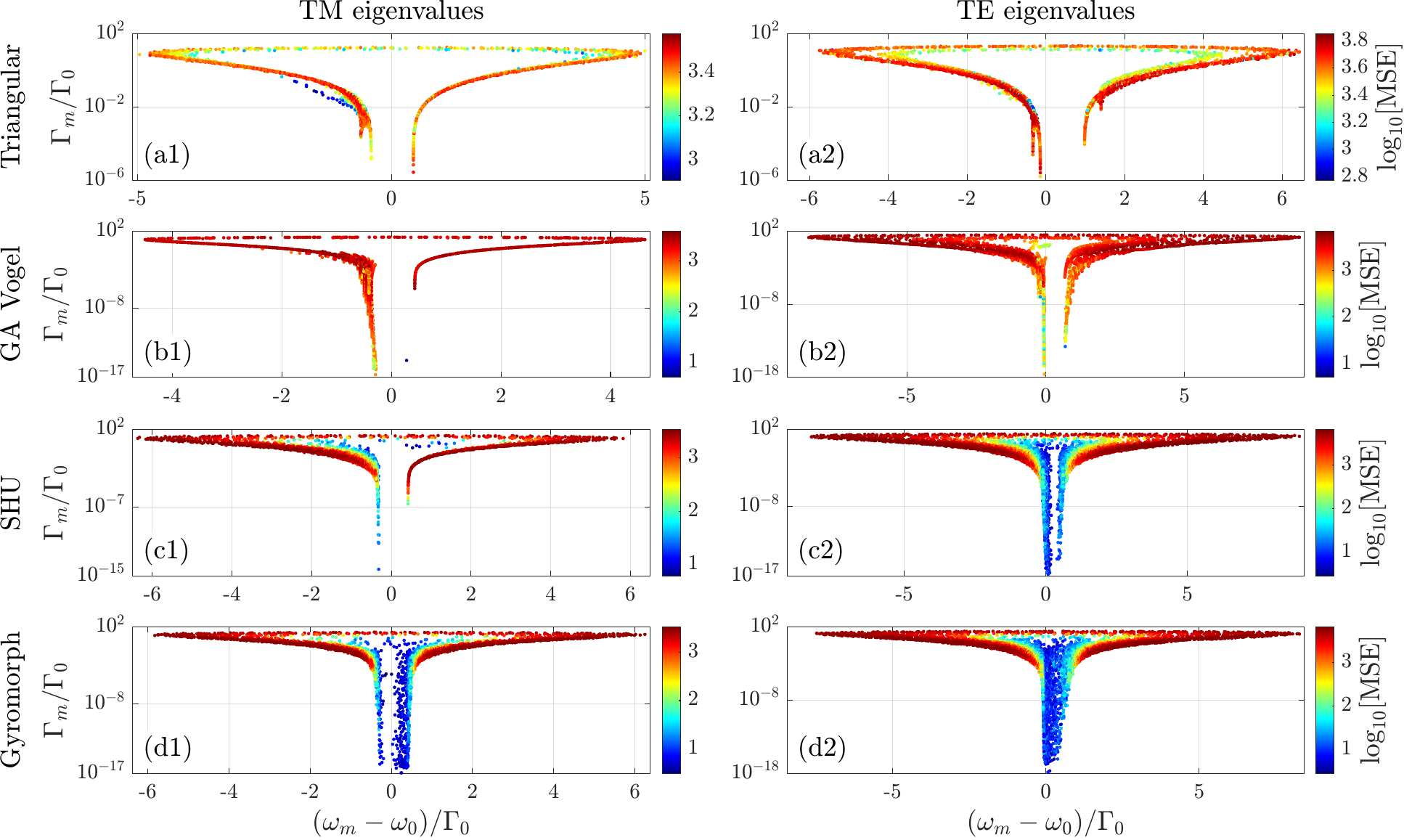}
    \caption{TM and TE Green's-matrix complex eigenvalue distributions for a system of $N=6500$ dipoles. Left-hand column [(a1), (b1), (c1), (d1)], eigenvalues for the TM polarization; right-hand column [(a2), (b2), (c2), (d2)], eigenvalues for the TE polarization. The rows correspond to the four investigated structures: [(a1), (a2)] triangular lattice, [(b1), (b2)] golden-angle Vogel spiral, [(c1), (c2)] stealthy hyperuniform, and [(d1), (d2)] gyromorph. The horizontal axis represents the normalized angular frequency $(\omega_m-\omega_0)/\Gamma_0$ (real part of the eigenvalues), and the vertical axis shows the normalized decay rates $\Gamma_m/\Gamma_0$ (imaginary part of the eigenvalues) on a logarithmic scale. The color scale indicates the $\log_{10}[\text{MSE}]$ values, where MSE denotes the modal spatial extent. \textcolor{black}{The $\mathrm{MSE}$ parameter quantifies the spatial extension of a given eigenstate of the system \cite{Da21,Sg20,monsarrat2022pseudogap} and is equal to the participation ratio ($\mathrm{PR}$).} For the TM polarization $k_0\rho^{-1/2}=2.5$ and for the TE polarization $k_0\rho^{-1/2}=4.4$.
    }
    \label{fig:TEandTM_Eigen}
\end{figure*}

\begin{figure*}
    \centering
    \includegraphics[width=\linewidth]{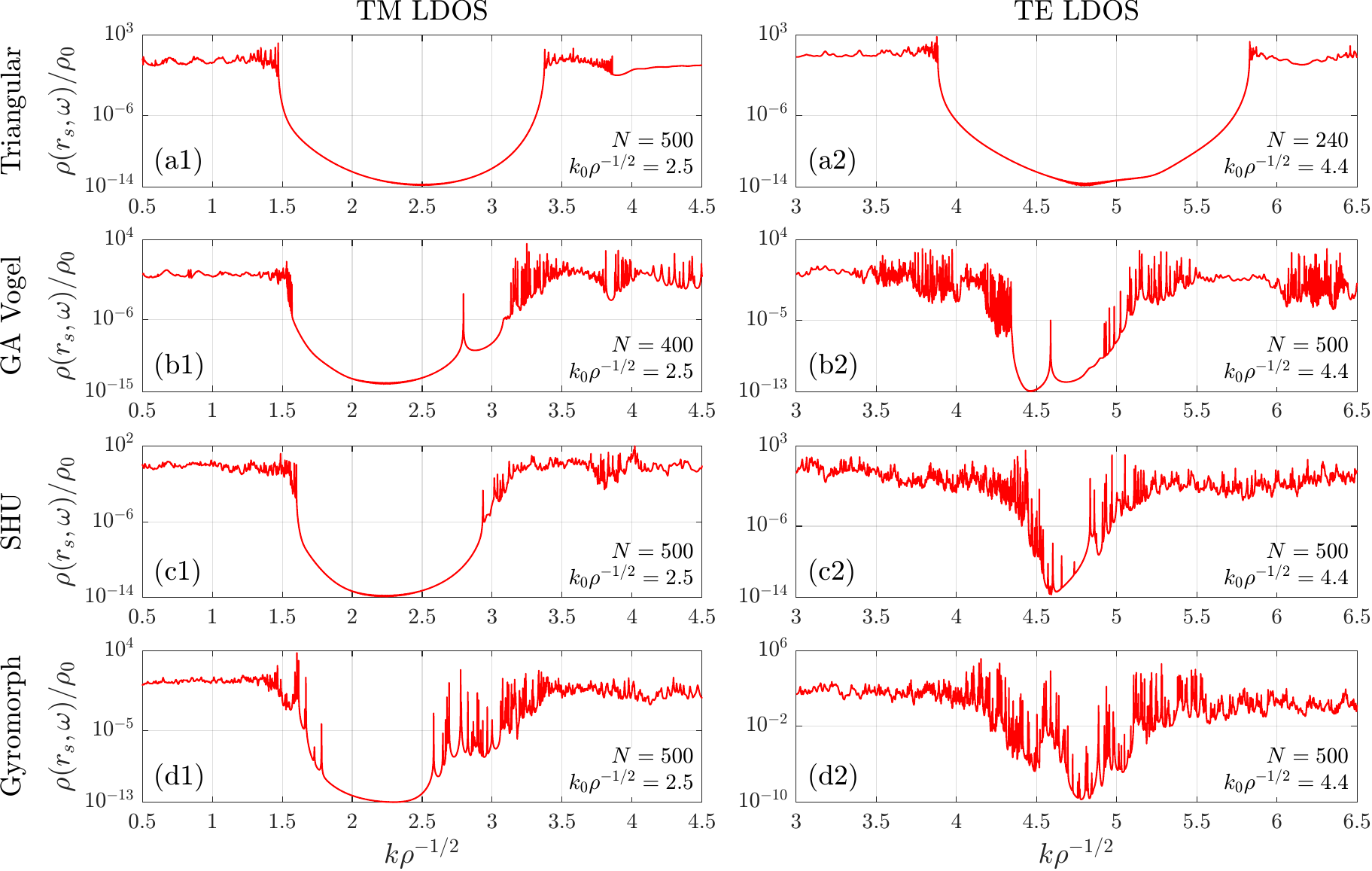}
    \caption{TM and TE LDOS distributions for the four investigated structures. Left-hand column [(a1), (b1), (c1), (d1)], TM LDOS; right-hand column [(a2), (b2), (c2), (d2)], TE LDOS. The rows correspond to the [(a1), (a2)] triangular lattice, [(b1), (b2)] golden-angle Vogel spiral, [(c1), (c2)] stealthy hyperuniform, and [(d1), (d2)] gyromorph structures. The vertical axis represents the normalized LDOS $\rho(r_s, \omega)/\rho_0$ on a logarithmic scale, plotted as a function of the parameter $k\rho^{-1/2}$.
    For the TM polarization, $k_0\rho^{-1/2}=2.5$ for all structures, with $N=500$ for the triangular, SHU, and gyromorph geometries, and $N=400$ for the golden-angle Vogel spiral.
    For the TE polarization, $k_0\rho^{-1/2}=4.4$ for all structures, and the number of dipoles is $N=240$ for the triangular lattice, and $N=500$ for the golden-angle Vogel spiral, SHU, and gyromorph structures. 
    For the TE LDOS calculations, the dipole orientation is set to $p = [p_0, p_0]/\sqrt{2}$.
    The resolution in $k\rho^{-1/2}$ is $4\times10^{-5}$.
    }
    \label{fig:TEandTM_LDOS}
\end{figure*}

The TE DOS results for the triangular lattice, golden-angle Vogel spiral, $\chi\approx0.49$ stealthy hyperuniform, and $G=60$ gyromorph point-dipole arrays are shown in Figs.~\ref{fig:TE_DOS_map} and \ref{fig:TE_DOSvsN_Tri_Vogel_SHU_G}. 
In this case, the dipoles lie in the two-dimensional plane and can have different orientations, making the problem fully vectorial.
Longitudinal vector-coupling effects are known to play an important role in disordered dipole arrays, where they can open additional propagation channels and hinder localization \cite{Sk14,skipetrov2016red,maximo2015spatial}.
\textcolor{black}{Comparing the DOS maps across the four geometries reveals a clear hierarchy in TE band-gap formation. The triangular lattice serves as the optimal ordered benchmark, exhibiting the widest and cleanest band gap with perfectly smooth edges. The golden-angle Vogel spiral also successfully opens a robust gap, featuring a distinctly smooth and fully depleted region around $k_0\rho^{-1/2}=4.4$. Similarly, the SHU array exhibits a clear TE gap with a strongly depleted DOS in this wave-number range, though it is noticeably narrower than those of the triangular lattice and Vogel spiral. Our TE DOS results for both the triangular lattice and SHU arrays are in excellent agreement with the findings previously reported in Ref.~\cite{monsarrat2022pseudogap}. By contrast, the gyromorph array completely fails to open a TE band gap over the same spectral range. Instead of a depleted region, its TE spectrum is densely populated with extremely narrow localized resonances.
The finite-size scaling analysis in Fig.~\ref{fig:TE_DOSvsN_Tri_Vogel_SHU_G} confirms that these structural distinctions persist and evolve as the number of dipoles, $N$, increases. For the triangular lattice, the band gap remains perfectly clean and structurally stable regardless of system size. In the golden-angle Vogel spiral, the overall gap structure is preserved at larger $N$, though the band edges exhibit slight structural deformations. For the SHU array, the strictly smooth, peak-free gap region slightly narrows as the system size increases; however, a prominent and fully depleted band gap clearly survives even at $N=6500$. Conversely, the gyromorph spectrum shows no convergence toward a true gap. As $N$ increases, the pseudogap region simply fills with an ever-denser spectral forest of sharp, narrow peaks.
This singular DOS structure in the gyromorph is reminiscent of the behavior expected in strongly disordered systems near a localization threshold \cite{caze2013strong}.}

\begin{figure*}
    \centering
    \includegraphics[width=.84\linewidth]{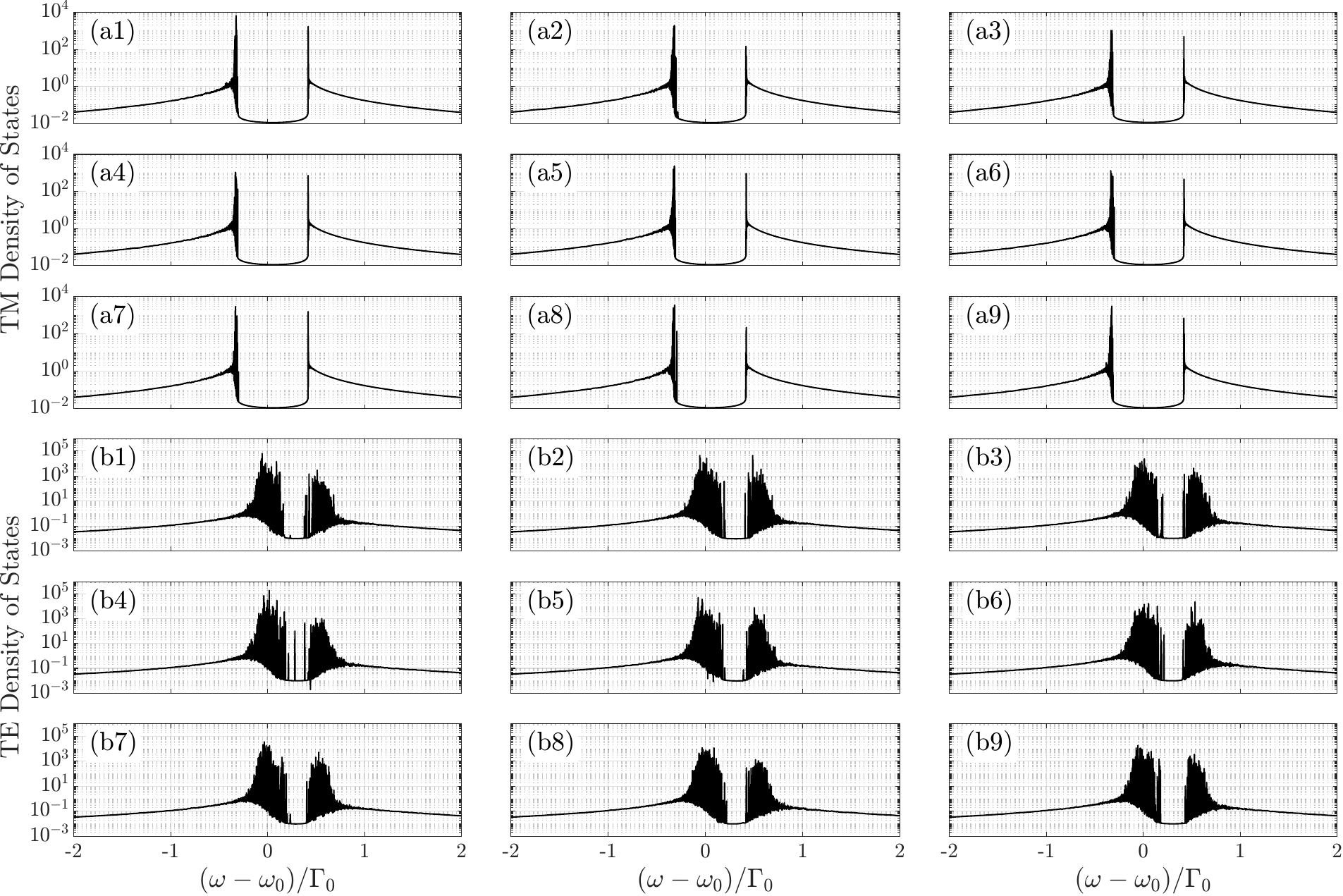}
    \caption{(a) TM DOS and (b) TE DOS as a function of the normalized angular frequency $(\omega-\omega_0)/\Gamma_0$ for different realizations of the SHU structure. The resolution in $(\omega-\omega_0)/\Gamma_0$ is $10^{-7}$. The number of dipoles is $N = 6500$ and the normalized wave number is set to $k_0\rho^{-1/2} = 2.5$ for the TM polarization and to $k_0\rho^{-1/2} = 4.4$ for the TE polarization.}
    \label{fig:DOS_SHU_realizations}
\end{figure*}

\begin{figure*}
    \centering
    \includegraphics[width=.84\linewidth]{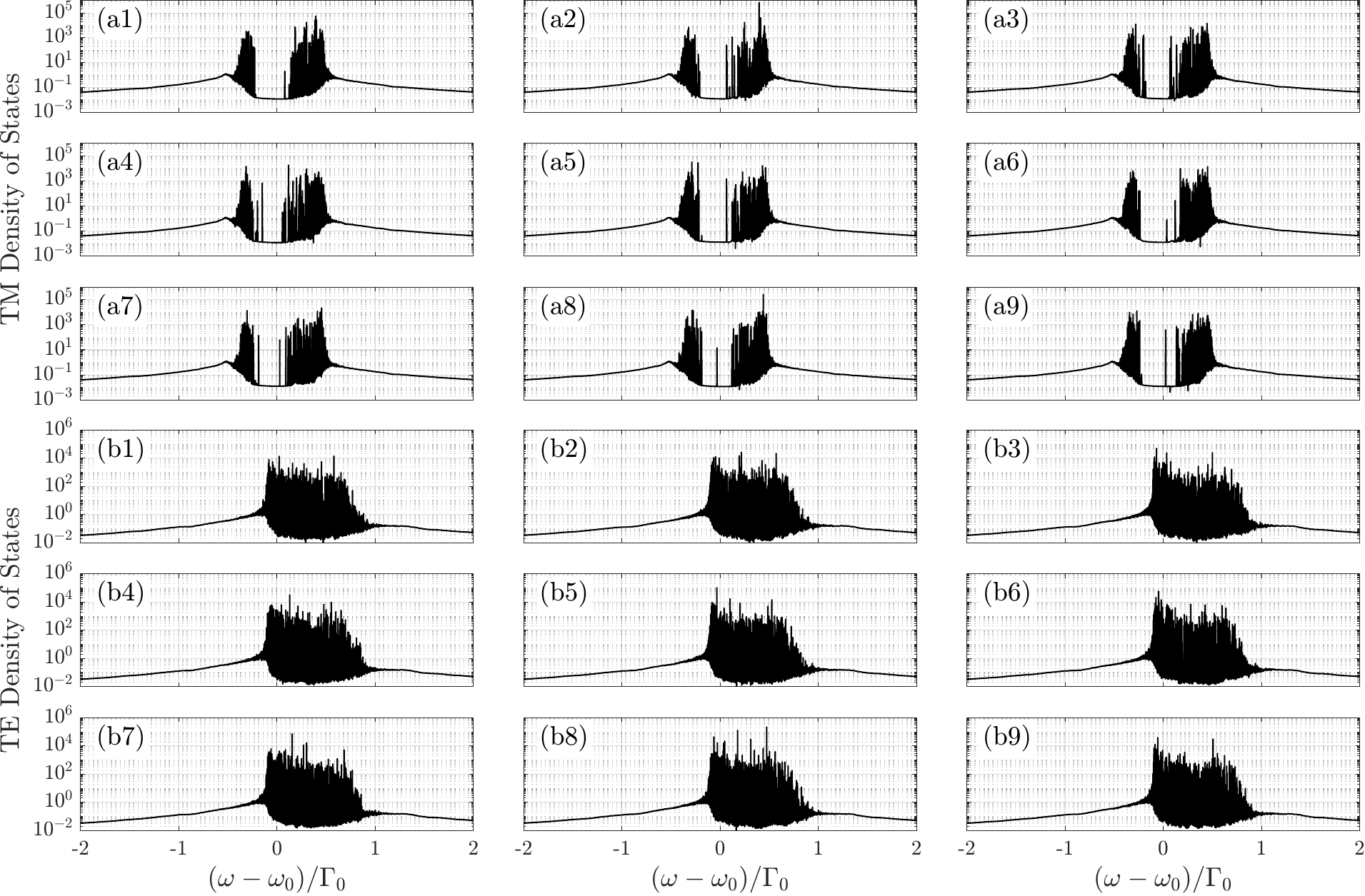}
    \caption{(a) TM DOS and (b) TE DOS as a function of the normalized angular frequency $(\omega-\omega_0)/\Gamma_0$ for different realizations of the gyromorph structure. The resolution in $(\omega-\omega_0)/\Gamma_0$ is $10^{-7}$. The number of dipoles is $N = 6500$ and the normalized wave number is set to $k_0\rho^{-1/2} = 2.5$ for the TM polarization and to $k_0\rho^{-1/2} = 4.4$ for the TE polarization.}
    \label{fig:DOS_Gyro_realizations}
\end{figure*}

\textcolor{black}{To gain deeper physical insight into the nature of these depleted spectral regions, we analyze the Green's-matrix complex eigenvalue distributions in Fig.~\ref{fig:TEandTM_Eigen}. The eigenvalues directly map the normalized decay rates $\Gamma_m/\Gamma_0$ (imaginary part) and detuned frequencies $(\omega_m-\omega_0)/\Gamma_0$ (real part) of the system's quasimodes. For both TM and TE polarizations, the triangular lattice, golden-angle Vogel spiral, and SHU arrays exhibit well-defined spectral gaps completely devoid of eigenvalues, with modes cleanly accumulating at the band edges. In stark contrast, the gyromorph distributions reveal a pseudogap region densely populated by modes exhibiting extremely small normalized decay rates ($\Gamma_m/\Gamma_0 \ll 1$). These long-lived, subradiant modes indicate severe spatial localization of light, fundamentally disrupting the formation of a true band gap.
These global spectral features are fully corroborated by the LDOS calculations presented in Fig.~\ref{fig:TEandTM_LDOS}, which act as a direct probe of the Purcell factor. Here, the exciting dipole is placed at the center of each array. To ensure numerical stability—given the extreme exponential suppression of the LDOS inside the gaps—we restrict these specific calculations to systems of at most $N=500$ dipoles. In the TM case, the triangular, Vogel, and SHU structures all feature deep, continuous, and smooth troughs spanning several orders of magnitude, confirming a robust suppression of local spontaneous emission. Conversely, the gyromorph TM LDOS exhibits a significantly smaller gap, with a high density of localized states populating the band edges. The TE polarization further exacerbates this trend: while the ordered and stealthy systems retain clearer LDOS depletion, the gyromorph TE response is entirely dominated by high-amplitude localized resonances across the entire bandwidth, effectively failing to act as an optical insulator.
Note that for the TE LDOS calculations presented in Fig.~\ref{fig:TEandTM_LDOS}, the source dipole orientation is set to $\mathbf{p} = [p_0, p_0]/\sqrt{2}$; however, the results remain equivalent when using $\mathbf{p} = [p_0, 0]$ or $\mathbf{p} = [0, p_0]$.}

Finally, Figs.~\ref{fig:DOS_SHU_realizations} and \ref{fig:DOS_Gyro_realizations} show that the gyromorph TM DOS fluctuates appreciably from realization to realization, consistent with the large ensemble fluctuations in the gyromorphs' hyperuniformity exponents $\alpha$ and their apparent lack of bounded holes. 
By contrast, the stealthy hyperuniform TM DOS is far more reproducible across realizations, consistent with the fixed small-$k$ constraints and bounded-hole property of stealthy systems. 
\textcolor{black}{For the TE spectra, while there are realization-to-realization fluctuations in the stealthy DOS, the gap width remains roughly fixed across ensembles. 
No such TE gap is observed in any of the gyromorph configurations considered here.}



\begin{thebibliography}{60}

\bibitem{To03a}
S. Torquato and F. H. Stillinger, Local density fluctuations, hyperuniformity, and order metrics, Phys. Rev. E \textbf{68}, 041113 (2003).

\bibitem{To18a}
S. Torquato, Hyperuniform states of matter, Phys. Rep. \textbf{745}, 1 (2018).

\bibitem{Le16}
O. Leseur, R. Pierrat, and R. Carminati, High-density hyperuniform materials can be transparent, Optica \textbf{3}, 763 (2016).

\bibitem{Fr17}
L. S. Froufe-P\'erez, M. Engel, J. J. S\'aenz, and F. Scheffold, Band gap formation and Anderson localization in disordered photonic materials with structural correlations, Proc. Natl. Acad. Sci. USA \textbf{114}, 9570 (2017).

\bibitem{To21a}
S. Torquato and J. Kim, Nonlocal effective electromagnetic wave characteristics of composite media: Beyond the quasistatic regime, Phys. Rev. X \textbf{11}, 021002 (2021).

\bibitem{Ki23}
J. Kim and S. Torquato, Effective electromagnetic wave properties of disordered stealthy hyperuniform layered media beyond the quasistatic regime, Optica \textbf{10}, 965 (2023).

\bibitem{Ki24}
J. Kim and S. Torquato, Theoretical prediction of the effective dynamic dielectric constant of disordered hyperuniform anisotropic composites beyond the long-wavelength regime [Invited], Opt. Mater. Express \textbf{14}, 194 (2024).

\bibitem{Ri25}
R. Riganti, Y. Zhu, W. Cai, S. Torquato, and L. Dal Negro, Multiscale physics-informed neural networks for the inverse design of hyperuniform optical materials, Adv. Opt. Mater. \textbf{13}, 2403304 (2025).

\bibitem{Kl26}
M. A. Klatt, P. J. Steinhardt, and S. Torquato, Transparency versus Anderson localization in one-dimensional disordered stealthy hyperuniform layered media, Opt. Mater. Express \textbf{16}, 2080 (2026).

\bibitem{Va26}
C. Vanoni, J. Karcher, M. C. Rechtsman, B. L. Altshuler, P. J. Steinhardt, and S. Torquato, Effective delocalization in the one-dimensional Anderson model with stealthy disorder, Phys. Rev. Lett. \textbf{136}, 150404 (2026).

\bibitem{To21d}
S. Torquato, Diffusion spreadability as a probe of the microstructure of complex media across length scales, Phys. Rev. E \textbf{104}, 054102 (2021).

\bibitem{Zh16b}
G. Zhang, F. H. Stillinger, and S. Torquato, Transport, geometrical, and topological properties of stealthy disordered hyperuniform two-phase systems, J. Chem. Phys \textbf{145}, 244109 (2016).

\bibitem{To18c}
S. Torquato and D. Chen, Multifunctional hyperuniform cellular networks: Optimality, anisotropy and disorder, Multifunct. Mater. \textbf{1}, 015001 (2018).

\bibitem{Fl09b}
M. Florescu, S. Torquato, and P. J. Steinhardt, Designer disordered materials with large, complete photonic band gaps, Proc. Natl. Acad. Sci. USA \textbf{106}, 20658 (2009).

\bibitem{Kl22}
M. Klatt, P. Steinhardt, and S. Torquato, Wave propagation and band tails of two-dimensional disordered systems in the thermodynamic limit, Proc. Natl. Acad. Sci. USA \textbf{119}, e2213633119 (2022).

\bibitem{To21c}
S. Torquato, Structural characterization of many-particle systems on approach to hyperuniform states, Phys. Rev. E \textbf{103}, 052126 (2021).

\bibitem{Uc04b}
O. U. Uche, F. H. Stillinger, and S. Torquato, Constraints on collective density variables: Two dimensions, Phys. Rev. E \textbf{70}, 046122 (2004).

\bibitem{Ba08}
R. D. Batten, F. H. Stillinger, and S. Torquato, Classical disordered ground states: Super-ideal gases and stealth and equi-luminous materials, J. Appl. Phys. \textbf{104}, 033504 (2008).

\bibitem{To15}
S. Torquato, G. Zhang, and F. H. Stillinger, Ensemble theory for stealthy hyperuniform disordered ground states, Phys. Rev. X \textbf{5}, 021020 (2015).

\bibitem{Ki24b}
J. Kim and S. Torquato, Extraordinary optical and transport properties of disordered stealthy hyperuniform two-phase media, J. Phys.: Condens. Matter \textbf{36}, 225701 (2024).

\bibitem{Sk23}
M. Skolnick and S. Torquato, Simulated diffusion spreadability for characterizing the structure and transport properties of two-phase materials, Acta Mater. \textbf{250}, 118857 (2023).

\bibitem{Sk25a}
M. Skolnick and S. Torquato, Effective transport and mechanical properties of two-phase materials across the order-disorder spectrum, Acta Mater. \textbf{290}, 120921 (2025).

\bibitem{Va25}
C. Vanoni, J. Kim, P. J. Steinhardt, and S. Torquato, Dynamical properties of particulate composites derived from ultradense stealthy hyperuniform sphere packings, Phys. Rev. E \textbf{112}, 015406 (2025).

\bibitem{Ca25}
M. Casiulis, A. Shih, and S. Martiniani, Gyromorphs: A new class of functional disordered materials, Phys. Rev. Lett. \textbf{135}, 196101 (2025).

\bibitem{MPB}
S. G. Johnson and J. D. Joannopoulos, Block-iterative frequency-domain methods for Maxwell's equations in a planewave basis, Opt. Express \textbf{8}, 173 (2001).

\bibitem{Chaik08}
L. Cort\'e, P. M. Chaikin, J. P. Gollub, and D. J. Pine, Random organization in periodically driven systems, Nat. Phys. \textbf{4}, 420 (2008).

\bibitem{He15}
D. Hexner and D. Levine, Hyperuniformity of critical absorbing states, Phys. Rev. Lett. \textbf{114}, 110602 (2015).

\bibitem{Ma19}
Z. Ma and S. Torquato, Hyperuniformity of generalized random organization models, Phys. Rev. E \textbf{99}, 022115 (2019).

\bibitem{To02a}
S. Torquato, \textit{Random Heterogeneous Materials: Microstructure and Macroscopic Properties} (Springer-Verlag, New York, 2002).

\bibitem{To10d}
S. Torquato, Reformulation of the covering and quantizer problems as ground states of interacting particles, Phys. Rev. E \textbf{82}, 056109 (2010).

\bibitem{Ki19b}
J. Kim and S. Torquato, Methodology to construct large realizations of perfectly hyperuniform disordered packings, Phys. Rev. E \textbf{99}, 052141 (2019).

\bibitem{Zh17a}
G. Zhang, F. H. Stillinger, and S. Torquato, Can exotic disordered {\lq\lq}stealthy{\rq\rq} particle configurations tolerate arbitrarily large holes? Soft Matter \textbf{13}, 6197 (2017).

\bibitem{Gh18}
S. Ghosh and J. L. Lebowitz, Generalized stealthy hyperuniform processes: Maximal rigidity and the bounded holes conjecture, Commun. Math. Phys. \textbf{363}, 97 (2018).

\bibitem{Zh15a}
G. Zhang, F. H. Stillinger, and S. Torquato, Ground states of stealthy hyperuniform potentials: I. Entropically favored configurations, Phys. Rev. E \textbf{92}, 022119 (2015).

\bibitem{Sh24}
A. Shih, M. Casiulis, and S. Martiniani, Fast generation of spectrally shaped disorder, Phys. Rev. E \textbf{110}, 034122 (2024).

\bibitem{note:SI}
See Supplemental Material below for a comprehensive quantitative validation of the present gyromorph structures; details include structure factors, pair correlation functions, gyromorphic correlation functions, and coupled-dipole calculations, which includes Refs.~\cite{Ca25,MAGREETE}.

\bibitem{skipetrov2016finite}
S. E. Skipetrov, Finite-size scaling analysis of localization transition for scalar waves in a three-dimensional ensemble of resonant point scatterers, Phys. Rev. B \textbf{94}, 064202 (2016).

\bibitem{skipetrov2020finite}
S. E. Skipetrov, Finite-size scaling of the density of states inside band gaps of ideal and disordered photonic crystals, Eur. Phys. J. B \textbf{93}, 70 (2020).

\bibitem{monsarrat2022pseudogap}
R. Monsarrat, R. Pierrat, A. Tourin, and A. Goetschy, Pseudogap and Anderson localization of light in correlated disordered media, Phys. Rev. Res. \textbf{4}, 033246 (2022).

\bibitem{dal2016structural}
L. Dal Negro, R. Wang, and F. A. Pinheiro, Structural and spectral properties of deterministic aperiodic optical structures, Crystals \textbf{6}, 161 (2016).

\bibitem{Sg20}
F. Sgrignuoli and L. Dal Negro, Subdiffusive light transport in three-dimensional subrandom arrays, Phys. Rev. B \textbf{101}, 214204 (2020).

\bibitem{Sg22}
F. Sgrignuoli, S. Torquato, and L. Dal Negro, Subdiffusive wave transport and weak localization transition in three-dimensional stealthy hyperuniform disordered systems, Phys. Rev. B \textbf{105}, 064204 (2022).

\bibitem{Sg21}
F. Sgrignuoli and L. Dal Negro, Hyperuniformity and wave localization in pinwheel scattering arrays, Phys. Rev. B \textbf{103}, 224202 (2021).

\bibitem{Sg19}
F. Sgrignuoli, R. Wang, F. A. Pinheiro, and L. Dal Negro, Localization of scattering resonances in aperiodic Vogel spirals, Phys. Rev. B \textbf{99}, 104202 (2019).

\bibitem{Da21}
L. Dal Negro, \textit{Waves in Complex Media} (Cambridge University Press, Cambridge, UK, 2022).

\bibitem{caze2013strong}
A. Caz\'e, R. Pierrat, and R. Carminati, Strong coupling to two-dimensional Anderson localized modes, Phys. Rev. Lett. \textbf{111}, 053901 (2013).

\bibitem{Li11}
X.-F. Li, X. Ni, L. Feng, M.-H. Lu, C. He, and Y.-F. Chen, Tunable unidirectional sound propagation through a sonic-crystal-based acoustic diode, Phys. Rev. Lett. \textbf{106}, 084301 (2011).

\bibitem{Tr12}
J. Trevino, S. F. Liew, H. Noh, H. Cao, and L. Dal Negro, Geometrical structure, multifractal spectra and localized optical modes of aperiodic Vogel spirals, Opt. Express \textbf{20}, 3015 (2012).

\bibitem{note:poisson_Hv}
Indeed, even a nonhyperuniform Poisson distribution of points (which possesses arbitrarily large holes in the thermodynamic limit) must necessarily have bounded holes in simulations of $10^5$ particles. One can infer from such finite simulations that Poisson point processes can support arbitrarily large holes in the thermodynamic limit due to large-$r$ tails in $H_V(r)$ for $r$ on the order of one-half of the box size.

\bibitem{To25}
S. Torquato and J. Kim, Existence of nonequilibrium glasses in the degenerate stealthy hyperuniform ground-state manifold, Soft Matter \textbf{21}, 4898 (2025).

\bibitem{Ki25}
J. Kim and S. Torquato, Ultradense sphere packings derived from disordered stealthy hyperuniform ground states, J. Chem. Phys. \textbf{163}, 024902 (2025).

\bibitem{Ad11}
J. A. Adam, \textit{A Mathematical Nature Walk} (Princeton University Press, Princeton, NJ, 2011).

\bibitem{pollard2009low}
M. E. Pollard and G. J. Parker, Low-contrast bandgaps of a planar parabolic spiral lattice, Opt. Lett. \textbf{34}, 2805 (2009).

\bibitem{Li11b}
S. F. Liew, H. Noh, J. Trevino, L. D. Negro, and H. Cao, Localized photonic band edge modes and orbital angular momenta of light in a golden-angle spiral, Opt. Express \textbf{19}, 23631 (2011).

\bibitem{trevino2011circularly}
J. Trevino, H. Cao, and L. Dal Negro, Circularly symmetric light scattering from nanoplasmonic spirals, Nano Lett. \textbf{11}, 2008 (2011).

\bibitem{dal2012analytical}
L. Dal Negro, N. Lawrence, and J. Trevino, Analytical light scattering and orbital angular momentum spectra of arbitrary Vogel spirals, Opt. Express \textbf{20}, 18209 (2012).

\bibitem{Sk14}
S. E. Skipetrov and I. M. Sokolov, Absence of Anderson localization of light in a random ensemble of point scatterers, Phys. Rev. Lett. \textbf{112}, 023905 (2014).

\bibitem{skipetrov2016red}
S. E. Skipetrov and J. H. Page, Red light for Anderson localization, New J. Phys. \textbf{18}, 021001 (2016).

\bibitem{maximo2015spatial}
C. E. M\'aximo, N. Piovella, P. W. Courteille, R. Kaiser, and R. Bachelard, Spatial and temporal localization of light in two dimensions, Phys. Rev. A, \textbf{92} 062702 (2015).

\bibitem{MAGREETE}
M. Casiulis, A. Shih, and S. Martiniani, MAGreeTe, \url{https://github.com/martiniani-lab/MAGreeTe} (2024), Materials Analysis through Green's Tensor.

\end{thebibliography}

%

\end{document}


\title{Supporting Information for ``Structural and physical  properties of gyromorphs and disordered stealthy hyperuniform media''}

\author{Murray Skolnick}
\affiliation{Princeton Materials Institute, Princeton University, Princeton, New Jersey 08544, USA}%

\author{Riccardo Franchi}
\affiliation{Department of Electrical \& Computer Engineering, Boston University, Boston, Massachusetts 02215, USA}

\author{Luca Dal Negro}
\email{dalnegro@bu.edu}
\affiliation{Department of Electrical \& Computer Engineering, Boston University, Boston, Massachusetts 02215, USA}
\affiliation{Department of Physics, Boston University, Boston, Massachusetts 02215, USA}
\affiliation{Division of Materials Science \& Engineering, Boston University, Brookline, Massachusetts 02446, USA}

\author{Paul J. Steinhardt}
\affiliation{Department of Physics, Princeton University, Princeton, New Jersey 08544, USA}

\author{Salvatore Torquato}
\email{torquato@electron.princeton.edu}
\affiliation{Department of Chemistry, Princeton University, Princeton, New Jersey 08544, USA}
\affiliation{Department of Physics, Princeton University, Princeton, New Jersey 08544, USA}
\affiliation{Princeton Materials Institute, Princeton University, Princeton, New Jersey 08544, USA}
\affiliation{Program in Applied and Computational Mathematics, Princeton University, Princeton, New Jersey 08544, USA}

\maketitle
\section{Validation of Gyromorph Structures}

In this section, we validate that our generated gyromorphs reproduce the structural and optical signatures reported in Ref.~\cite{Ca25}.
For the structural diagnostics, we follow the procedures described in the Supplemental Material of Ref.~\cite{Ca25}: the two-dimensional structure factors are computed from circular cutouts using a Hamming window, while the pair correlation functions are computed directly in real space by binning pair separations.
We also compute the gyromorphic correlation function $g_G(r)$, which probes the imposed $G$-fold orientational order.
For the optical validation, we use the MAGreeTe coupled-dipoles code~\cite{MAGREETE} and the same gyromorph parameters used in Ref.~\cite{Ca25}, namely $G=60$, $KL/(2\pi)=100$, and $N\sim 10^4$. 

\subsection{Structure Factors}

\begin{figure}
\centering
\subfloat[\label{fig:1D_Sk_G0008}]{\includegraphics[width=0.33\textwidth]{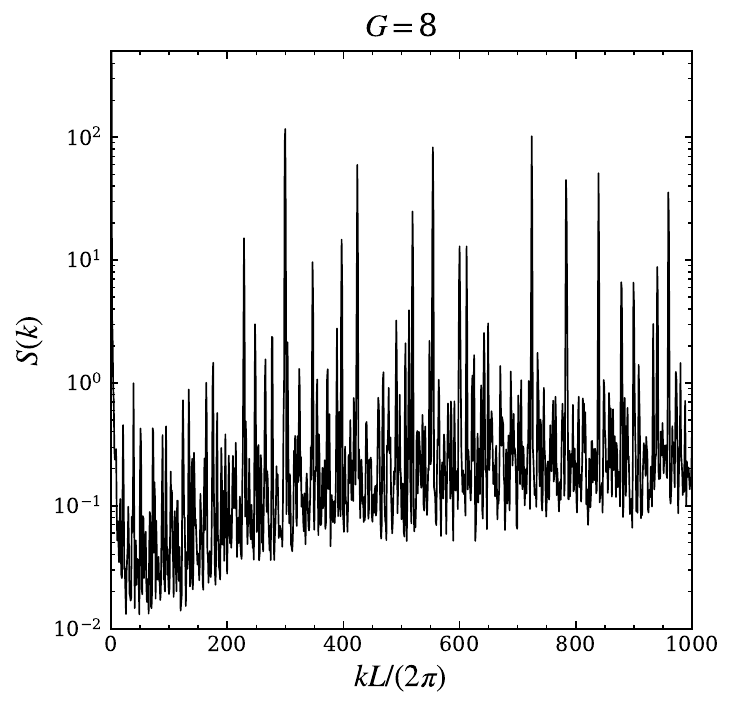}}%
\subfloat[\label{fig:1D_Sk_G0010}]{\includegraphics[width=0.33\textwidth]{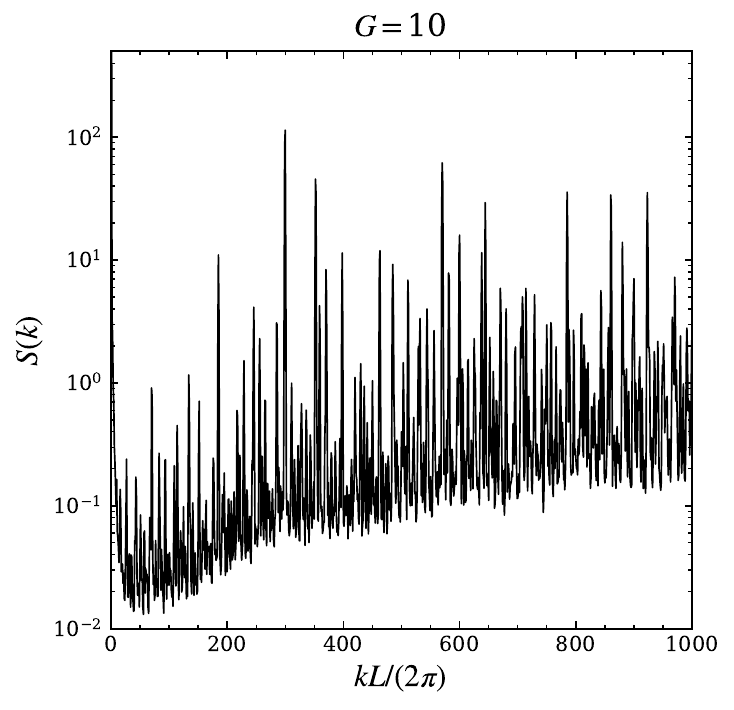}}%
\subfloat[\label{fig:1D_Sk_G0014}]{\includegraphics[width=0.33\textwidth]{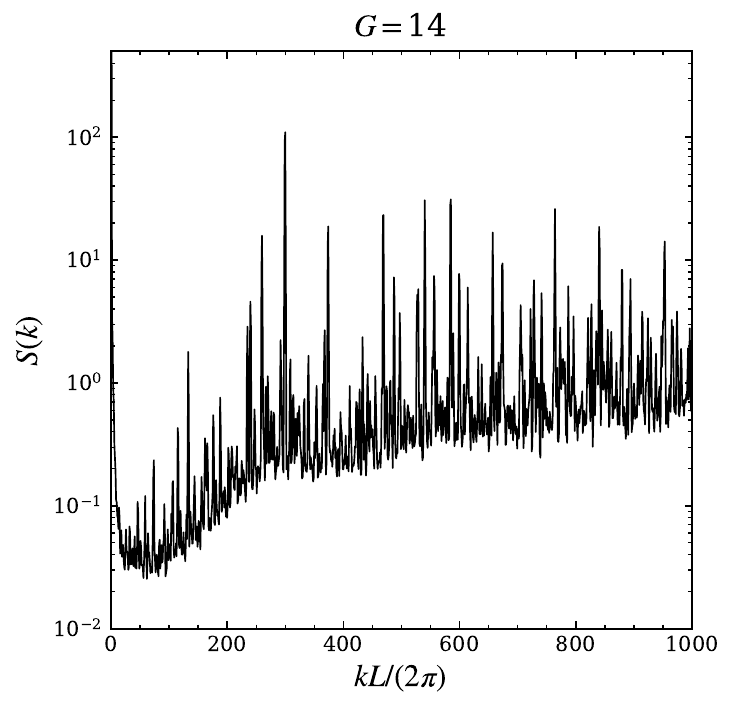}}\
\subfloat[\label{fig:1D_Sk_G0018}]{\includegraphics[width=0.33\textwidth]{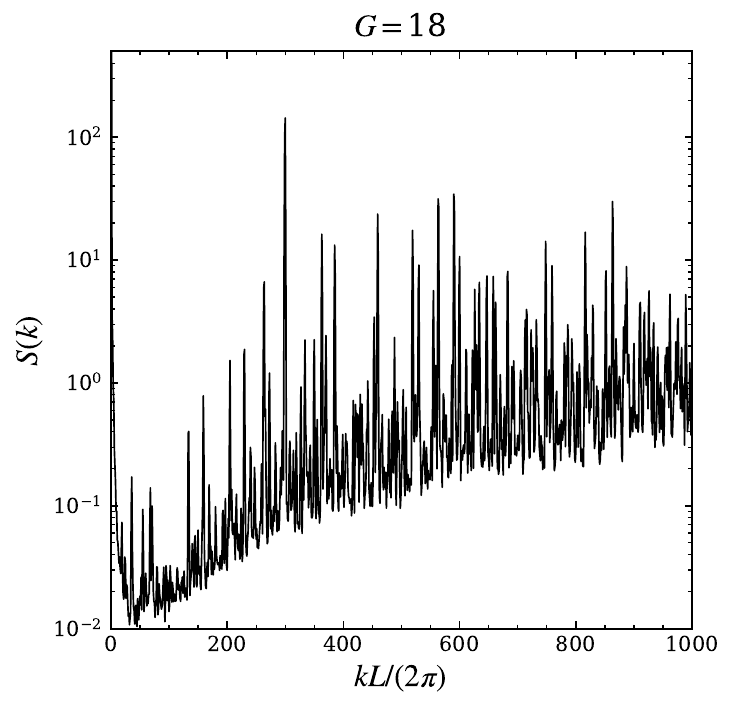}}%
\subfloat[\label{fig:1D_Sk_G0024}]{\includegraphics[width=0.33\textwidth]{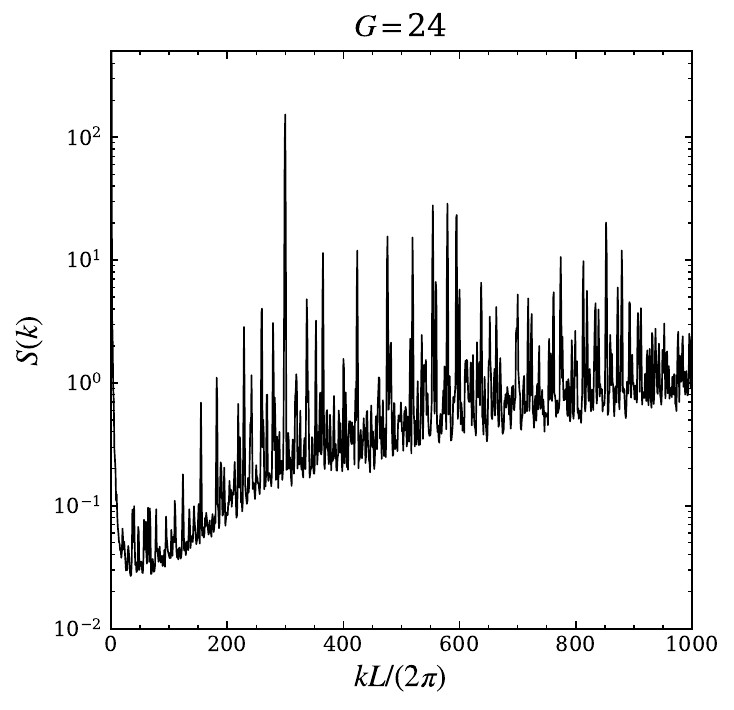}}%
\subfloat[\label{fig:1D_Sk_G0060}]{\includegraphics[width=0.33\textwidth]{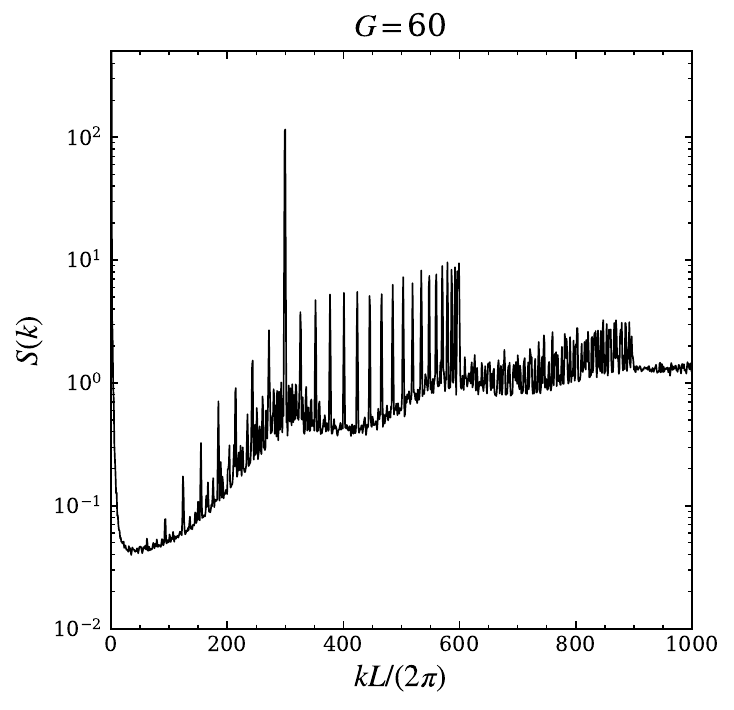}}\
\subfloat[\label{fig:1D_Sk_G0100}]{\includegraphics[width=0.33\textwidth]{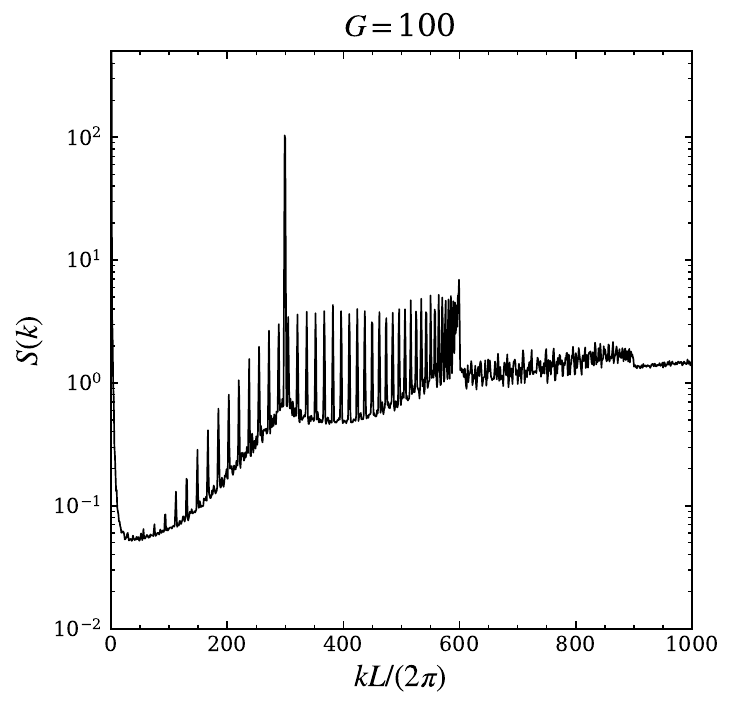}}%
\subfloat[\label{fig:1D_Sk_G0120}]{\includegraphics[width=0.33\textwidth]{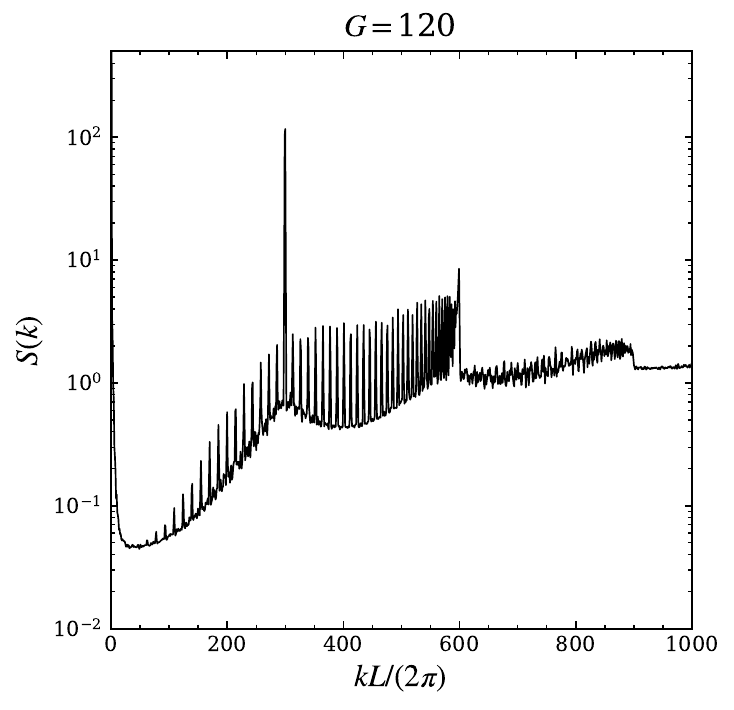}}%
\subfloat[\label{fig:1D_Sk_G0200}]{\includegraphics[width=0.33\textwidth]{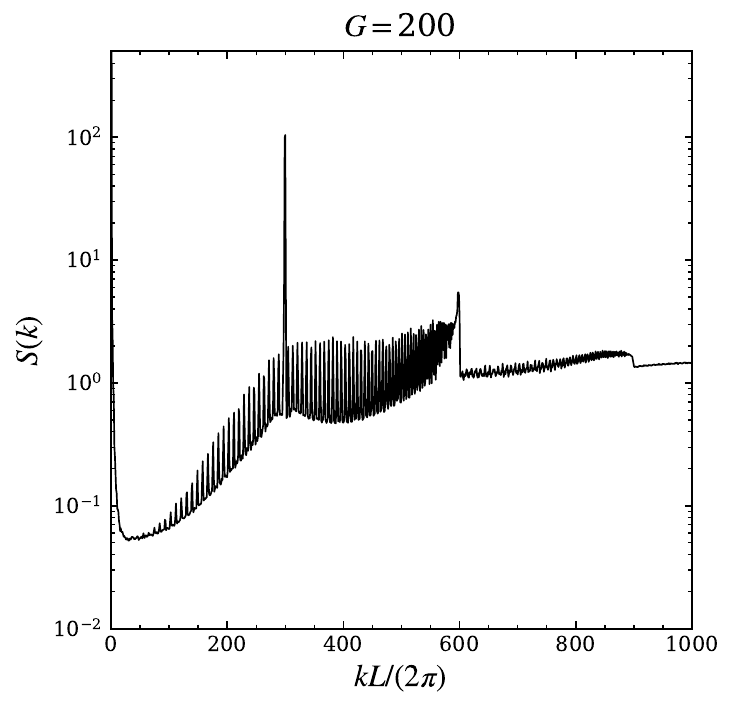}}\
\caption{Angularly averaged structure factors $S(k)$ of $N\sim 10^5$ gyromorphs with rotational orders $G=$ (a) 8, (b) 10, (c) 14, (d) 18, (e) 24, (f) 60, (g) 100, (h) 120, and (i) 200.
The structure factors are computed from circular cutouts using the same Hamming-window procedure used for the corresponding plots in Ref.~\cite{Ca25}.
For each $G$, the plotted curve is an ensemble average over 50 independent configurations.
The intense peaks at the imposed Fourier radius and their weaker echoes are consistent with those seen in the gyromorph structure-factors reported in Ref.~\cite{Ca25}.}
\label{figSI:1D_Sk}
\end{figure}

\begin{figure}
\centering
\subfloat[\label{fig:2D_Sk_G0008}]{\includegraphics[width=0.33\textwidth]{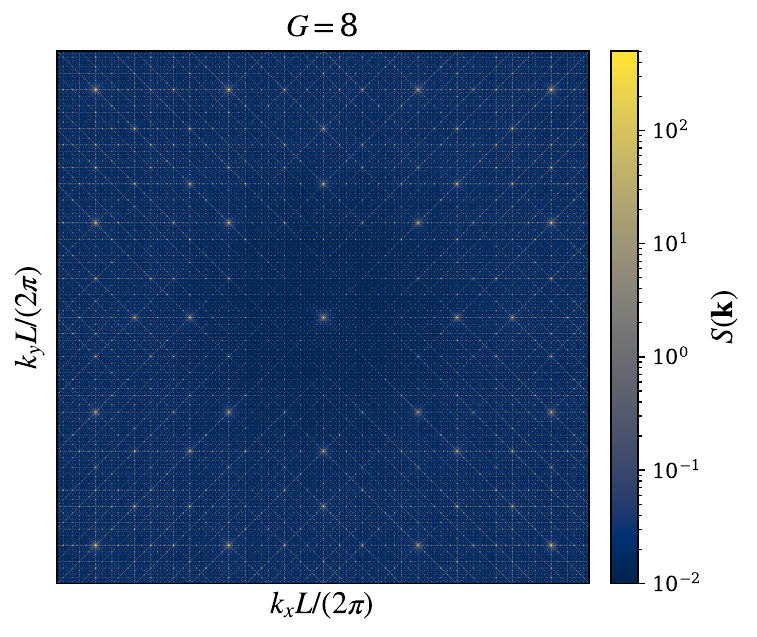}}%
\subfloat[\label{fig:2D_Sk_G0010}]{\includegraphics[width=0.33\textwidth]{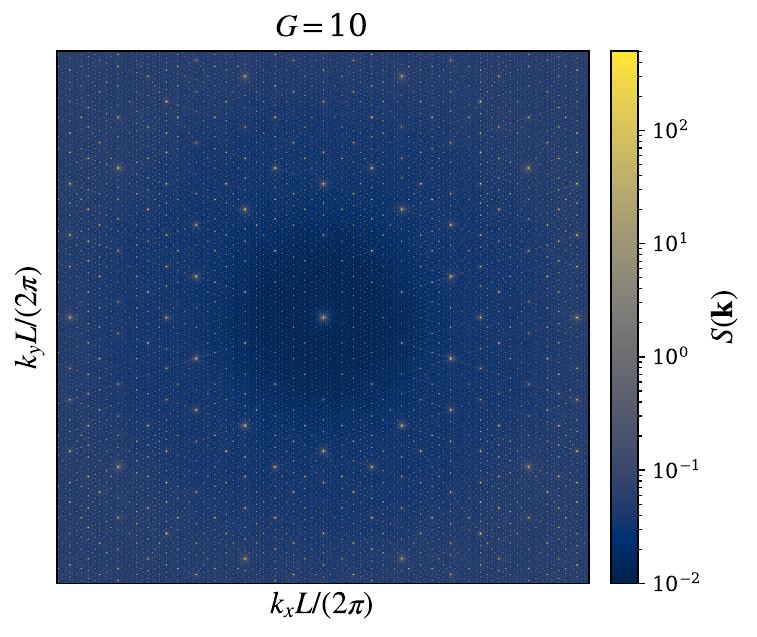}}%
\subfloat[\label{fig:2D_Sk_G0014}]{\includegraphics[width=0.33\textwidth]{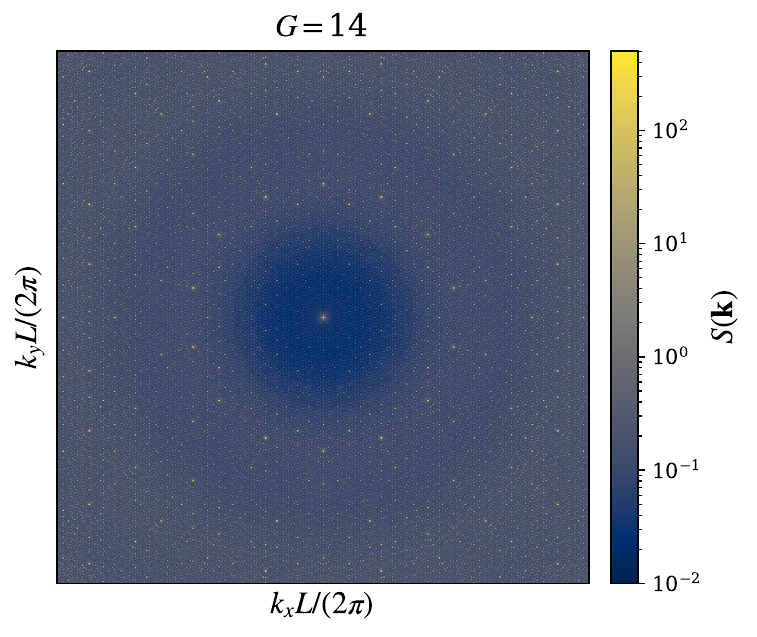}}\
\subfloat[\label{fig:2D_Sk_G0018}]{\includegraphics[width=0.33\textwidth]{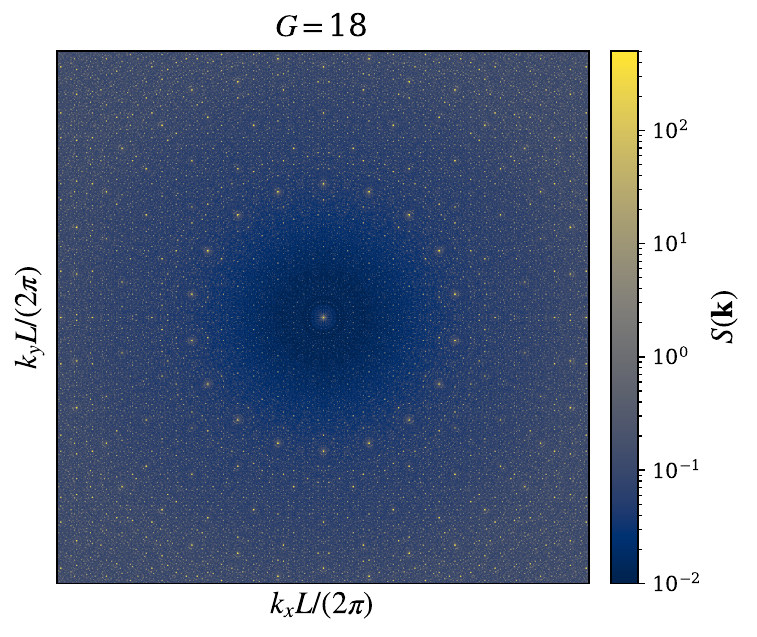}}%
\subfloat[\label{fig:2D_Sk_G0024}]{\includegraphics[width=0.33\textwidth]{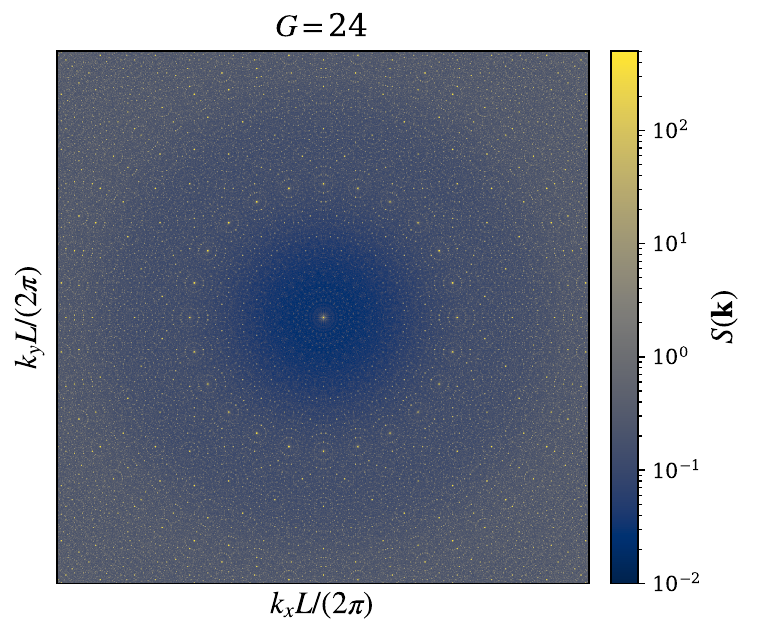}}%
\subfloat[\label{fig:2D_Sk_G0060}]{\includegraphics[width=0.33\textwidth]{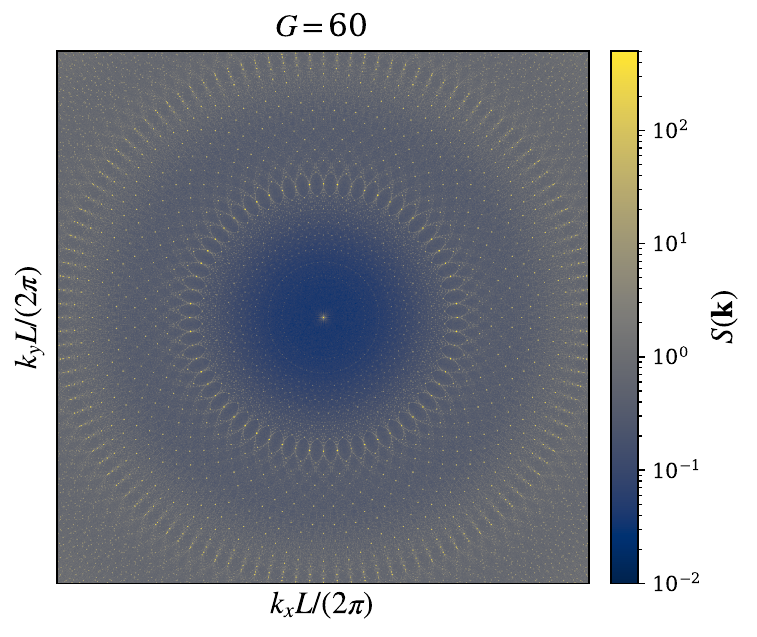}}\
\subfloat[\label{fig:2D_Sk_G0100}]{\includegraphics[width=0.33\textwidth]{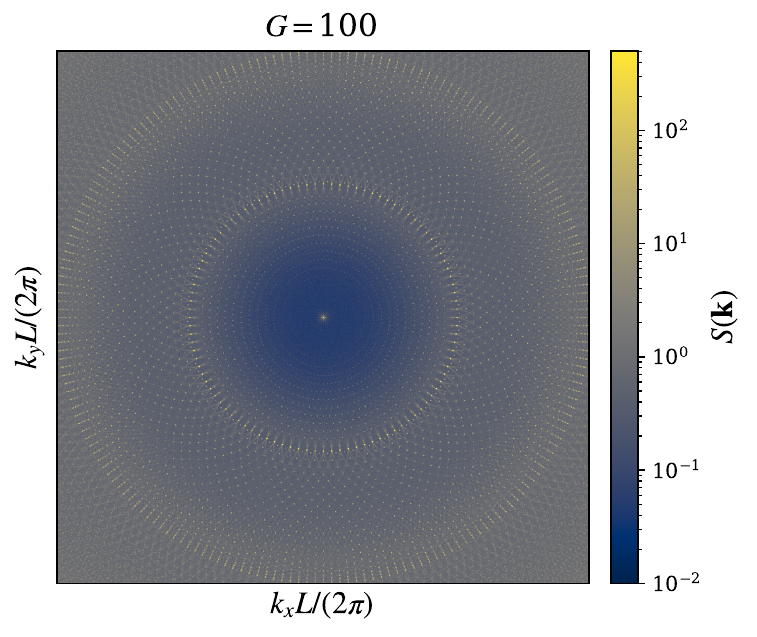}}%
\subfloat[\label{fig:2D_Sk_G0120}]{\includegraphics[width=0.33\textwidth]{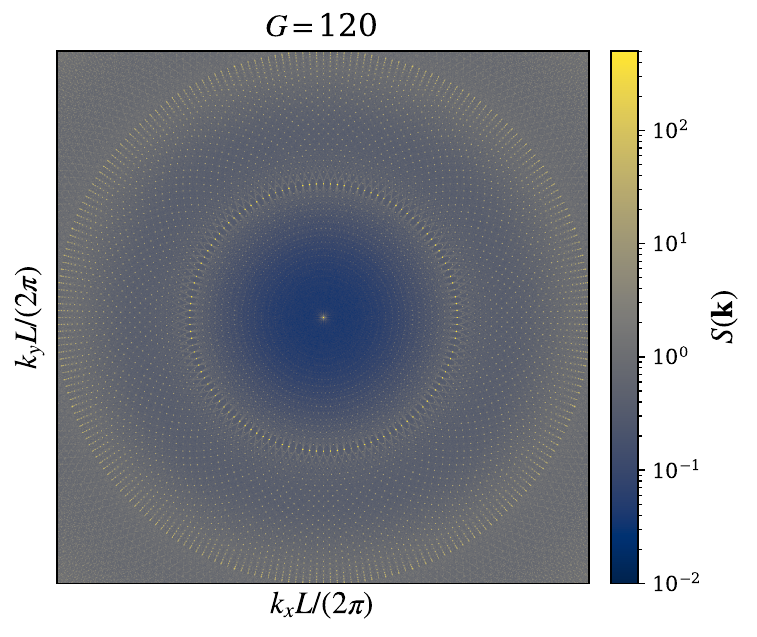}}%
\subfloat[\label{fig:2D_Sk_G0200}]{\includegraphics[width=0.33\textwidth]{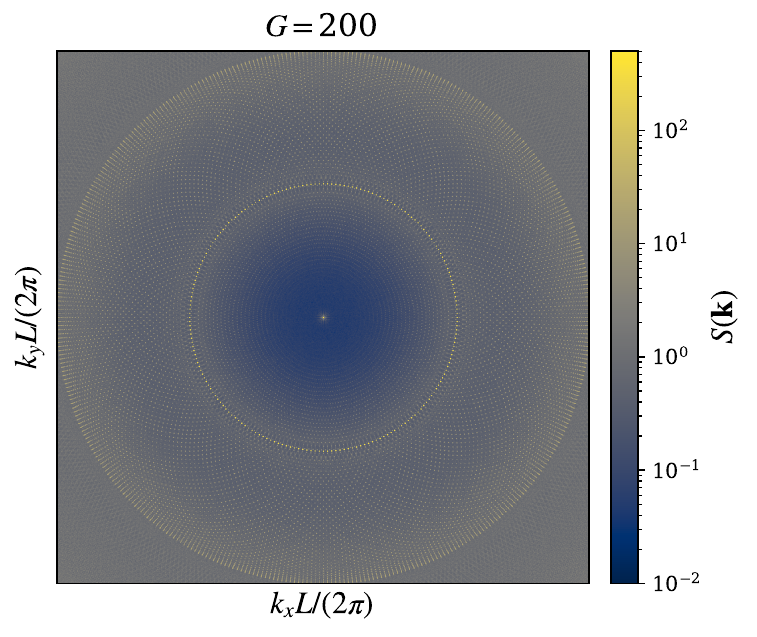}}\
\caption{Two-dimensional structure factors $S(\mathbf{k})$ of $N\sim 10^5$ gyromorphs with rotational orders $G=$ (a) 8, (b) 10, (c) 14, (d) 18, (e) 24, (f) 60, (g) 100, (h) 120, and (i) 200.
The plotted $S(\mathbf{k})$ are ensemble averages over 50 independent configurations and are computed from circular cutouts using a Hamming window, following Ref.~\cite{Ca25}.
The imposed $G$-fold ring of Bragg-like peaks is clearly visible for all $G$, and the ring becomes increasingly dense and effectively isotropic as $G$ increases.}
\label{figSI:2D_Sk}
\end{figure}

Figures~\ref{figSI:1D_Sk} and \ref{figSI:2D_Sk} show that the generated structures exhibit the defining reciprocal-space signature of gyromorphs: a ring of intense Bragg-like peaks at the imposed Fourier radius, together with weaker echoes at larger wave numbers.
The peak locations, peak heights, and increasing angular isotropy with increasing $G$ are in quantitative and qualitative agreement with the structure factors reported in Ref.~\cite{Ca25}.
Small visual differences are expected because our plots are ensemble averages over 50 configurations and because our structural-validation systems contain $N\sim 10^5$ points, roughly an order of magnitude larger than those used in the corresponding validation plots of Ref.~\cite{Ca25}. 

We emphasize that the structure factors shown in Figures~\ref{figSI:1D_Sk} and \ref{figSI:2D_Sk} cannot be used to analyze the hyperuniformity of gyromorphs. 
In particular, the angularly averaged structure factors shown below are computed from finite circular cutouts with a windowing function, which produces an effectively inhomogeneous weighted density field.
Hyperuniformity is not defined for such windowed, inhomogeneous configurations, and the apparent divergence of the plotted $S(k)$ near the origin should not be interpreted as a small-$k$ hyperuniformity diagnostic.
Our hyperuniformity classification in the main text is instead based on direct-space local number statistics computed from the unwindowed point configurations. 

\subsection{Pair Correlation Functions}

\begin{figure}
\centering
\subfloat[\label{fig:1D_g2_G0008}]{\includegraphics[width=0.33\textwidth]{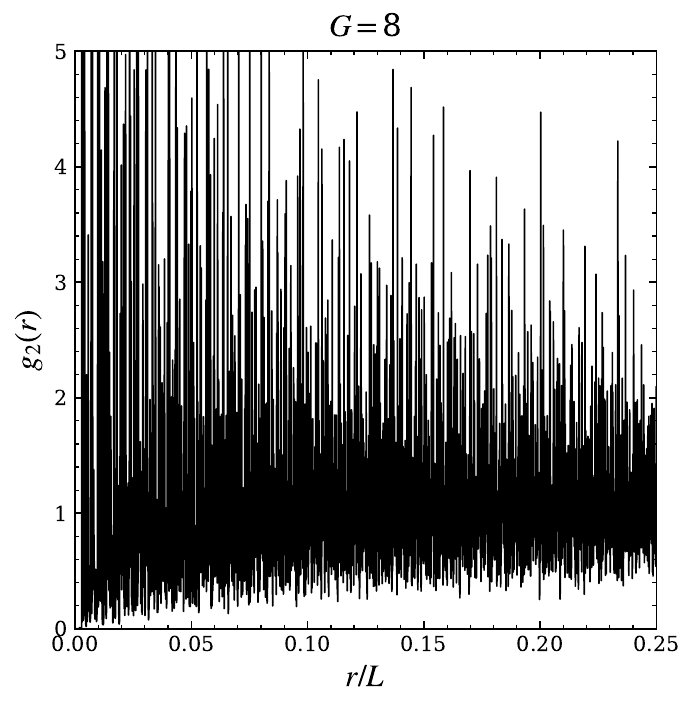}}%
\subfloat[\label{fig:1D_g2_G0010}]{\includegraphics[width=0.33\textwidth]{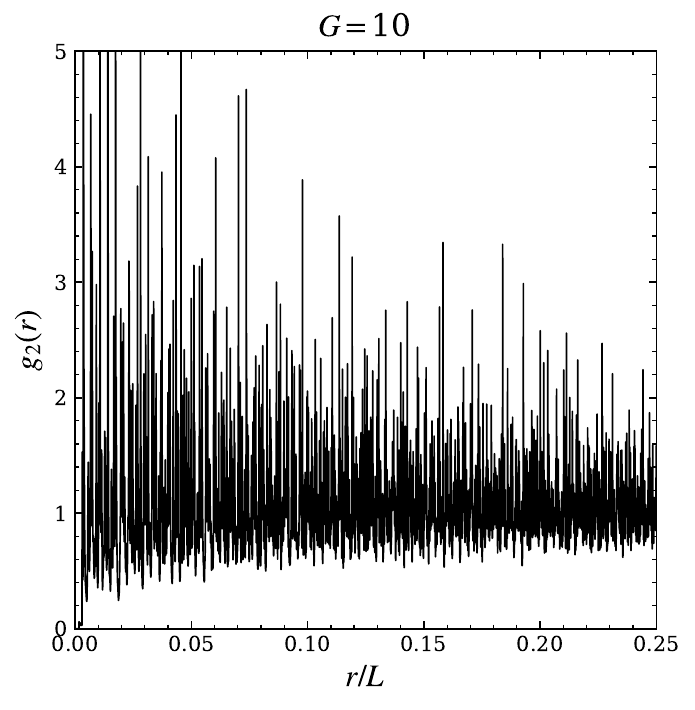}}%
\subfloat[\label{fig:1D_g2_G0014}]{\includegraphics[width=0.33\textwidth]{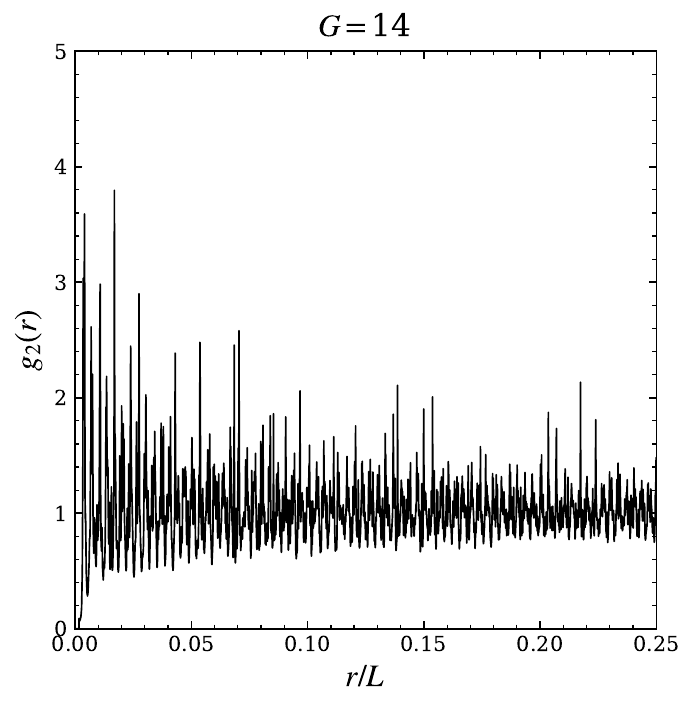}}\
\subfloat[\label{fig:1D_g2_G0018}]{\includegraphics[width=0.33\textwidth]{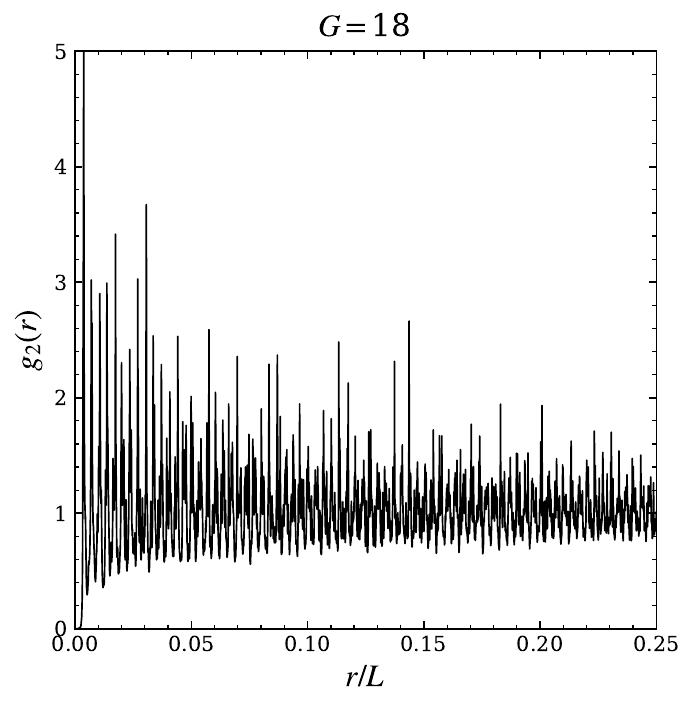}}%
\subfloat[\label{fig:1D_g2_G0024}]{\includegraphics[width=0.33\textwidth]{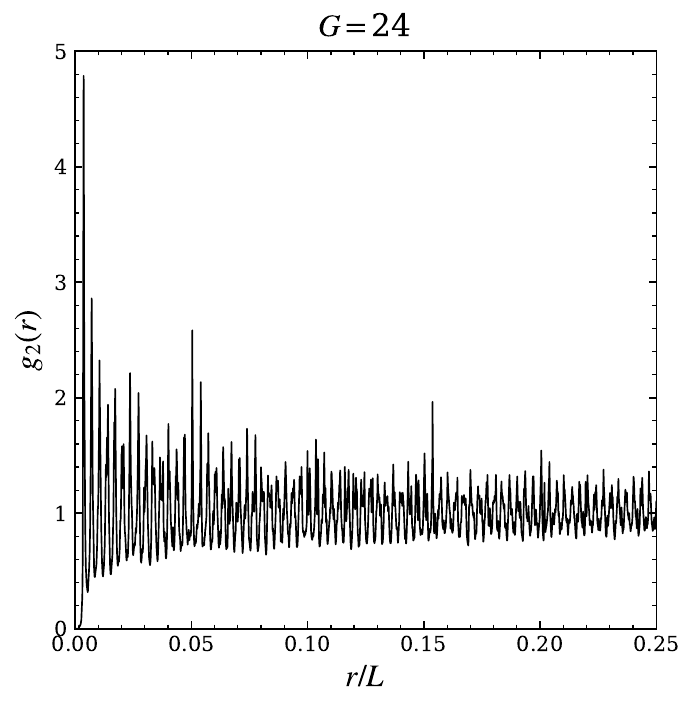}}%
\subfloat[\label{fig:1D_g2_G0060}]{\includegraphics[width=0.33\textwidth]{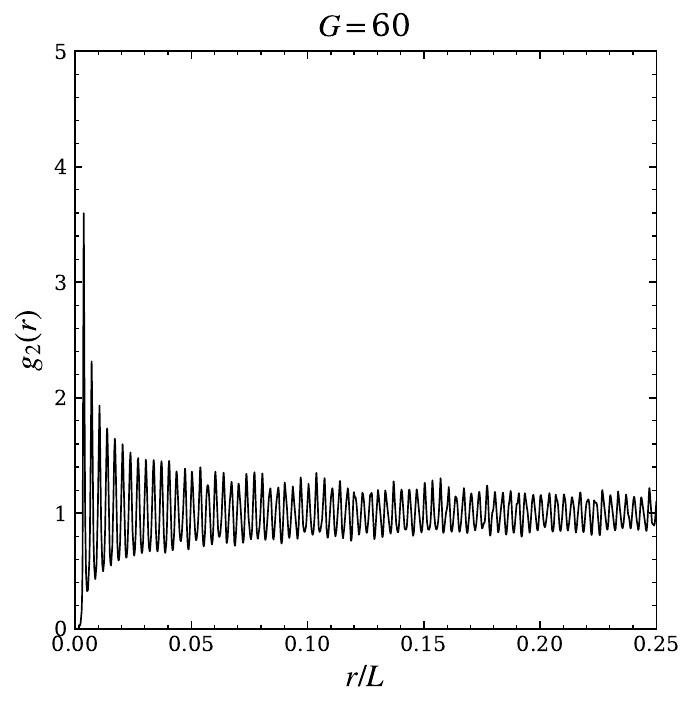}}\
\subfloat[\label{fig:1D_g2_G0100}]{\includegraphics[width=0.33\textwidth]{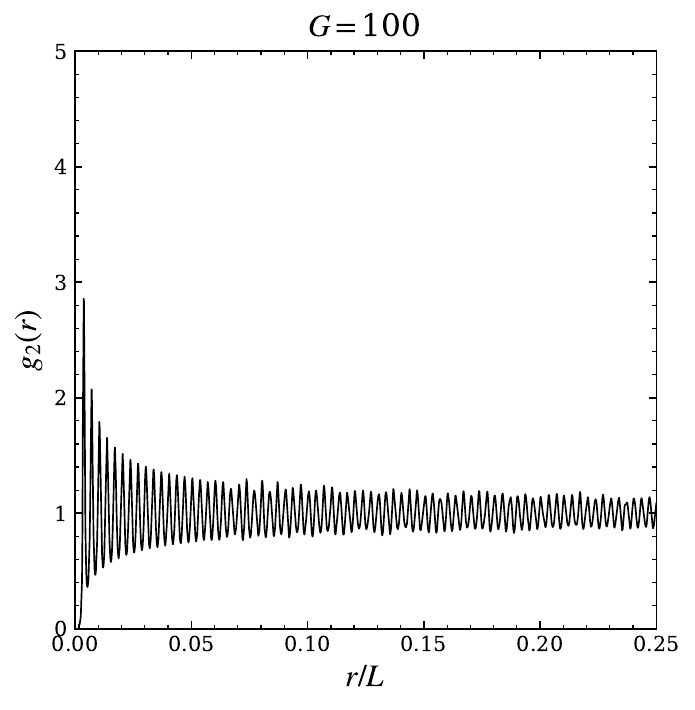}}%
\subfloat[\label{fig:1D_g2_G0120}]{\includegraphics[width=0.33\textwidth]{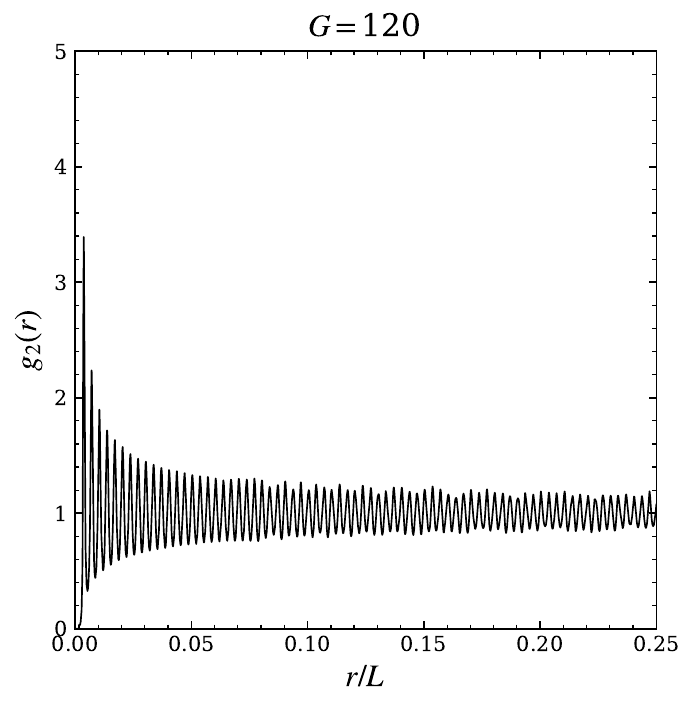}}%
\subfloat[\label{fig:1D_g2_G0200}]{\includegraphics[width=0.33\textwidth]{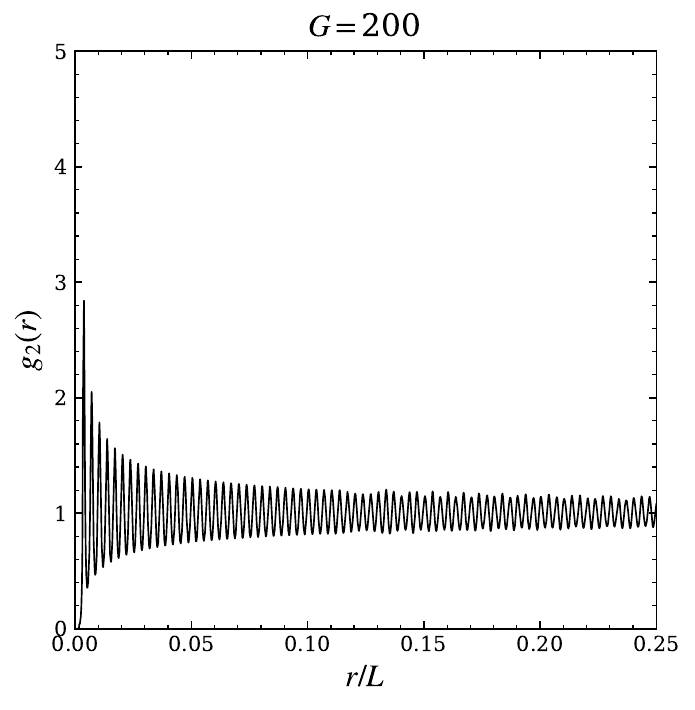}}\
\caption{Angularly averaged pair correlation functions $g_2(r)$ of $N\sim 10^5$ gyromorphs with rotational orders $G=$ (a) 8, (b) 10, (c) 14, (d) 18, (e) 24, (f) 60, (g) 100, (h) 120, and (i) 200.
The functions are computed directly in real space by binning pair separations, following the procedure of Ref.~\cite{Ca25}.
The short-distance exclusion region and long-lived oscillations in $g_2(r)$ are consistent with those seen in the gyromorph pair statistics reported in Ref.~\cite{Ca25}.}
\label{figSI:1D_g2}
\end{figure}

\begin{figure}
\centering
\subfloat[\label{fig:2D_g2_G0008}]{\includegraphics[width=0.33\textwidth]{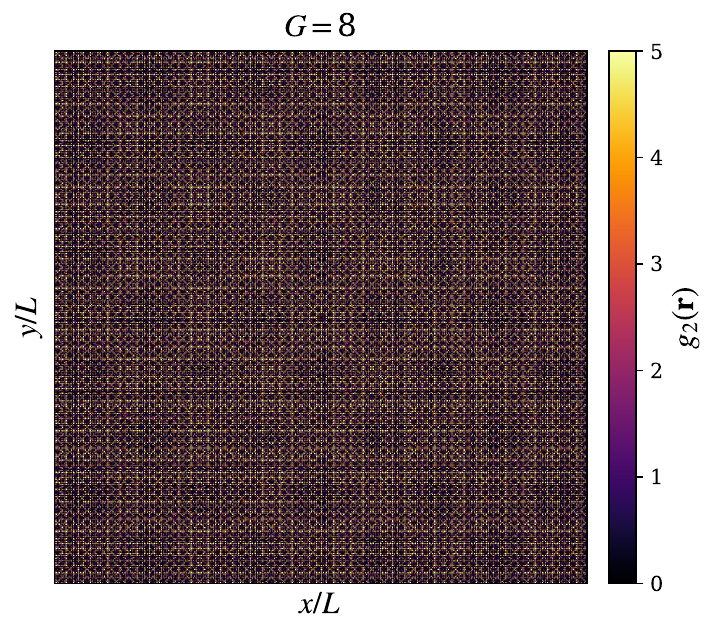}}%
\subfloat[\label{fig:2D_g2_G0010}]{\includegraphics[width=0.33\textwidth]{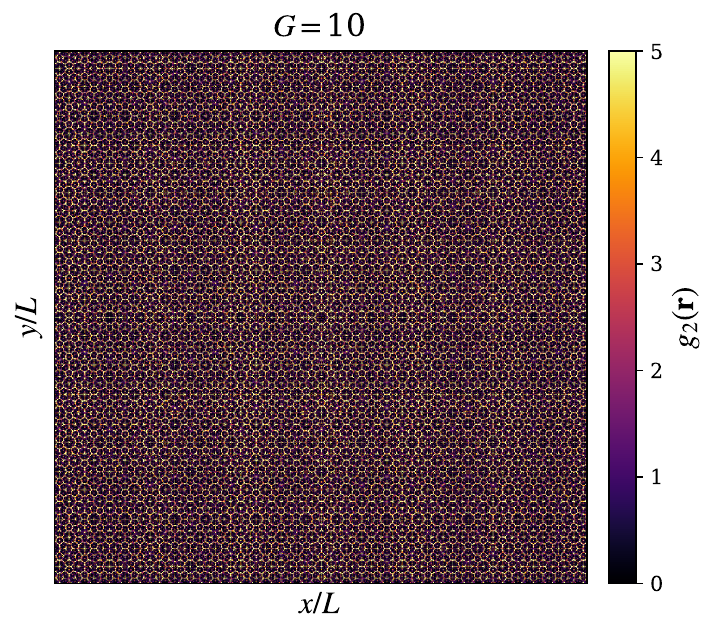}}%
\subfloat[\label{fig:2D_g2_G0014}]{\includegraphics[width=0.33\textwidth]{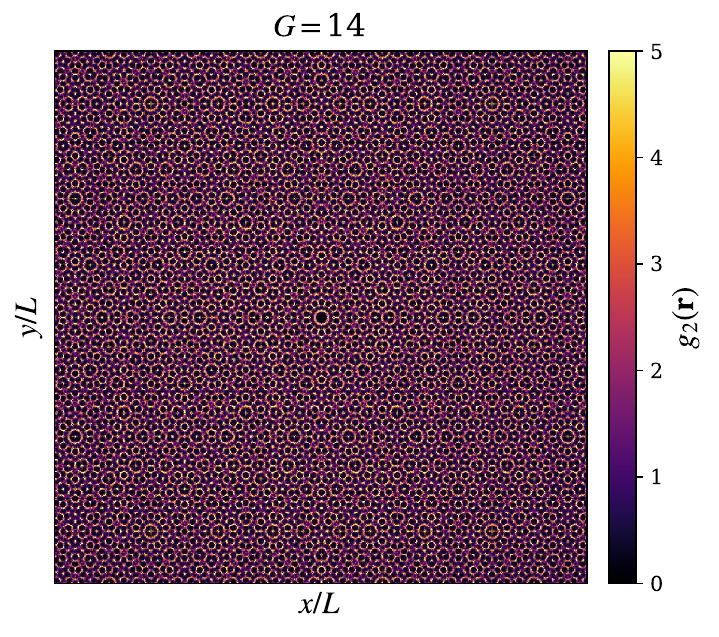}}\
\subfloat[\label{fig:2D_g2_G0018}]{\includegraphics[width=0.33\textwidth]{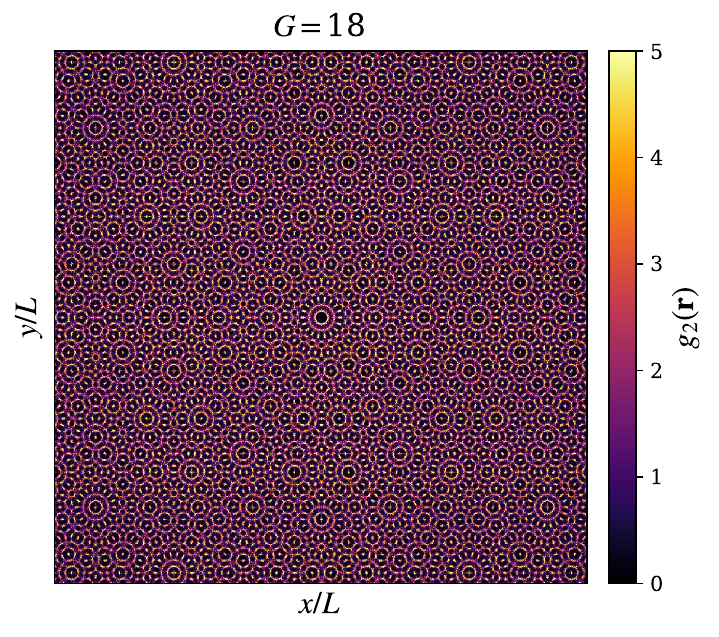}}%
\subfloat[\label{fig:2D_g2_G0024}]{\includegraphics[width=0.33\textwidth]{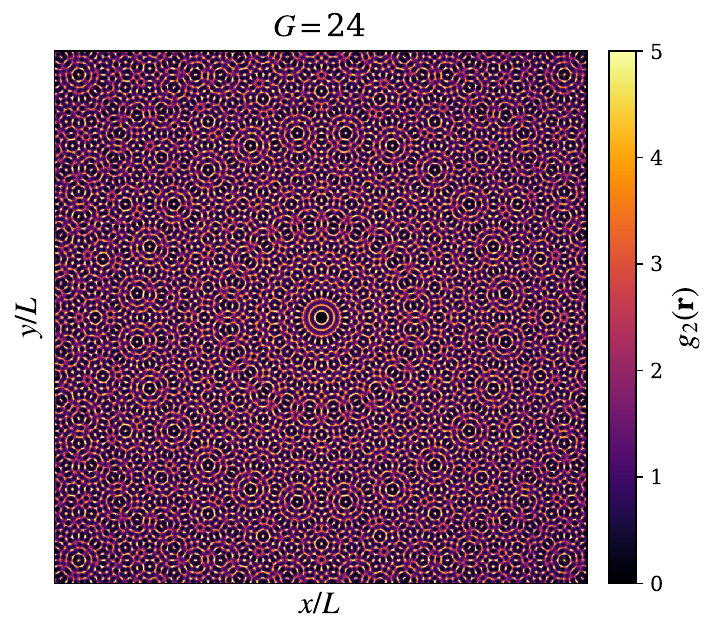}}%
\subfloat[\label{fig:2D_g2_G0060}]{\includegraphics[width=0.33\textwidth]{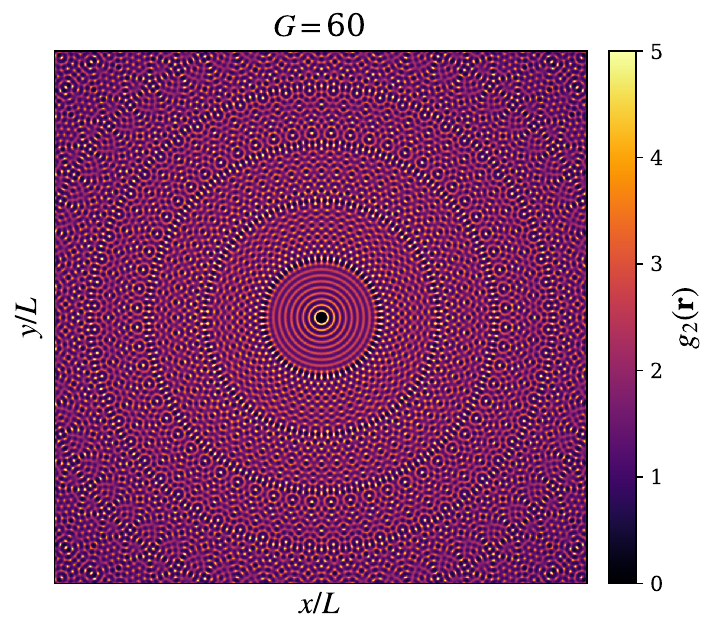}}\
\subfloat[\label{fig:2D_g2_G0100}]{\includegraphics[width=0.33\textwidth]{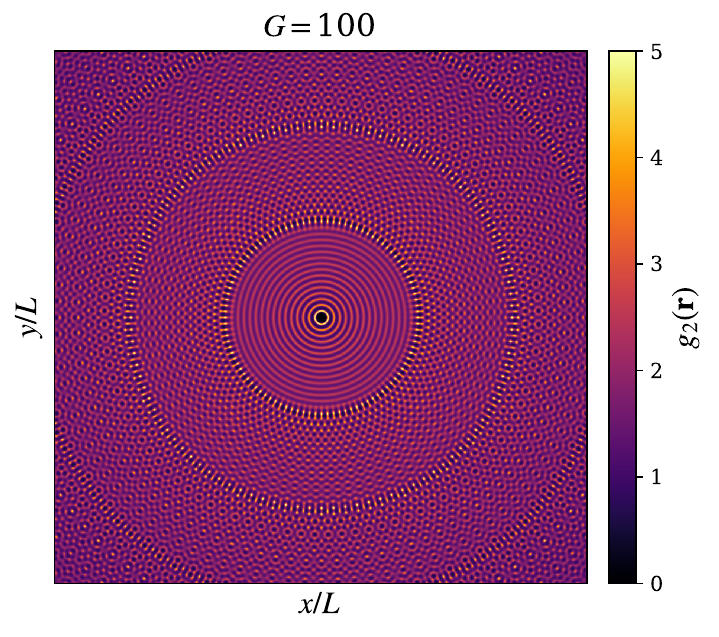}}%
\subfloat[\label{fig:2D_g2_G0120}]{\includegraphics[width=0.33\textwidth]{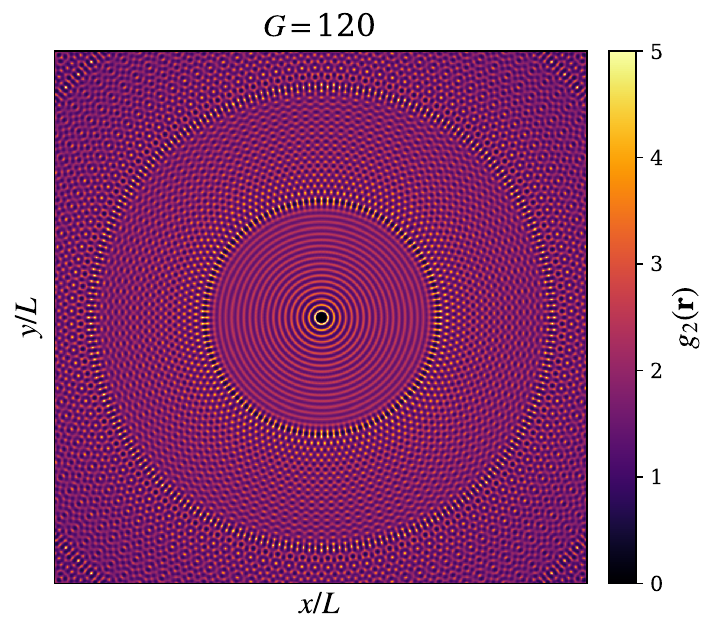}}%
\subfloat[\label{fig:2D_g2_G0200}]{\includegraphics[width=0.33\textwidth]{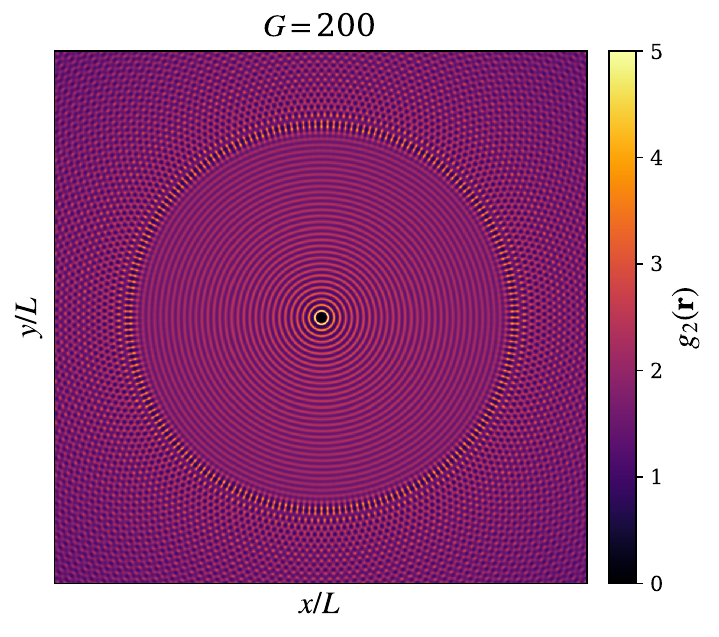}}\
\caption{Two-dimensional pair correlation functions $g_2(\mathbf{r})$ of $N\sim 10^5$ gyromorphs with rotational orders $G=$ (a) 8, (b) 10, (c) 14, (d) 18, (e) 24, (f) 60, (g) 100, (h) 120, and (i) 200.
For each $G$, the plotted $g_2(\mathbf{r})$ is averaged over 50 independent configurations.
The gear-like angular correlations and their evolution with $G$ are consistent with those observed in the corresponding $g_2(\mathbf{r})$ plots reported in Ref.~\cite{Ca25}.}
\label{figSI:2D_g2}
\end{figure}

Figures~\ref{figSI:1D_g2} and \ref{figSI:2D_g2} show the corresponding pair correlation functions.
The real-space correlations reproduce the characteristic gyromorph structure reported in Ref.~\cite{Ca25}: a short-distance exclusion region, oscillatory radial correlations, and an emergent gear-like pattern whose angular order reflects the imposed value of $G$.

\subsection{Gyromorphic Correlation Functions}

\begin{figure}
\centering
\includegraphics[width=0.5\linewidth]{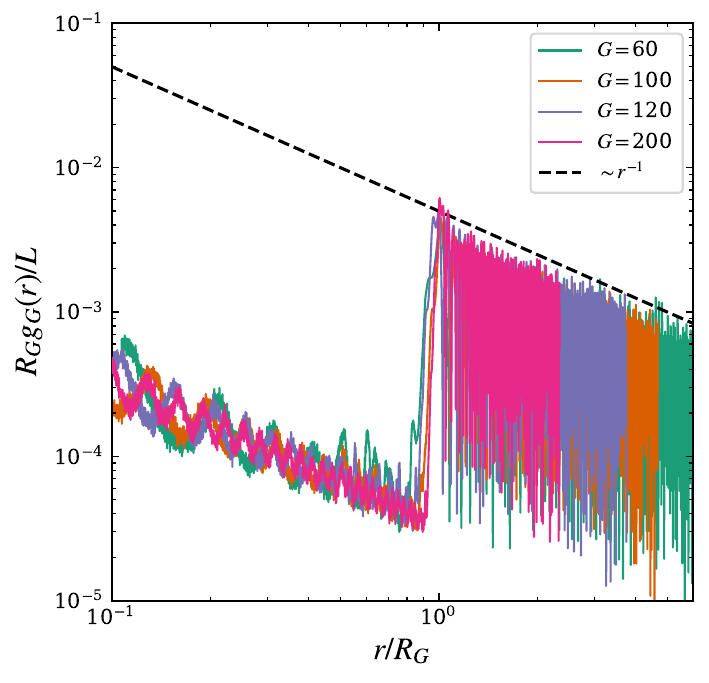}
\caption{Gyromorphic correlation function $g_G(r)$ for $G=60$, 100, 120, and 200 gyromorphs with $KL/(2\pi)=300$.
The data are rescaled as $R_G g_G(r)/L$ and plotted against $r/R_G$, where $R_G$ is the characteristic radius at which $G$-fold gyromorphic order first appears.
The dashed line indicates $r^{-1}$ decay.
The observed collapse and approximate $1/r$ scaling are consistent with the gyromorphic correlations reported in Ref.~\cite{Ca25}.}
\label{fig:gG}
\end{figure}

To quantify the imposed rotational order more directly, we compute the gyromorphic correlation function introduced in Ref.~\cite{Ca25},
\begin{equation}
g_G(r)\equiv \left| \frac{L^d}{N^2} \frac{1}{2\pi} \sum_{p\neq q} e^{iG\theta_{pq}}\delta(r-r_{pq}) \right| ,
\label{eqn:gG}
\end{equation}
where $r_{pq}$ and $\theta_{pq}$ are the polar coordinates of the separation vector between particles $p$ and $q$.
As shown in Fig.~\ref{fig:gG}, the rescaled gyromorphic correlations decay approximately as $r^{-1}$ for the large-$G$ gyromorphs, in agreement with Ref.~\cite{Ca25}.
This confirms that our configurations reproduce the quasi-long-range rotational order that distinguishes gyromorphs from ordinary disordered point patterns.

\subsection{Coupled-Dipoles Calculations}

The structural diagnostics above validate the generated gyromorphs at the two-point level through $S(k)$ and $g_2(r)$, and at the orientational-correlation level through $g_G(r)$.
As a further validation, we reproduce the coupled-dipoles calculations of Ref.~\cite{Ca25} using the MAGreeTe code~\cite{MAGREETE} with the same gyromorph parameters, $G=60$, $KL/(2\pi)=100$, $N\sim10^4$, filling fraction $\phi=0.05$, and refractive index $n=3$.
Unlike $S(k)$ and $g_2(r)$, the coupled-dipoles calculation accounts for all orders of multiple scattering among the particles and therefore probes structural information beyond two-point correlations.
The agreement of the resulting relative DOS changes $\delta\rho$ with those reported in Ref.~\cite{Ca25} provides a stringent validation that our generated point patterns are genuine gyromorphs.

\begin{figure}
\centering
\subfloat[\label{fig:CD_TM_0}]{\includegraphics[width=0.25\textwidth]{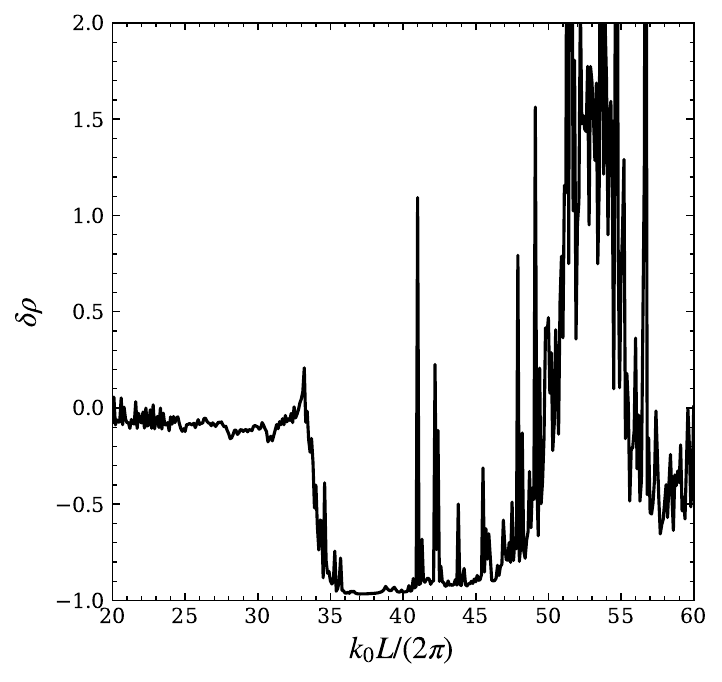}}%
\subfloat[\label{fig:CD_TM_1}]{\includegraphics[width=0.25\textwidth]{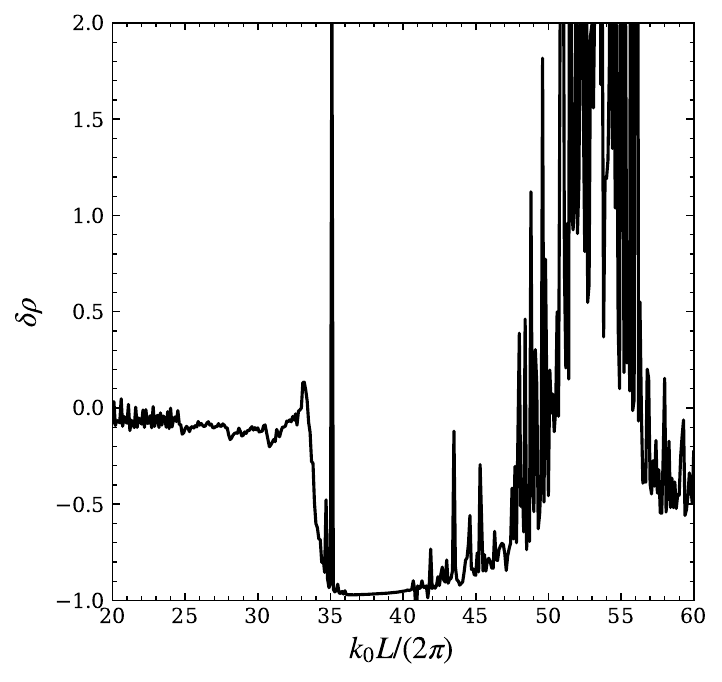}}%
\subfloat[\label{fig:CD_TM_2}]{\includegraphics[width=0.25\textwidth]{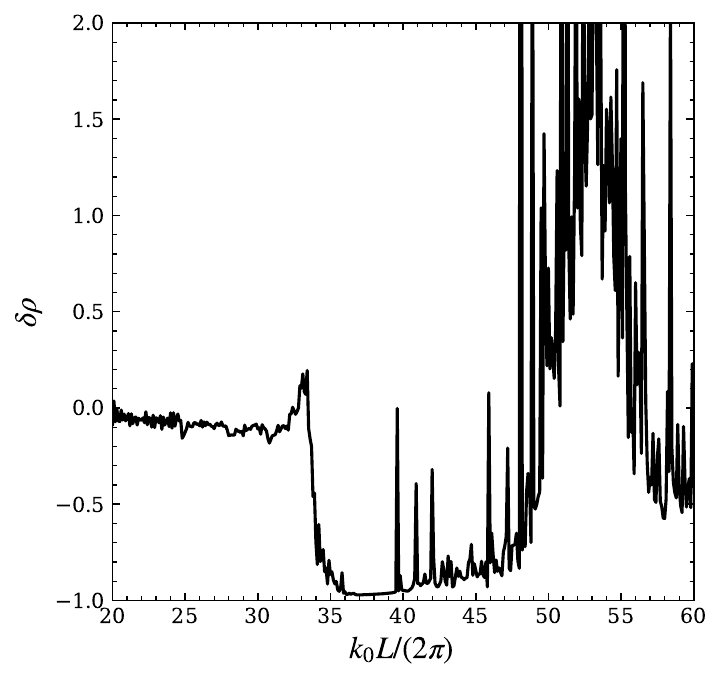}}%
\subfloat[\label{fig:CD_TM_A}]{\includegraphics[width=0.25\textwidth]{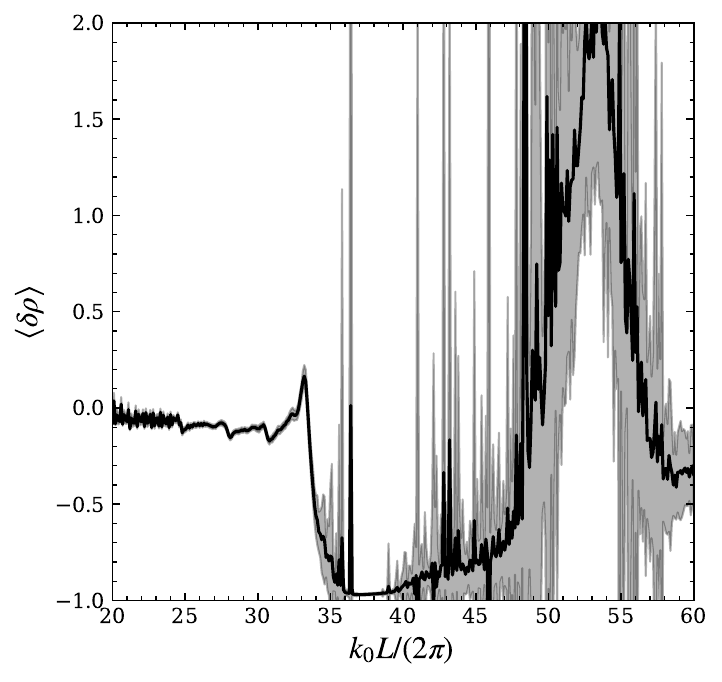}}\
\subfloat[\label{fig:CD_TE_0}]{\includegraphics[width=0.25\textwidth]{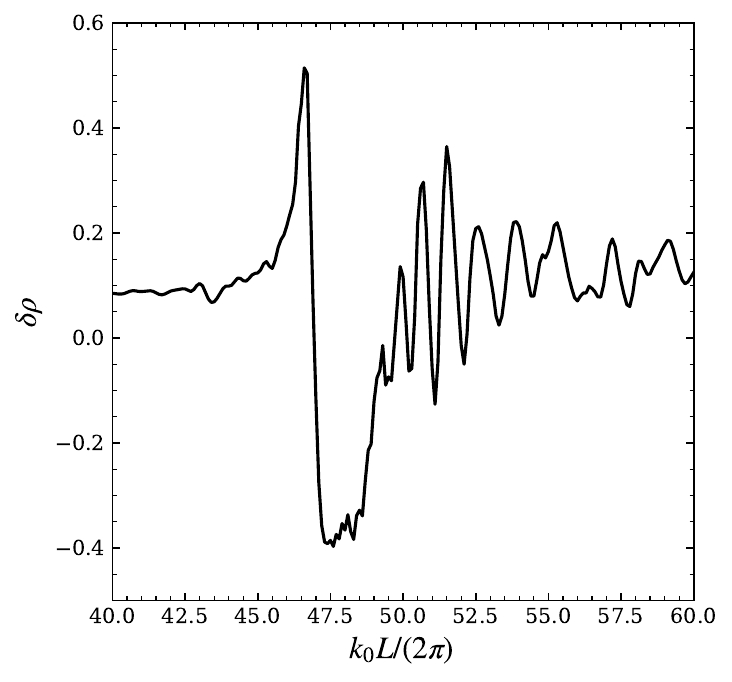}}%
\subfloat[\label{fig:CD_TE_1}]{\includegraphics[width=0.25\textwidth]{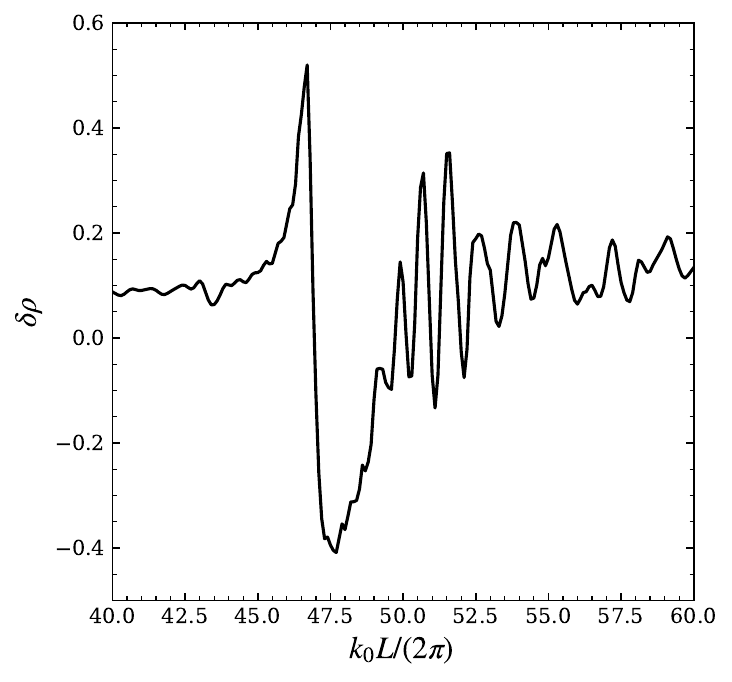}}%
\subfloat[\label{fig:CD_TE_2}]{\includegraphics[width=0.25\textwidth]{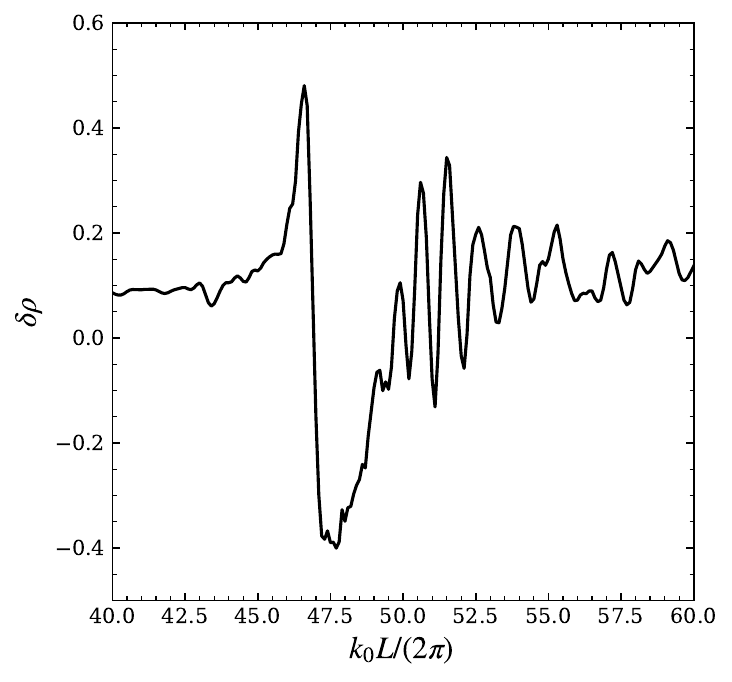}}%
\subfloat[\label{fig:CD_TE_A}]{\includegraphics[width=0.25\textwidth]{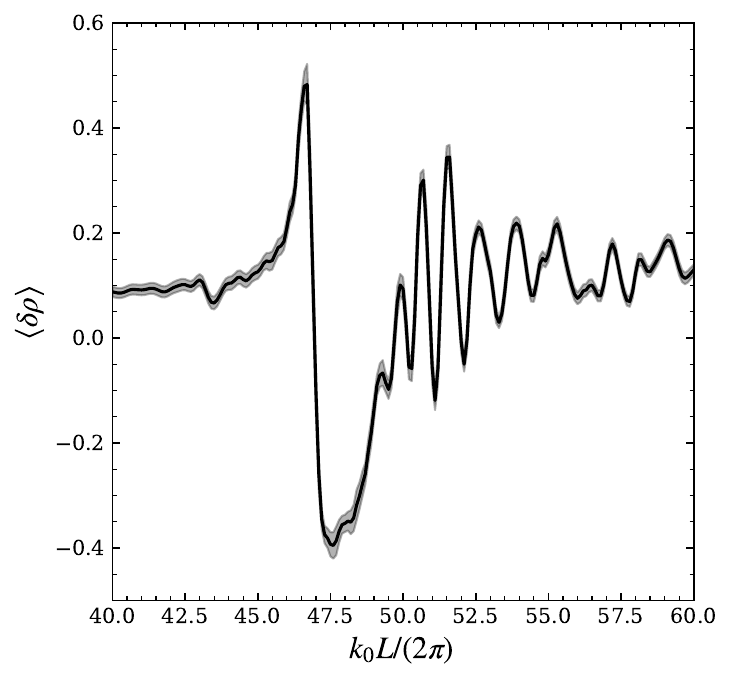}}\
\caption{Coupled-dipoles validation calculations for $G=60$ gyromorphs with $KL/(2\pi)=100$ and $N\sim\mathcal{O}(10^4)$ prior to taking the circular cutout.
Panels (a)--(c) show representative individual scalar/TM calculations and panel (d) shows the corresponding ensemble-averaged result.
Panels (e)--(g) show representative individual vector/TE calculations and panel (h) shows the corresponding ensemble-averaged result.
The plotted quantity is the relative DOS change $\delta\rho$ computed using MAGreeTe with the same optical parameters as Ref.~\cite{Ca25}. 
In panels (d) and (h), the gray shaded regions represent the ensemble standard deviation of $\delta\rho$. 
Both the individual-configuration fluctuations and the ensemble-averaged depletion features are consistent with the coupled-dipoles results reported in Ref.~\cite{Ca25}.}
\label{fig:CD}
\end{figure}
\color{black}


 \newcommand{\noop}[1]{}